 \documentclass[final,5p,times,twocolumn,authoryear]{elsarticle}
 
\usepackage{natbib}

\usepackage{rotating}
\usepackage{lscape}
\usepackage{amssymb}

\usepackage[round]{natbib}

\journal{Molecular Astrophysics}

\begin{document}

\begin{frontmatter}



\title{H$_2$ formation on interstellar dust grains: the viewpoints of theory, experiments, models and observations}


\author[1]{Valentine Wakelam} 
\author[2]{Emeric Bron}
\author[3]{Stephanie Cazaux}
\author[4]{Francois Dulieu}
\author[5]{C\'ecile Gry}
\author[6]{Pierre Guillard}
\author[7]{Emilie Habart}
\author[8]{Liv Hornek{\ae}r} 
\author[9]{Sabine Morisset}
\author[10]{Gunnar Nyman}
\author[11]{Valerio Pirronello}
\author[12]{Stephen D. Price}
\author[13]{Valeska Valdivia}
\author[14]{Gianfranco Vidali}
\author[15]{Naoki Watanabe}

\address[1]{Laboratoire d'astrophysique de Bordeaux, Univ. Bordeaux, CNRS, B18N, all\'ee Geoffroy Saint-Hilaire, 33615 Pessac, France}
\address[2]{Instituto de Ciencias de Materiales de Madrid (CSIC), 28049, Madrid, Spain\\LERMA, Obs. de Paris, PSL Research University, CNRS, Sorbonne Universit\'es, UPMC Univ. Paris 06, ENS, F-75005, France}
\address[3]{Faculty of Aerospace Engineering, Delft University of Technology, Delft, Netherlands\\Leiden Observatory, Leiden University, P.O. Box 9513, NL 2300 RA Leiden, The Netherlands}
\address[4]{LERMA, Universit\'e de Cergy Pontoise, Sorbonne Universit\'es, UPMC Univ. Paris 6, PSL Research University, Observatoire de Paris,UMR 8112 CNRS,    5 mail Gay Lussac 95000 Cergy Pontoise, France}
 \address[5]{Aix Marseille Univ, CNRS, LAM, Laboratoire d'Astrophysique de Marseille, Marseille, France}
 \address[6]{Sorbonne Universit\'es, UPMC Univ. Paris 6 \& CNRS, UMR 7095, Institut d'Astrophysique de Paris, 98 bis bd Arago, 75014 Paris, France}
 \address[7]{Institut d'Astrophysique Spatiale, Univ. Paris-Sud \& CNRS, Univ. Paris-Saclay - IAS, b\^atiment 121, univ Paris-Sud, 91405 Orsay, France}
\address[8]{Dept. Physics and Astronomy, Aarhus University, Ny Munkegade 120, 8000 Aarhus C, Denmark}
 \address[9]{Institut des Sciences Mol\'eculaires d'Orsay, ISMO, CNRS, Universit\'e Paris-Sud, Universit\'e Paris Saclay, F-91405 Orsay, France}
 \address[10]{Department of Chemistry and Molecular Biology, University of Gothenburg, SE 412 96 Gothenburg, Sweden}
 \address[11]{Dipartimento di Fisica e Astronomia, Universit\'a di Catania, Via S. Sofia 64, 95123 Catania, Sicily, Italy}
 \address[12]{Chemistry Department, University College London, 20 Gordon Street, London WC1H 0AJ UK}
 \address[13]{Laboratoire AIM, Paris-Saclay, CEA/IRFU/DAp - CNRS - Universit\'e Paris Diderot, 91191, Gif-sur-Yvette Cedex, France}
\address[14]{Syracuse University, 201 Physics Bldg., Syracuse, NY 13244 (USA)}
\address[15]{Institute of Low Temperature Science, Hokkaido University, Sapporo, Hokkaido 060-0819, Japan}

\begin{abstract}
 Molecular hydrogen is the most abundant molecule in the universe. It is the first one to form and survive photo-dissociation in tenuous environments. Its formation involves catalytic reactions on the surface of interstellar grains. The micro-physics of the formation process has been investigated intensively in the last 20 years, in parallel of new astrophysical observational and modeling progresses. In the perspectives of the probable revolution brought by the future satellite JWST, this article has been written to present what we think we know about the H$_2$ formation in a variety of interstellar environments.
\end{abstract}

\end{frontmatter}


\section{Introduction}\label{intro}

Molecular hydrogen is, by a few orders of magnitude, the most abundant molecule in the Universe. The first detection  of this molecule in the interstellar medium (ISM) was obtained via a rocket flight in 1970 \citep{Carruthers1970}, three decades after the first interstellar detection of  CH, CH$^+$ and CN  \citep[see][and references therein]{2006ARA&A..44..367S}. Since H$_2$ is a symmetric and homonuclear diatomic molecule, electric dipole driven ro-vibrational  transitions are forbidden and only weak electric-quadrupole transitions are allowed, making its detection extremely difficult in emission\footnote{H$_2$ is however easily detected in absorption in the far-UV electronic bands, provided a far-UV spectrum of a background target is available}, unless the emission is from energized environments such as those with, for example, high temperature or high luminosity. 


In diffuse molecular clouds, which are regions characterized by molecular fractions  $f_\mathrm{H_2} = 2n_\mathrm{H_2}/n_\mathrm{H}> 0.1$ ($n_\mathrm{H_2}$ being the number density of H$_2$ molecules and $n_\mathrm{H}$ the total proton number density), the first molecule to form is H$_2$ \citep{2006ARA&A..44..367S}. In Photo-Dissociation Regions (PDRs), which are predominantly neutral regions bathed in far ultraviolet light, the emission  of H$_2$ is a tracer of the physical conditions of the cloud edge \citep{Hollenbach99}. In such environments H$_2$  can be dissociated by ultraviolet radiation, and therefore an efficient route for molecular formation  must be present \citep{1974ApJ...191..375J,jura75}. Furthermore, molecular hydrogen, either in its neutral or ionized form,  controls much of the chemistry in the ISM. 
In dense clouds where UV penetration is greatly reduced, most of the hydrogen is in molecular form, and most of the Universe's molecular hydrogen resides in these dense clouds.


It has been recognized   for a long time that under ISM conditions  H$_2$  cannot be formed efficiently enough in the gas-phase to explain its abundance. 
 Indeed, even in the 40's \citet{vandehulst1949} had proposed his dirty ice model of dust, where molecules form by combination of atoms on the surface.
The link between the presence of H$_2$ and dust was noted a long time ago \citep{Hollenbach:1971b}. Indeed, it is now well established that H$_2$ formation occurs via catalytic reactions on surfaces of interstellar dust grains. 

The aim of this paper is to provide the current status of the understanding of the formation of H$_2$ on interstellar dust grains and identify the important questions that still remain to be answered in this field. This account is motivated by the new observational possibilities that the James Webb Space Telescope (JWST) should provide. In addition, over the last ten years great progress in the modeling of astrophysical media, as well as in the understanding of the associated molecular physics, has been made. Sometimes this progress is directly linked to specific experiments \cite[e.g.][]{Pirronello:1997a, Pirronello:1997b, 2006JChPh.124k4701C, 2010ApJ...714L.233W} or calculations and simulations \citep[e.g.][]{Katz:1999, Cuppen2010, Cazaux2011};  at other times progress results from an intrinsic change in the treatment of one specific aspect of the formation process, such as stochastic effects \cite[e.g.]{Green:2001,Biham:2001}. Given the nature of this progress, earlier works in the literature and values used in models can rapidly become outdated, leading to potentially significant differences in the predictions of models if the most up-to-date values are not used.  Given this issue, this paper presents, in a unified account, the current viewpoint regarding the formation of molecular hydrogen on interstellar dust grains from the perspective of observers, modelers and chemical physicists. To this end, a group of specialists from these three disciplines gathered for 3 days in Arcachon (France) in June 2016.  This paper is the result of this meeting and aims to present the ``state of the art'' in characterizing and understanding interstellar H$_2$ formation.

The paper is organized as follows:
Section~\ref{state_of_the_art}  gives an overview of the properties of H$_2$ and the challenges involved in observing H$_2$ in space. Section~\ref{state_of_the_art} also presents a summary of theoretical and laboratory work aimed at understanding the processes involved in H$_2$ formation on dust grain analogs (silicates, carbonaceous materials and ices): sticking, diffusion, reaction, desorption and energy the partitioning of the nascent H$_2$ as it leaves the surface. Several astrophysical models used to study the chemistry of H$_2$ in various environments are also briefly described in this section. In section 3, we provide a list of values for the physico-chemical quantities necessary to describe the sticking, diffusion and reactivity of H$_2$ that can be used in astrochemical models. Section 4 gives an in-depth view of the formation of H$_2$ in different interstellar environments. A summary and a set of conclusions is then provided at the end of the paper.

\section{State of the art}\label{state_of_the_art}
\subsection{Methods and tools to observe H$_2$ in the Universe }

\subsubsection{Properties of the H$_2$ molecule}

Containing two identical hydrogen atoms linked by a covalent bond, the hydrogen molecule is homonuclear and thus highly symmetric. Due to this symmetry, the molecule has no permanent dipole moment and so all the observed ro-vibrational transitions are forbidden electric quadrupole transitions ($\Delta J=\pm 2$) with low values of the spontaneous emission coefficient ($A$). Since H$_2$ is the lightest possible molecule it has a low moment of inertia, and hence a large rotational constant ($B / k_{\rm B} = 85.25$~K), leading to widely spaced energy levels even when the rotational quantum number $J$ is small. 
In addition, there are no radiative transitions between ortho-H$_2$ (spins of H nuclei parallel, odd $J$) and para-H$_2$ (spins antiparallel, even $J$), so the ortho and para molecules constitute two almost independent states of H$_2$. The first accessible rotational transition is therefore  $J=2 \rightarrow 0$, which has an associated energy of $\Delta E/k_\mathrm{B}\sim$510~K. Even so, the lowest excited rotational levels of molecular hydrogen are not easily populated, making H$_2$ one of the most difficult molecules to detect in space via emission. In absorption, the situation is different since Lyman ($B^1 \Sigma ^1 _u$) and Werner ($C^1 \Sigma _u$) electronic bands in the far-UV (from 912 \AA\ to 1155\AA) provide a very sensitive tool to detect even very diffuse H$_2$, down to column densities as low as a few 10$^{12}$ cm$^{-2}$ -- provided a space-born far-UV spectroscopic facility, as well as a UV-bright background source, are available.

\subsubsection{Excitation mechanisms}
H$_2$ may be excited via several mechanisms as described below. 
The relative population of the H$_2$ levels depends on the exciting sources and the physical conditions of the gas.  

\begin{description}
\item[-] Inelastic collisions:
If the gas density and temperature are high enough, inelastic collisions with H$^0$, He, H$_2$ and e$^-$ can be the dominant excitation mechanism, at least for the lower rotational energy levels \citep[e.g.][]{LeBourlot1999}. 
\item[-] Radiative pumping:
In the presence of far-ultraviolet radiation (FUV, $\lambda > 912$ \AA), the molecule is radiatively pumped into its electronically excited states.  
As it decays back into the electronic ground
state, it populates the high vibrational levels,
and the subsequent cascade to $v=0$ gives rise to
a characteristic distribution of level populations
and fluorescent
emission in the visible and infrared (IR) regions of the spectrum 
\citep[e.g.][]{Black1987, Sternberg1989}. This excitation mechanism is observed in PDRs where
it is the dominant pathway for excitation of ro-vibrational and high rotational levels. UV pumping could also contribute significantly to the excitation of the pure rotational 0-0 S(2)-S(5) lines, since their upper states ($v$=0, $J$=4-7) are relatively high in energy and their critical densities are high even at moderate temperatures ($n_\mathrm{crit}\ge 10^4$~cm$^{-3}$ for $T\le 500~\mathrm{K}$).
\item[-] By formation:
The internal energy of the nascent H$_2$ can also specifically affect the level populations. However, 
of all the UV photons absorbed by H$_2$ only 10 to 15\% lead to dissociation. Then, for an equilibrium between photo-dissociation and formation, the ratio of the rates of formation pumping and fluorescent pumping of the high-excitation levels in the electronic ground state is $\sim 15/85$. Fluorescent pumping should therefore dominate over formation pumping by a factor five. Thus, unless the level distribution of newly formed H$_2$ is strongly concentrated toward a small number of high energy levels, the H$_2$ formation excitation will not specifically affect the H$_2$ spectrum (see e.g. \citealt{Black1987,LeBourlot:1995b} for models and e.g. \citealt{Burton:2002} for possible observational signatures).
\item[-] X-ray photons and cosmic rays:
 In X-ray emitting environments (such as active galactic nuclei or 
young stellar objects), X-rays
which are capable of penetrating deeply into zones opaque to UV photons, can influence the  excitation of H$_2$ \citep[e.g.][]{Maloney1996, Tine1997}. H$_2$ excitation may also occur by collisions with secondary electrons generated by cosmic ray ionization. 

\end{description}

\subsubsection{H$_2$ excitation diagrams: what information can we get?}


\begin{figure}
\includegraphics[scale=0.4]{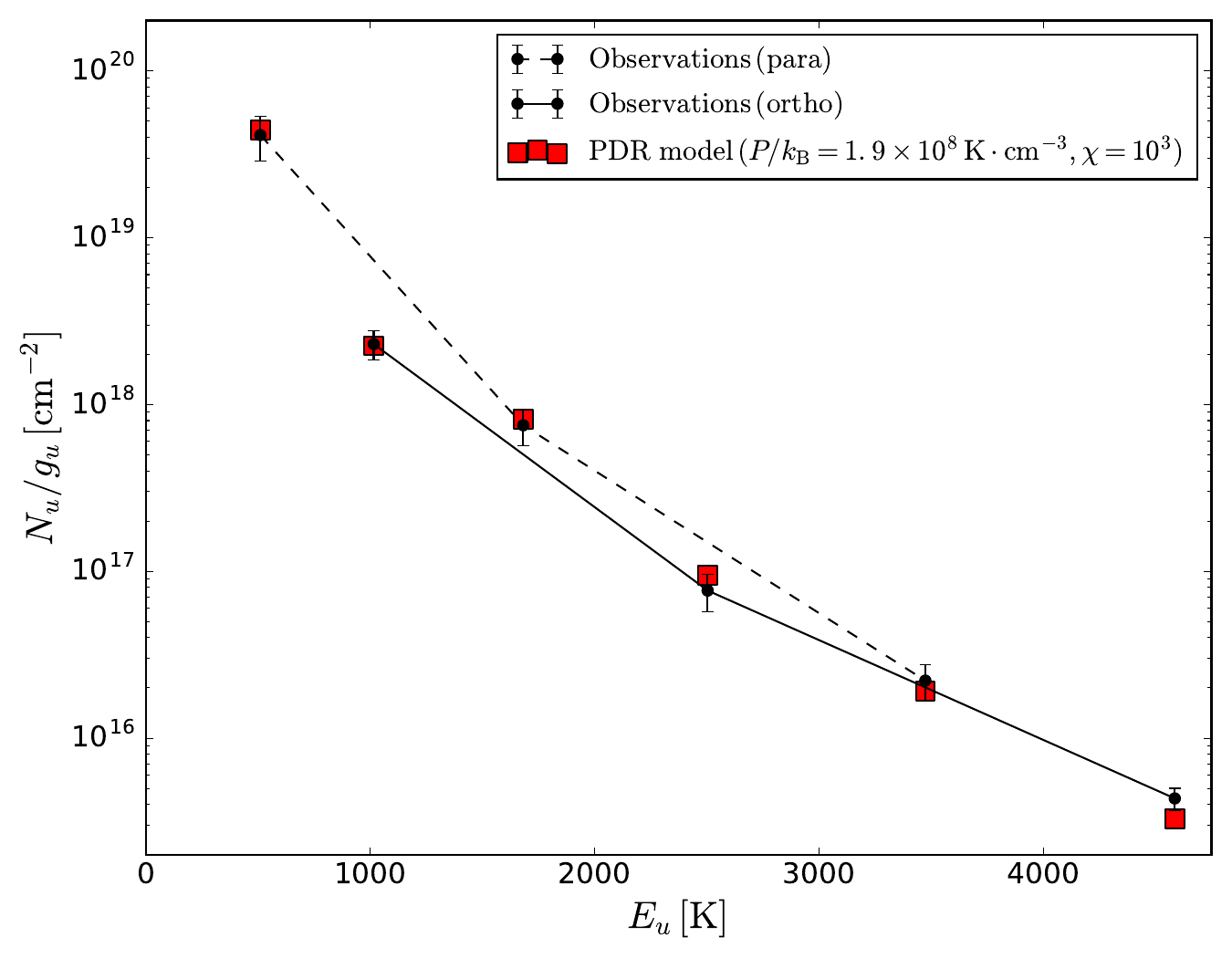}
\caption{Rotational diagram of H$_2$ in the NGC7023 NW PDR, comparing the observations \citep{fuente99} with PDR models \citep[with the Meudon PDR Code,][]{2006ApJS..164..506L}.
Ortho and para transitions are distinguished to highlight the non-LTE ortho-para ratio.\label{plot_H2_excit}}
\end{figure}
\begin{figure}
\includegraphics[scale=0.4]{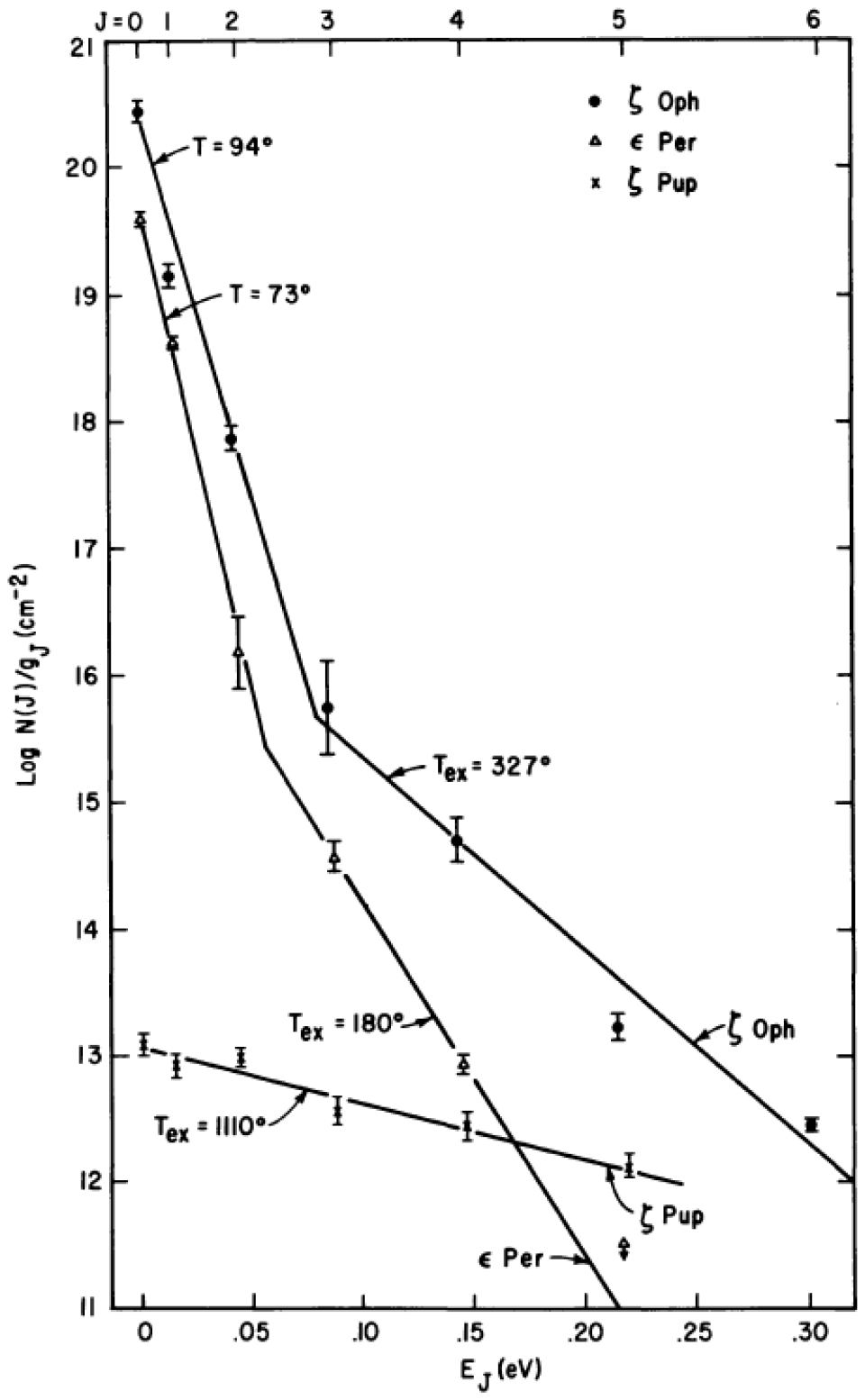}
\caption{First H$_2$ excitation diagram published for three stars observed with {\it Copernicus} \citep{Spitzer.Cochran1973}. This diagram illustrates the fact that two distinct temperatures are needed to fit all J levels, except for low H$_2$ column densities (N(H$_2$) $<$ 10$^{15}$ cm$^{-2}$).\label{spitzer-cochranfig2}}
\end{figure}

H$_2$ excitation diagrams are commonly used to show the population distribution of the molecules across the available levels. Assuming the mid-IR lines are optically thin, the column density of the upper level of each pure rotational transition is measured from the spectral line flux $F_{\nu}$ of a given transition according to $N_u = 4 \pi F_{\nu} /(h \nu A \Omega)$, where $h$ is Planck's constant, $\nu$ is the frequency of the transition, $A$ is the Einstein coefficient for the transition, and $\Omega$ is the solid angle of the observed region. In Local Thermodynamic Equilibrium (LTE), the upper level column density is related to both the excitation temperature $T$, and the total column density $N_\mathrm{tot}$ via, $N_u/g_u =  N_\mathrm{tot} \exp(-E_u/k_\mathrm{B} T) / Z(T)$, where $E_u$ is the energy of the upper level of the transition, $k_\mathrm{B}$ is the Boltzmann constant and $Z(T)$ is the partition function\footnote{An analytical approximation is given by $Z(T) = 0.0247 T/(1-\exp(-6000/T)$, where $T$ is in K \citep{Herbst1996}.}, and $g_u = (2S+1)(2J+1)$ is the degeneracy of the upper level of the transition. In this last expression $S$ is the spin quantum number for a given $J$ transition. The spin value is $S= 0$ for even $J$ (para-H$_2$), and $S = 1$ for odd J (ortho-H$_2$). The H$_2$ excitation diagram is usually presented as a plot of $\log _e (N_u / g_u)$ versus $E_u /k$ (see Fig.~\ref{plot_H2_excit}). For a single excitation temperature the slope of a line fit to these points would be proportional to $T^{-1}$.   

 Two approaches to fit the H$_2$ excitation data referred to above will now be discussed.
The first is a traditional method of fitting single or multiple temperature components to the excitation diagrams. 
This method was first used for the local diffuse ISM detected in absorption in {\it Copernicus} spectra of a few bright stars \citep{Spitzer.Cochran1973} (see Fig.~\ref{spitzer-cochranfig2}) and has been generalized to many {\it Copernicus} \citep{Savage1977} and FUSE \citep{Rachford2002,Rachford2009} lines of sight. For translucent lines of sight generally studied in absorption, the excitation diagrams yield  mean gas temperatures around 55--80~K from the first  excitation levels $J=0$ to $J=2$, and  excitation temperatures above 180~K from the higher $J$ levels.
This method is commonly
used to study H$_2$ studies in other galaxies. It is generally assumed that, for the lower pure rotational H$_2$ transitions, the ortho and para-H$_2$ species should be in collisional equilibrium. As shown by \citet{Roussel2007} for H$_2$ densities $\gtrsim$ 10$^3$ cm$^{-3}$, most of the lower rotational transitions 
should be thermalized, and temperatures derived from fits to the ortho- and para-H$_2$ transitions should yield consistent temperatures. 
After normalizing by the ortho-para ratio (OPR), significant deviations from LTE would appear as an offset between the odd- and even-$J$ H$_2$ transitions when plotted on an excitation diagram. 

A second method of fitting the excitation data is an extension of the first method, by assuming that the molecular gas temperatures can be modeled as a single power-law distribution, again assuming that the gas is in thermal equilibrium \citep{Togi2016, Appleton2017}.

A non-LTE ortho-para ratio appears in excitation diagrams as a systematic offset between the data for ortho and para levels (see Fig.~\ref{plot_H2_excit}). Such non-thermalized OPRs (for the rotational levels) are commonly observed in PDRs \citep{fuente99,Moutou1999,Habart03,habart2011,fleming2010}, and can either indicate that other conversion mechanisms dominate over reactive collisions \citep[e.g. dust surface conversion,][]{Lebourlot2000,2016A&A...588A..27B}, or that H$_2$ doesn't have time to thermalize because of fast advection through the dissociation front.
Non-LTE OPRs are also commonly seen in the excitation diagrams associated with ro-vibrational transitions, but these ratios are not indicative of the actual OPR of the gas because of preferential pumping effects affecting the populations of the vibrational states \citep{sternberg99}.

\subsubsection{H$_2$ transitions and specific diagnostic power}
The radiative and collision properties of the H$_2$ molecule make it a diagnostic probe of unique capability in a variety of environments (See Sect. \ref{sect:environments} for a discussion of these environments). 

\begin{description}
\item[-]A unique probe of gas excitation: 
Many competing mechanisms can contribute to the excitation of molecular hydrogen. Since we understand reasonably well the radiative and collisional properties of this molecule we can construct realistic models of the response of H$_2$ to its surrounding to probe the dominant heating processes taking place in a given environment (e.g., photon heating, shocks, dissipation of turbulence, X-rays).

\item[-] A thermometer and mass scale of the warm gas: The lowest rotational transitions of H$_2$, generally promoted by collisions, provide a wonderful thermometer for the bulk of the gas above $\sim 80~\mathrm{K}$. 
The rotational excitation of H$_2$ becomes important only for temperatures $T \gtrsim 80$~K because the $ J = 2$ state lies 510~K above the $ J = 0$ state ($J = 3$ lies 845~K above $J = 1$). 
Due to the low $A$ values of the associated optical transitions, any optical depth effects are usually unimportant for these spectroscopic lines. 
H$_2$ lines are optically thin up to column densities as high as 10$^{23}$~cm$^{-2}$. 
Furthermore, H$_2$ is the principal constituent of the molecular gas.
Thus, these spectral lines provide accurate probes of the mass of the cool/warm ($T\gtrsim 80$~K) gas. 

\item[-] A unique probe of the warmest photo-dissociation layers subject to photo-evaporation: 
Self-shielding of H$_2$ against photo-dissociation is efficient from low H$_2$ column densities. 
H$_2$ can then be present when other molecules, such as CO, would already be photo-dissociated.
Thus H$_2$ can probe, in a unique way, the outer warmest photo-dissociation layers of clouds or proto-planetary disks which are subject to photo-evaporation.
\end{description}

\begin{figure}
\includegraphics[scale=0.36]{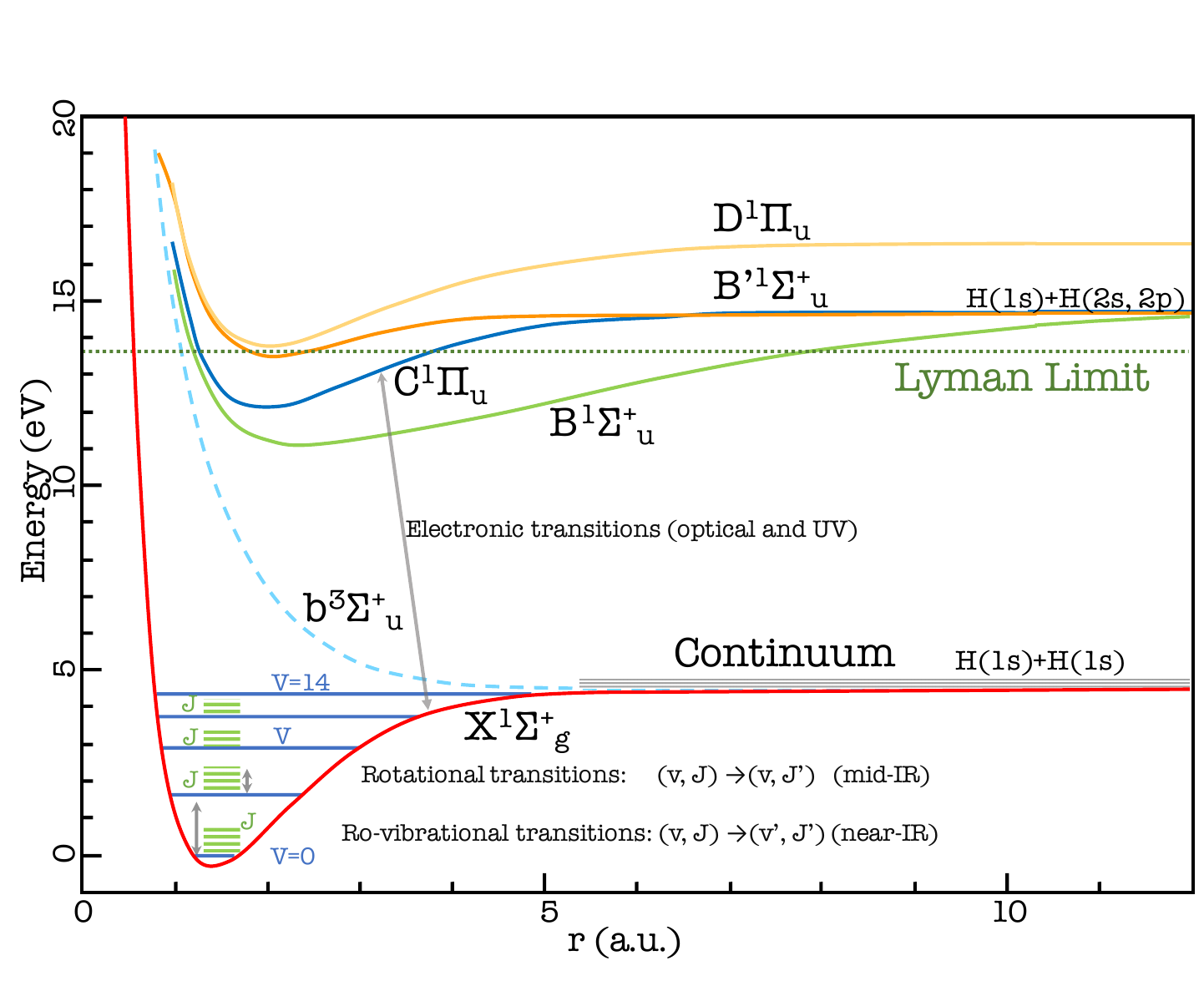}
\caption{Electronic potentials of H$_2$ as a function of the  separation between both atoms. The subscripts $\mathrm{g}$ and  $\mathrm{u}$ stand for \emph{gerade} (even) and \emph{ungerade} (odd) symmetries. Vibrational and rotational levels are indicated schematically for the lowest electronic level. Energy levels are indicated with respect to the ground state ($v = 0$). Levels with vibrational excitation $v>14$, in the continuum region, lead to the dissociation of the molecule. Adapted from \citet{Lepetit_2002_th}. 
\label{plot_H2_levels}}
\end{figure}

Three types of spectroscopic transitions can be observed for H$_2$ \citep[shown in Fig.~\ref{plot_H2_levels}, see also][]{Field:1966}: 
the electronic bands in the UV (shown in Fig.~\ref{shull2000fig1}) , the ro-vibrational transitions in the near-IR (shown in Fig.~\ref{fig_spectrum_NIR_MIR_PDR}), and the pure rotational transitions in the mid-IR (shown in Figs.~\ref{fig_spectrum_NIR_MIR_PDR} and \ref{fig_extragalactic_H2_spectra}).
Electronic transitions of H$_2$, in the UV, can be used as probes of two gas regimes: 
(i) in absorption  to probe cold gas ($T\sim$50-100~K, such as in the diffuse ISM); 
(ii) in emission to probe highly excited gas ($T\sim$few 1000~K such as in outflows or inner disks). 
UV absorption measurements of vibrationally excited interstellar H$_2$ can also be used as probes of highly excited gas.
H$_2$ electronic transitions in absorption occur between the ground vibrational level of the ground electronic state ($\mathrm{X^1\Sigma^+_g}$) and the vibrational levels of the first ($\mathrm{B^1\Sigma^+_u}$) or the second ($\mathrm{C^1\Pi_u}$) excited electronic states. 
In the $\mathrm{X^1\Sigma^+_g}$ state the $v = 1$ vibration level is $\approx 6000$~K above the
ground state, so that ro-vibrational excitation (such as that associated with the 2.12$\,\mu$m line) requires kinetic temperatures $T > 1000$~K or FUV pumping excitation. 
The main utility of these near-IR H$_2$ lines lies in their applicability for probing  very small quantities of hot gas.
H$_2$ pure-rotational emission in the mid-IR traces the bulk of the warm gas, generally at temperatures from $100~\mathrm{K}$ up to $1000~\mathrm{K}$. 

\subsubsection{Observational challenges: how and where can we observe the H$_2$ molecule in space?}
\begin{figure*}
\includegraphics[scale=0.94]{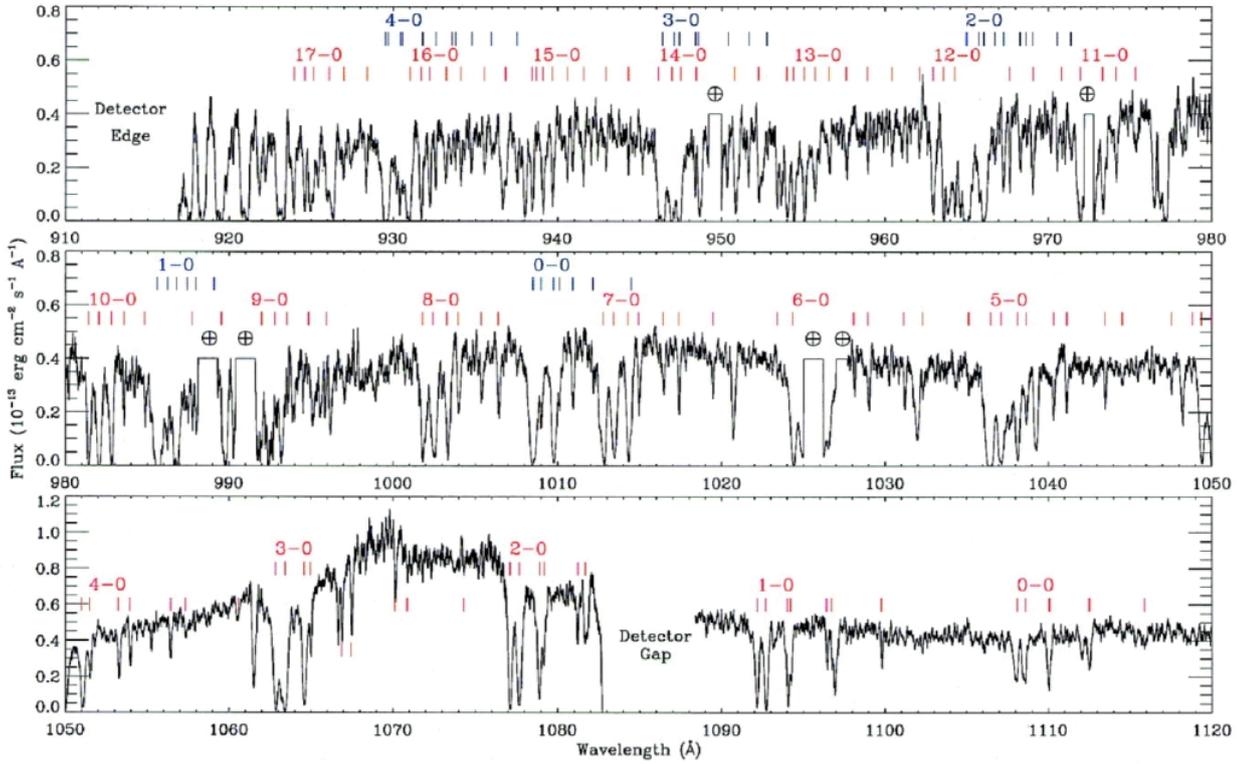}
\caption{Full FUSE spectrum of ESO 141-G55, which illustrates diffuse Galactic H$_2$ detected in absorption in the spectrum of a Seyfert galaxy.  $N(H_2) = 1.9\,10^{19}cm^{-2}$ ; $N(HI) = 3.5\,10^{20}cm^{-2}$. 
This spectrum has a resolution of R $\approx$ 12,000 and S/N $\approx$ 15 per smoothed (30 km s$^{-1}$) bin (1040$-$1050 \AA) and S/N $\approx$ 25 at 1070 \AA. Lower (red) and upper (blue) ticks mark the detected Lyman and Werner lines of H$_2$, respectively. Bright terrestrial airglow lines superimposed on the interstellar HI lyman absorption lines have been truncated. From \citet{shull2000}.\label{shull2000fig1}}
\end{figure*}

As noted above, the electronic transitions of H$_2$ occur at ultraviolet wavelengths, a region of the spectrum to which the Earth's atmosphere is opaque; hence, observations in this spectral region can only be made from space. 
The first detection of H$_2$ beyond the Solar System was made by \citet{Carruthers1970} via UV absorption spectroscopy employing a rocket-borne spectrometer. 
This discovery was followed by UV observations with the Copernicus space mission that confirmed the presence of the hydrogen  molecule in diffuse interstellar clouds \citep[for a first review on this subject see][and references therein]{Spitzer1975}.
The H$_2$ absorption lines from the diffuse ISM, i.e. those arising from the low-lying rotational levels of the lowest vibrational level of the ground electronic state (as mentioned in section 2.1.2), can only be observed in the far UV, below $1130$ \AA, accessible to Copernicus, ORPHEUS and FUSE  (see Fig.~\ref{shull2000fig1}), as well as HST/COS after 2010 (but only at low resolution with R $\approx 2000$). 
Only the excited vibrational levels have lines above $1150$ \AA, accessible to IUE, and GHRS and STIS on board HST, but they are detected only in a few ISM lines of sight of very high excitation \citep[see][for an absorption spectrum of vibrationally excited H$_2$ toward HD~37903, the star responsible for the illumination of NGC~2023]{Meyer2001}. In emission those lines appear only in circumstellar regions like the cited case of the accretion disk observed with HST by \citet{France2010}, or in many T~Tauri stars observed with IUE, HST or FUSE.

\begin{figure*}
\includegraphics[scale=0.3]{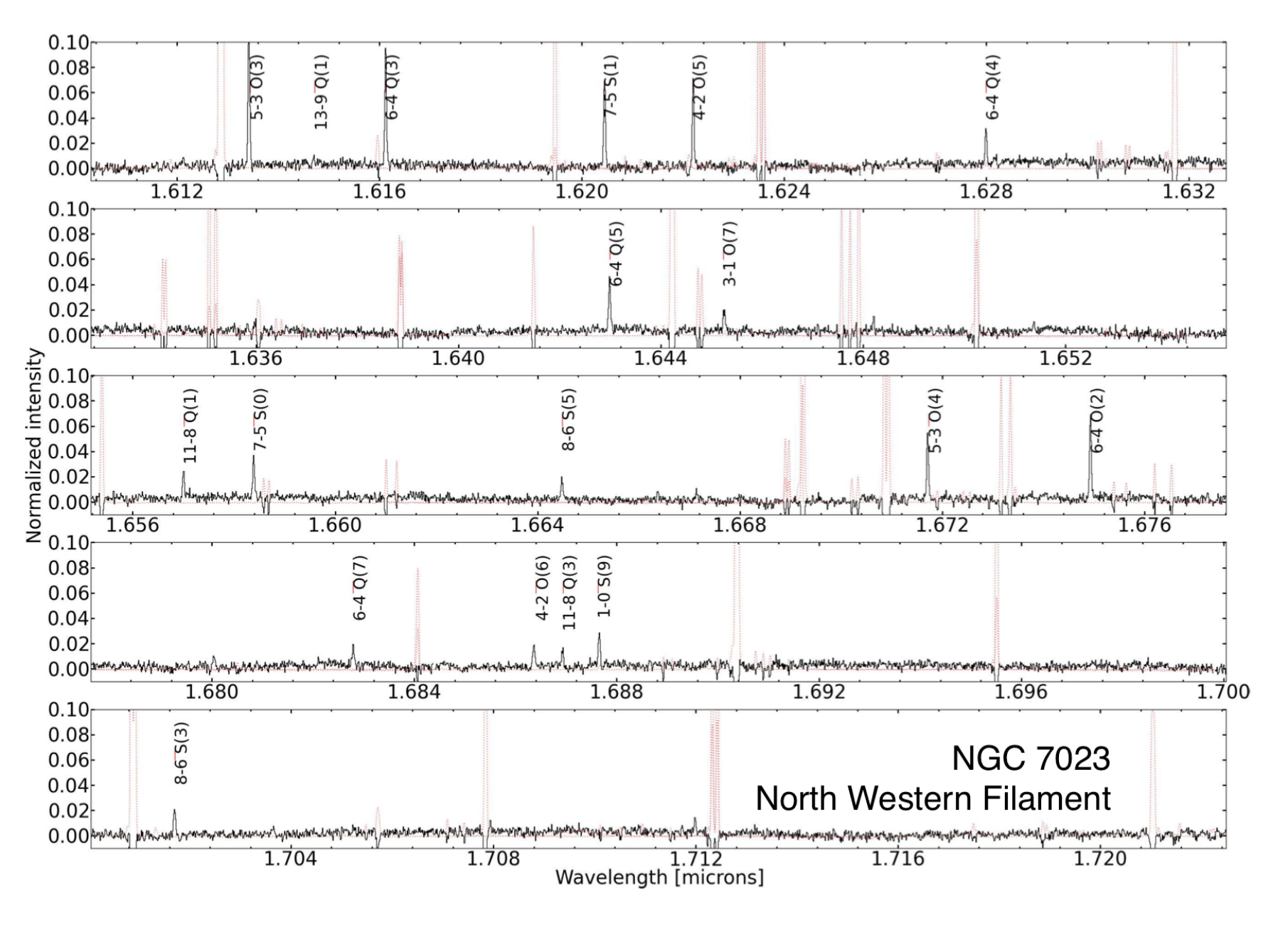}
\includegraphics[scale=0.3]{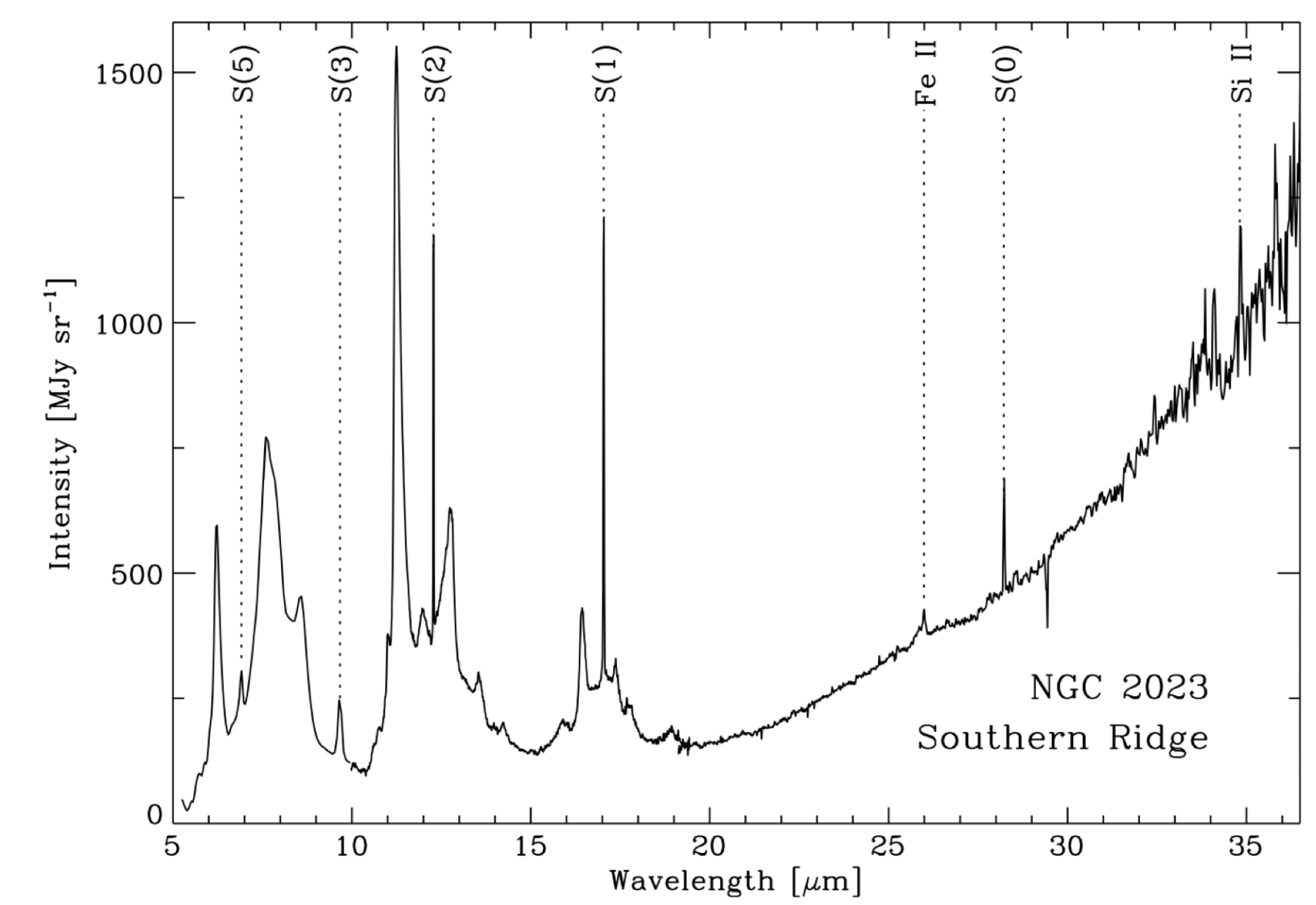}
\caption{
 Left pannel: Part of the near-IR spectra  
from the north western filament of the reflection nebula NGC 7023, which illustrates H$_2$ rovibrationnally excited detected in emission in PDRs. This spectrum obtained with the Immersion GRating INfrared Spectrograph (IGRINS), has a resolution of $R\simeq$45,000. 
The spectra show here are into the wavelength ranges 1.610-1.722 $\mu$m. 
The intensity has been normalized by the peak of the 1-0 S(1) line. The dash-red lines display OH airglow emission lines, observed at "off" position 120" to the north from the target. Within the 1"$\times$15" slit and the total wavelength coverage 1.45-2.45 $\mu$m, 68 H$_2$ emission lines from rovibrationnally excited H$_2$ have been detected. From \citet{Le2017}.
 Right pannel: Spitzer mid-IR spectra toward the reflection nebula NGC 2023, which illustrates H$_2$ rotationally excited detected in emission in PDRs. Full spectral coverage from the four Spitzer/IRS modules (SL2, SL1, SH, and LH with a resolution of $R\sim$60-120 and 600), as obtained by averaging 15 pixels that sample the Southern Ridge emission of the nebula. H$_2$ pure rotational and atomic fine structure emission lines are identified over strong PAH features and dust continuum. From \cite{sheffer2011}.
\label{fig_spectrum_NIR_MIR_PDR}}
\end{figure*}

Ro-vibrational and rotational transitions of H$_2$ are faint because of their quadrupolar origin, as noted above. Moreover, these lines lie, most of the time, on top of a very bright continuum due to the emission of interstellar dust (e.g., see Fig~\ref{fig_spectrum_NIR_MIR_PDR}); hence, observations at high spectral resolution are needed to disentangle these weak molecular lines. Ground based high-resolution spectrographs (e.g., VLT, Gemini, Subaru) are commonly used to probe the near-IR H$_2$ ro-vibrational lines. 
For the case of rotational lines which occur in the mid-IR, the Earth's atmosphere is again, at best, only partially transparent. 
The mid-IR window with high sky background is a very challenging region of the spectrum in which to perform high sensitivity observations from the ground. Thus, H$_2$ mid-IR emission studies from the ground (e.g., VISIR, TEXES) are, to date,  limited to relatively bright sources (with fluxes typically larger than $1~\mathrm{Jy}$). 
Space-based platforms are needed to observe fainter infrared sources in the mid-IR, but here spectral and spatial resolution are limited (e.g. ISO, Spitzer).

Finally, most of the interstellar H$_2$ can lie hidden in cool, shielded regions \citep[e.g.][]{Combes2000} where the molecular excitation could be too low for to H$_2$ to be seen via emission lines, and the local extinction is too high to allow the detection of lines resulting from UV pumping. In these regions, a way to estimate indirectly the molecular fraction has been proposed by \citet{Li2003} by measuring the residual atomic hydrogen fraction via HI Narrow Self-Absorption (HINSA) observations.

 In the near future, mid-IR instrumentation such as the high-resolution mid-IR spectrograph EXES in the airborne observatory SOFIA, and the mid-IR spectrograph MIRI in the James Webb Space Telescope will greatly increase the critical observational sensitivity, spatial and spectral resolution, and will provide stringent tests of our current understanding of H$_2$ in space. 

In the following section we give a few examples of multi-wavelength observations of H$_2$ transitions in Galactic and extra-galactic environments.

\paragraph{Galactic environments} 
H$_2$ lines have been detected from Galactic sources as diverse as photo-dissociation regions (PDRs), shocks associated with outflows or supernovae remnants, circumstellar envelopes and proto-planetary disks (PPDs) around young stars, planetary nebulae (PNe), diffuse ISM, and the galactic center. 
UV absorption lines measured with FUSE, a very sensitive experiment which detected H$_2$ down to $N(\mathrm{H_2}) <10^{14}$ cm$^{-2}$,
show  that H$_2$ is in fact ubiquitous in our Galaxy \citep[e.g.][]{shull2000
}.
UV absorption lines enable us to measure the column densities of H$_2$ in the rotational $J$ levels of the ground vibrational and electronic states in diffuse and translucent lines of sight, with visual extinctions ($A_V$) up to about 3 to 5, and to measure the molecular fractions $f_\mathrm{H_2}$. In the diffuse ISM, with visual extinctions $A_V \le 1~\mathrm{mag}$, the molecular fractions range from $10^{-6}$ at low HI column density up to $\sim$40\% \citep{Savage1977, Spitzer1975,Gillmon2006}. 
In translucent sight lines (which are lines of sight with greater extinction $A_V = 1-5~\mathrm{mag}$), the molecular fraction can be as high as $70\%$ \citep{Rachford2002, Rachford2009}, but is never too close to $1$. UV absorption lines also enable us to estimate the H$_2$ formation rate in the diffuse ISM \citep[e.g.][]{jura75, Gry2002}, as well as characterize the gas temperature and excitation \citep[e.g.][]{Spitzer.Cochran1973, Gry2002, Nehme:2008bv, 2016A&A...588A..27B}.
H$_2$ absorption lines have also been detected towards circumstellar envelopes of young stellar objects, YSOs, \citep[e.g.][]{martin-zaidi2008}. 
The main limitations here are the restricted number of sources against which H$_2$ can be detected in absorption (which is limited to interstellar gas with $A_V \leq 5$ and intercepting the line of sight toward a bright UV source, thus it prohibits the observation of dense molecular clouds). In practice, UV absorption observations only allow the study of the molecular gas in the Solar Neighborhood or in cirrus and molecular clouds at high Galactic latitude \citep{Gillmon.Shull2006, Gillmon2006}, in intermediate-velocity clouds in the Galactic halo, in the Magellanic Clouds \citep{Tumlinson2002}, in a few external galaxies, and in objects with a specific geometry. 
On the other hand, as mentioned above, far-UV H$_2$ emission lines have unveiled the presence of hot gas in the disks of many T-Tauri stars.
In cases where mid-IR CO spectra, or traditional accretion diagnostics,  suggest that the inner gas disk has dissipated, far-UV H$_2$ observations offer unambiguous evidence for the presence of a molecular disk \citep[e.g.][]{France2012}.

Infrared emission of H$_2$ was first observed via the 2 $\mu$m ro-vibrational lines (most notably the 1-0 S(1) line at 2.12 $\mu$m) towards Galactic shocks, PDRs, and PNe \citep[e.g.,][]{Gatley1986, Pak1996, burton90b,lemaire99,walmsley2000,cox2001}. \cite{luhman:1997} provided the first combined IR/UV picture of an H$_2$ fluorescence cascade in a single object (a PDR).
Deep near-infrared spectra of bright PDRs, taken with high resolution, enable us to detect many emission lines from ro-vibrationally excited molecular hydrogen that arise from transitions out of many upper ro-vibrational levels of the electronic ground state \citep[e.g.,][see Fig.\ref {fig_spectrum_NIR_MIR_PDR}]{kaplan:2017,Le2017}. 
Since atmospheric transmission in the K band is relatively good,  
H$_2$ lines in the near-IR have been searched for in relatively faint objects using large telescopes \citep[such as disks, e.g.][]{carmona2011}.

ISO and Spitzer provided a fundamental step forward, in that they enabled us to exploit the potential of the  H$_2$ pure rotational lines in the mid-infrared, probing the bulk of the warm gas.
This gave access to the H$_2$ rotational diagram in various sources (PDRs, shocks, YSOs, PNe, SNR, low UV excited clouds, diffuse ISM), as well as its spatial variation in some extended sources \citep[e.g.][]{Neufeld:1998,draine99,cesarsky99,rosenthal2000,vandenancker2000,lefloch2003,neufeld2009,Maret2009,goldsmith2010,fleming2010,habart2005a,habart2011,Hewitt2009, Rho2017,sheffer2011,mata2016}. 
H$_2$ data have allowed us to better characterize the shocks (e.g., measure the temperature history, and age) associated with outflows from young stars or supernova remnants and the different possible H$_2$ excitation mechanisms at different evolutionary stages of young stellar objects and planetary nebulae  \citep[e.g.,][]{Neufeld:1998,pineaudesforets99,cesarsky99,vandenancker2000,lefloch2003}.
This work also showed that H$_2$ is a major contributor to the cooling of astrophysical media where physical conditions lie in between those of hot molecular gas and cold molecular gas. It has proved possible to estimate the gas temperatures and densities of the transitional layers of the ISM which separate ionized and neutral molecular gas. However, by comparing the observations with the PDR model predictions, 
the model can account well for the H$_2$ rotational line intensities and excitation temperature in strongly irradiated PDRs \citep[e.g.][see Fig.~\ref{fig_spectrum_NIR_MIR_PDR}]{sheffer2011}, but underestimates the H$_2$ excitation temperature and intensity of the excited rotational lines in low/moderate UV irradiated regions \citep[e.g.][]{goldsmith2010,habart2011}. This underlines that our understanding of the warm H$_2$ gas is incomplete and could suggest additional excitation of H$_2$ or gas, or out-of-equilibrium processes.



\paragraph{Extra-galactic environments} 

\begin{figure}
\includegraphics[width=0.5\textwidth]{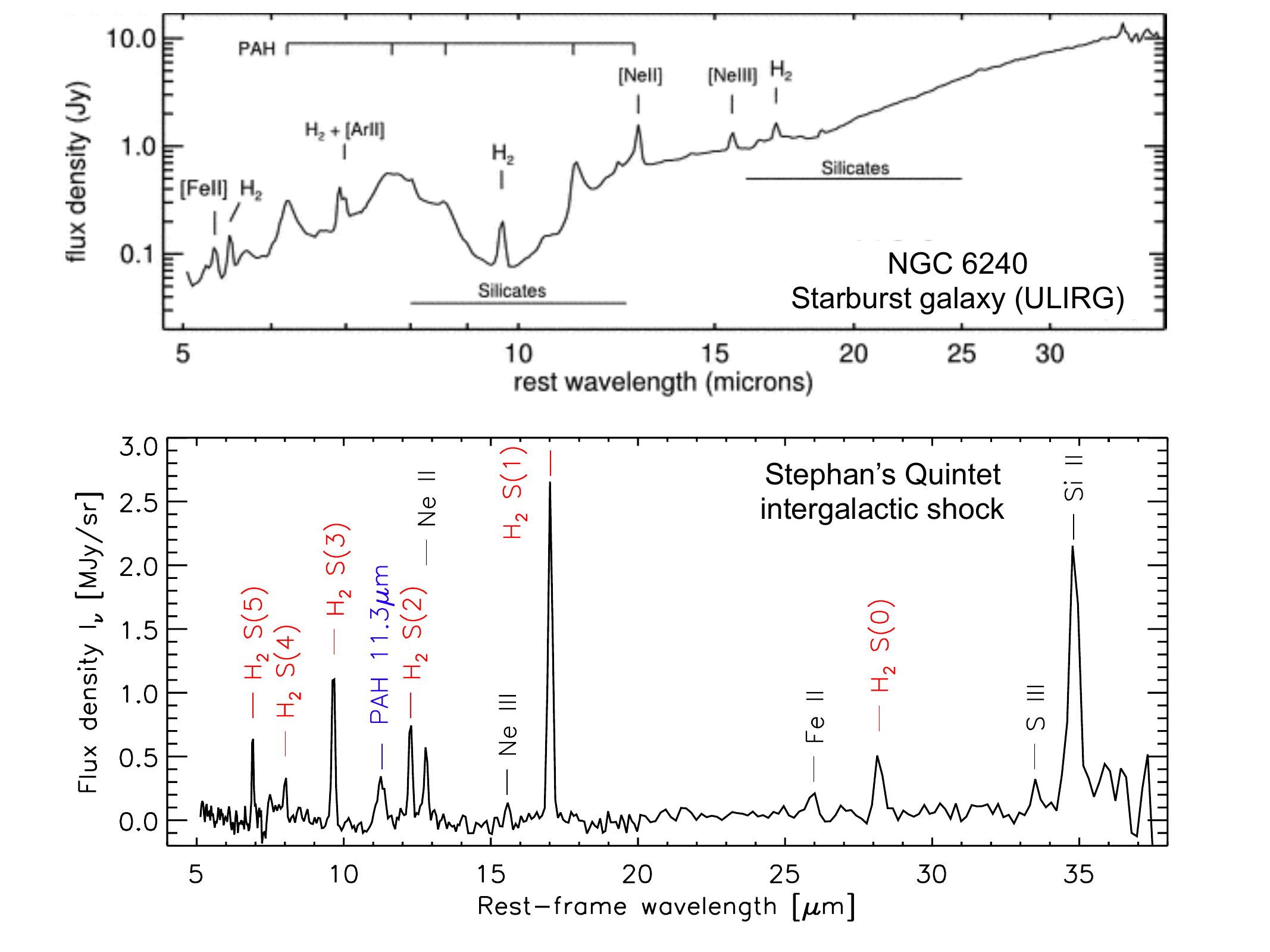}
\caption{
Two examples of extragalactic mid-infrared spectra (taken with the Spitzer IRS) showing prominent rotational lines of H$_2$. \textit{Top panel:} spectrum from \citet{Armus2006} of NGC 6240, a nearby ($z=0.0245$)  merging galaxy that has a powerful starburst, a buried (pair of) AGN, and a superwind. Prominent emission lines and absorption bands (horizontal bars) are marked. \textit{Bottom:} spectrum from \citet{Guillard2010} of the Stephan's Quintet intragroup medium, taken in between two colliding galaxies. The shocked medium is rich in H$_2$ but with very weak star formation and UV radiation field. Note the strength of the H$_2$ lines (marked in red) compared to the dust continuum, as opposed to the star-forming galaxy NGC~6240 shown above.  
\label{fig_extragalactic_H2_spectra}}
\end{figure}

Observations of molecular hydrogen emission from external galaxies started with the detection of near-IR emission coming from hot molecular gas found in photo-dissociation regions or shocks, especially in the central regions hosting AGN or major starbursts \citep[e.g.][]{Wright1993, Mouri1994, Goldader1997}. Because they are difficult to observe from the ground, the observations of the pure rotational lines of H$_2$ from external galaxies started with the Infrared Space Observatory (ISO), and continued with the \textit{Spitzer} infrared (IR) satellite \citep[e.g.][]{Valentijn1999, Lutz2003, Verma2005}. In star forming galaxies, rotational H$_2$ line emission is thought to come from PDRs \citep{Rigopoulou2002, Higdon2006, Roussel2007}. A tight correlation between the H$_2$ and IR luminosity is inferred for star-forming galaxies and the H$_2$ to polycyclic aromatic hydrocarbon (PAH) luminosity ratio in this kind of galaxies is within the range of values that are expected from PDR emission, suggesting that UV photons are the main H$_2$ excitation source. The Active Galactic Nuclei galaxies exhibit a stronger H$_2$ to PAH ratio than dwarf and star-forming  galaxies \citep{Roussel2007}, suggesting that the radiation from the AGN is not sufficient to drive H$_2$ emission.

More recently \textit{Spitzer} IRS observations have shown that our census of the warm H$_2$ gas in galaxies may be severely incomplete, revealing a new class of galaxies (including elliptical galaxies, AGN, galaxy groups, and galaxy clusters) with strongly enhanced H$_2$ rotational emission lines, while classical star formation indicators (far-infrared continuum emission, ionized gas lines, polycyclic aromatic hydrocarbons, PAHs) are strongly suppressed \citep{Appleton2006, Ogle2007, Guillard2009, Cluver2010, Ogle2010, Guillard2012, Ogle2012, Peterson2012, Guillard2015a}. Among the sample of H$_2$-luminous objects, the Stephan's Quintet is certainly the object where the astrophysical context is clear enough to identify the dominant source of energy that powers the H$_2$ emission and to associate it with the mechanical energy released in a galactic collision \citep{Guillard2009, Guillard2012a, Appleton2013, Appleton2017}. 
In these sources, the luminosity of the H$_2$ lines cannot be accounted for by UV or X-ray excitation, and their properties suggest that the dissipation of turbulence is the main heating mechanism for the warm H$_2$ gas. The strong H$_2$ line emission  is a dominant gas cooling channel and traces the turbulent cascade of energy associated with the formation of multiphase gas. The dynamical interaction between gas phases drives a cycle where H$_2$ gas is formed out of shocked atomic gas \citep{Guillard2009}. 
In the M82 starburst galaxy, the galactic wind is observed to be loaded with H$_2$ gas with dust entrained \citep{Beirao2015}. Because the timescale to accelerate molecular material from the galactic disk to tens of kpc in the wind is longer than the dynamical timescale of the outflow, and because the H$_2$ excitation is consistent with models of slow shocks, it has been argued that the H$_2$ gas in the outflow is formed by post-shock cooling during the interaction of the wind with the gas in the galactic halo.

\subsubsection{Observational constraints on the H$_2$ formation rate}
The determination of the formation rate and abundance of H$_2$ for a given region of the ISM is crucial, as it controls most of the subsequent development of the chemical complexity, as well as a substantial part of physics of the region, and can allow us to discriminate between the H$_2$ formation mechanisms that may be operating. Early studies \citep{gould63,hollenbach71,jura75} provided the first estimates of H$_2$ formation rates in the diffuse ISM, concluding that grain surface chemistry is an unavoidable route for efficient molecular hydrogen formation. 

The observationally determined H$_2$ formation rate coefficient in the diffuse ISM, $R_\mathrm{H_2}\sim 3-4\times 10^{-17}~\mathrm{cm^3 s^{-1}}$ \citep{Gry2002}, appears to be rather invariant. Nevertheless, \citet{Habart04} estimate considerably higher H$_2$ formation rates at high gas temperatures in PDRs. 
Using as a diagnostic the ratio of the rotational to ro-vibrational lines of H$_2$, as observed and as predicted by PDR models, they determined H$_2$ formation rates similar or higher (factor of 5 for moderately excited PDRs) than that measured in diffuse clouds. H$_2$ appears to form efficiently in PDRs with gas and grains at high temperatures ($T_{gas}\sim$300~K and $T\sim$30~K for a grain at thermal equilibrium with the radiation field). However, it must be underlined that these results are based on the assumptions that PDRs are static, in equilibrium, while propagation of the ionization and photo-dissociation fronts will bring fresh H$_2$ into the zone emitting line radiation. These rate values are thus upper limits. Finally, as we mention later, no  observational signatures of the excitation state or ortho-to-para ratio of the newly-formed (nascent) H$_2$ have been obtained yet. This last point remains observationally challenging.


\subsection{Experiments and quantum calculations}

The formation of molecular hydrogen in interstellar space occurs primarily on the surface of dust grains. In diffuse clouds, grains are bare and are usually classified as silicates or carbonaceous materials. In the silicate class we have olivines ((Mg$_x$,Fe$_{1-x}$)$_2$ SiO$_4$) and pyroxines (Mg$_x$,Fe$_{1-x}$ SiO$_3$). Observational evidence shows that these particles are sub-micron in size and are largely in an amorphous form \citep{Draine:2003, jones2013}. In the carbonaceous class we have sp$^3$  (nanodiamonds), sp$^2$ (graphite, PAHs) and mixed sp$^3$-sp$^2$ (amorphous carbon) carbon bonded materials. Again, observational evidence shows that these species are nano to sub-micron in size and largely amorphous \citep[]{jones2013}.  Further information about dust grains and their laboratory analogs can be found in \citet{Draine:2011,Krugel:2007,Henning:2010a,Henning:2010b,jones2013}.

Following  the first experiments studying H$_2$ formation  on a polycrystalline olivine sample \citep{Pirronello:1997a,Pirronello:1997b}, laboratory investigations focused on the various processes involved in H$_2$ formation on surfaces, in order to find rate limiting process(es) for the specifically chosen conditions (kinetic energy, material, morphology, temperature, etc.). 
In the next subsection, we list the processes relevant to the formation of molecular hydrogen on dust grains and  mention the most common experimental and theoretical techniques that have been used to study such processes.  There is a vast literature on these processes, but most of it is for well-characterized systems; that is, for processes occurring on clean and well-characterized surfaces, usually single crystal surfaces \citep{Kolasinski2008}. 

The application of results from the surface science literature to astrophysics environment should be performed very carefully, since the chemical and morphological compositions of actual ISM dust grains are largely unknown, and the processes described below (sticking, binding, diffusion, etc.) depend on many parameters, such as the kinetic energy of the incoming atom and  the chemical and morphological composition of the surface.
For a recent detailed review of experimental and theoretical work on the formation of H$_2$ on dust grains, relevant to the ISM, see \citet{2013ChRv..113.8762V}. For descriptions of apparatus and measuring methods, see \citet{Fraser:2002,Fraser:2004b,Perry:2002,Vidali:2005a,2008PrSS...83..439W,Lemaire:2010} as well as the reviews by \citet{Hama:2013,2013ChRv..113.8762V}. 

\subsubsection{Relevant processes: sticking, binding, diffusion, reaction, and desorption}

\begin{figure*}
   \centering
    \includegraphics[width=0.80\textwidth]{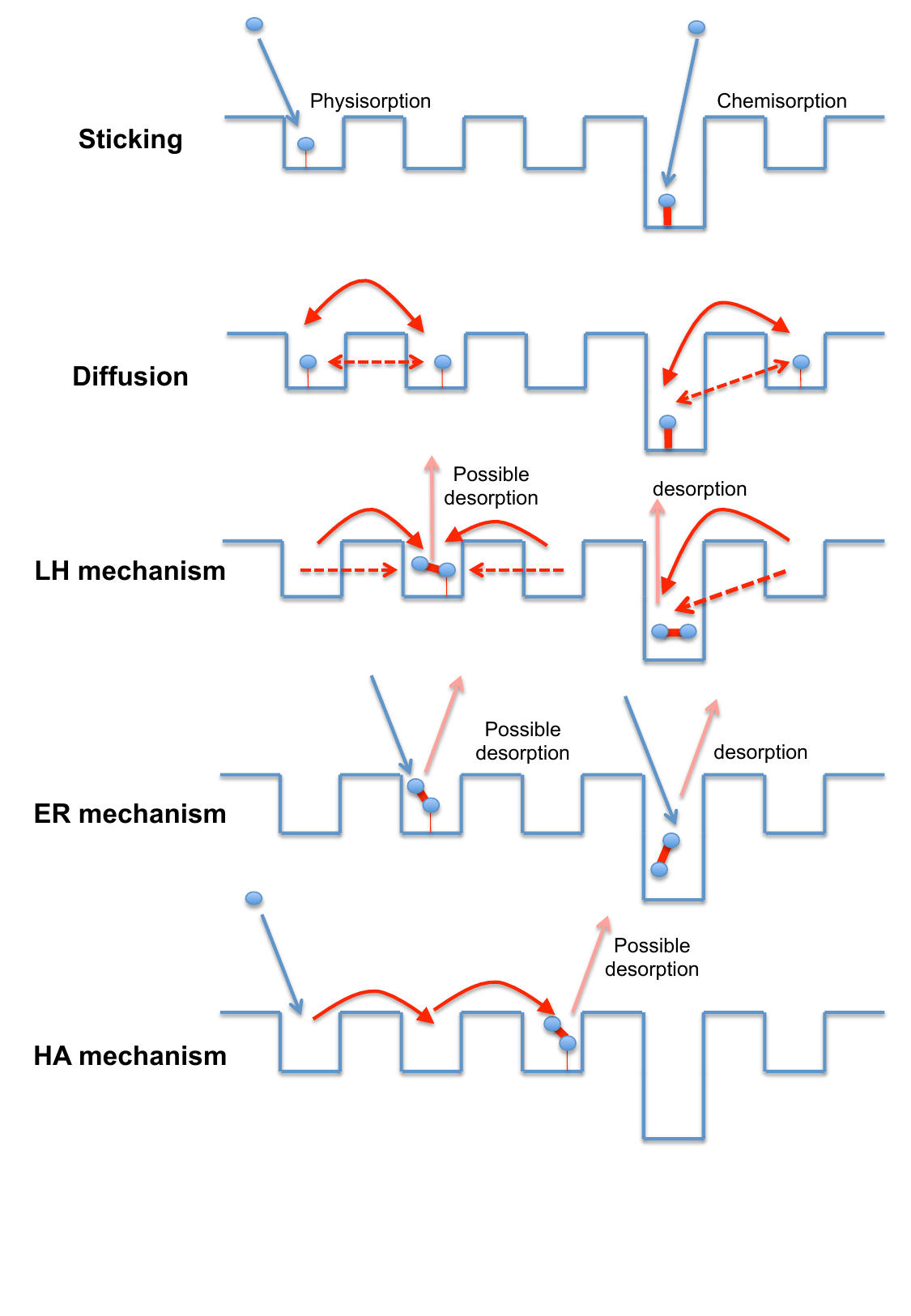}
      \caption{Schematic illustrating Key surface processes involved in the formation of H$_2$. } 
       \label{figure_processes}
\end{figure*}

The formation of molecular hydrogen on a solid surface involves a few key physical processes: trapping, binding, diffusion, reaction and desorption. It helps to make the distinction between weak, long-range interactions with the surface, and strong, localized interactions. In the former, called physisorption, the particle approaching from the gas phase interacts with the surface via long-range van der Waals forces \citep{Bruch:2007a}. This binding energy is of the order of tens of meV \citep{Vidali:1991}. Experiments studying H atoms interacting with silicate, water ice and graphite surfaces at low kinetic energy and low sample temperatures suggest that physisorption is the class of interaction that is pertinent in these cases, as discussed below. 

For the case of a strong interaction between the adsorbate and the surface, chemisorption, a strong bond ($\sim$ eV) is formed between the incoming atom and the surface \citep{Kolasinski2008}. This class of interaction is important, for example, in experiments involving energetic H atoms interacting with the basal plane of graphite  \citep{2002CPL...366..188Z,2006PhRvL..97r6102H,2006PhRvL..96o6104H} or of thermal H atoms encountering PAHs \citep{1998Natur.391..259S,2008ApJ...679..531R,2012ApJ...752....3T,2012ApJ...745L...2M,2016NatSR...619835C}.

In the trapping of an atom on a surface, the atom from the gas phase has to lose enough of its kinetic energy to remain confined to the surface. Trapping and sticking are often used interchangeably, but here we will define trapping as the temporary residence of the atom on the surface; that is, the atom is not necessarily fully energetically accommodated. We contrast ``trapping'' with ``sticking'' where the particle is fully accommodated  (i.e. thermalized) on the surface. The residence time on the surface is then determined by the strength of the bond to the surface and the surface temperature. Sticking has been measured for hydrogen molecules on a variety of surfaces and at different incident kinetic energies \citep{Matar:2010,Chaabouni:2012}. Due to technical difficulties, there are only few experimental measurements of atomic hydrogen sticking on analogs of interstellar dust grains \citep{Pirronello:2000}. Computationally, the sticking process has been investigated 
for H/H$_2$ on graphite
\citep{2005JChPh.122a4709S,Kerwin2008,Morisset2010,Cazaux2011,2011JChPh.134k4705L,PhysRevLett.107.236102} and on water ice \citep{1970JChPh..53...79H,Buch:1989,1991JChPh..95.6026B,Masuda:1997,1998A&A...330..773M,AlHalabi:2002,2007MNRAS.382.1648A,2014ApJ...790....4V} using methods ranging from molecular dynamics simulations to fully quantum mechanical calculations.

The binding and diffusion of atoms on surfaces are particularly important processes in the formation of molecular hydrogen on dust grains. Binding regulates the time a hydrogen atom or molecule resides on a grain via the relationship $$\tau=\tau_0 \; \exp{\; \frac{E_\mathrm{b}}{k_\mathrm{B}T_\mathrm{dust}}} \ ,$$ where $\tau_0$ is related to the fundamental vibrational frequency $\nu_0$ of an atom in a potential energy well describing the motion perpendicular to the surface,  $E_\mathrm{b}$ is the binding energy of the atom on the surface, $k_\mathrm{B}$ the Boltzmann constant and $T_\mathrm{dust}$ the temperature of the surface. 

For a crystalline surface of a given material, only a handful of binding sites need to be considered. For example, in the case of physisorption of H on the basal plane of graphite, the deepest binding energy site is at the center of the graphitic hexagon \citep{Petucci:2013}, while in the case of H chemisorption on graphite, the preferred binding site is on top of a carbon atom \citep{1999CPL...300..157J}. However, actual dust grains are amorphous, and therefore a range of binding sites, with potentially different binding energies, need to be considered. 

For long-distance diffusion, the atoms suffer a larger risk of being trapped at deep potential sites, hence long-distance diffusion tends to be limited by higher activation barriers than short-distance diffusion (see section~\ref{water_section}).
Except for cases where adsorption is activated, the desorption energy of a particle from the surface is the same as the binding energy.  

Using the technique of temperature programmed desorption (TPD) in which the temperature of the surface is increased rapidly and the desorbing particles collected, the distribution of desorption energies, and therefore of binding energy sites, has been obtained for many atom/molecule - surface systems \citep{Amiaud:2006,He:2011,Amiaud:2015}. Figure 8 shows the desorption of D$_2$ that has been deposited on an amorphous silicate surface. The peaks are rather wide, indicating that D$_2$ is desorbing from from sites with a wide distribution of binding energies. For a comparison with desorption from a single silicate crystal, see \citep{Vidali2010,He:2011}. Because of technical limitations in detecting atomic hydrogen, the distribution of the energy sites available on analogs of dust grain surfaces is known primarily for molecular hydrogen and its isotopologues, rather than for atomic hydrogen. 
 \begin{figure}
\includegraphics[scale=0.4]{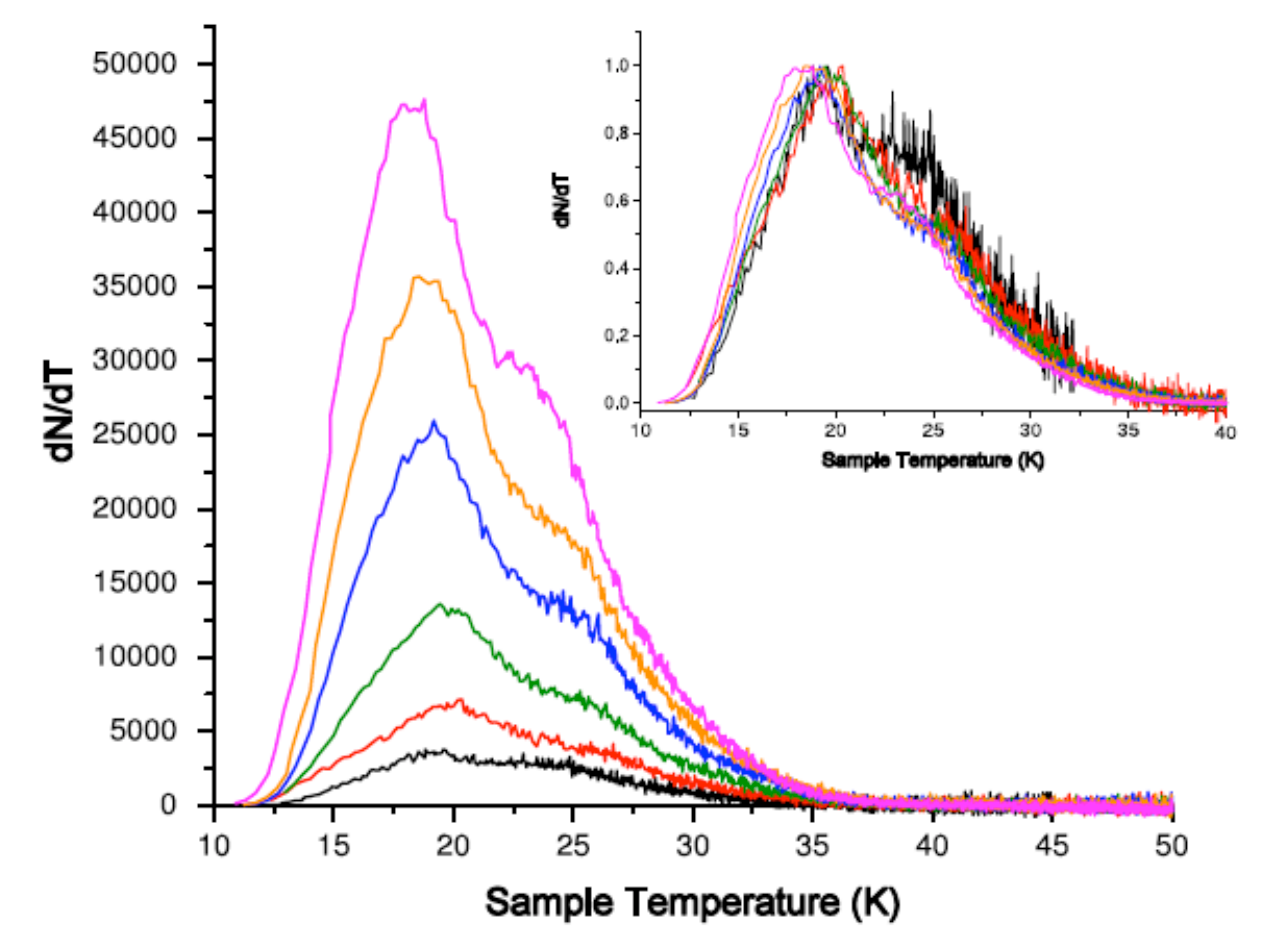}
\caption{TPD traces of D$_2$ after irradiation of an amorphous silicate sample at 12 K for different
lengths of time (2, 4, 8, 16 and 32 min). Inset: normalized traces; for
clarity, the trace of 2 min irradiation (black line) is not shown for
T $>$ 32 K. From \citet{Vidali2010} - GV}
\end{figure}
In the TPD experiments, information of atomic diffusion is derived from H$_2$ formation (recombination) rates coupled with H-atom diffusion. Therefore, the obtained activation energy should depend on the initial coverage of H atoms. That is, experiments at high coverage, where only short-distance diffusion is required for recombination, tend to yield lower activation energies for diffusion (see section~\ref{water_section}).

Given the low surface temperature of dust grains (10-20~K) in some ISM environments where molecules are formed, the motion of an atom that has landed on a grain surface may be restricted. The rate of hydrogen atoms landing on a sub-micron interstellar dust grain is very low. 
For a sub-micron-sized grain of cross-sectional area $\sigma\sim 10^{-10}$ cm$^2$, the number of hydrogen atoms landing on the grain per second is given by $\dot{N}=\frac{1}{4}n_\mathrm{H} \times~v~\times~\sigma$,  where $n_\mathrm{H}$ is the H number density in the gas and $v$ its speed. For $n=10^4$ atoms/cm$^3$ and $v=5\times 10^4~\mathrm{cm\ s^{-1}}$, $\dot{N} = 10^{-2}$ per second. 

For a successful H+H $\rightarrow$ H$_2$ reaction, either the surface needs to be saturated with H atoms, or an H atom has to sample a large part of the grain before encountering another H atom. Thus, experimental and theoretical works have aimed at characterizing H atom diffusion on morphologically complex surfaces and at finding the conditions required to obtain a high coverage of H atoms on the surface. 

For an H atom on a surface, diffusion can proceed  via tunneling or thermal hopping. The surface temperature strongly regulates the thermally activated hopping rate, as in the Arrhenius expression $$\Gamma=\nu  \exp{\left[ \frac{-E_\mathrm{d}}{k_\mathrm{B} T_\mathrm{dust}} \right]},$$ where $E_\mathrm{d}$ is the energy barrier for diffusion and $\nu=\nu_0~\exp{\left(\frac{\Delta S}{k_\mathrm{B} T}\right)},$  with $\Delta S$ being the change in entropy between the saddle point and the adsorption site \citep{Tsong:2005}. In practice, the approximation $\nu \sim \nu_0$ (the characteristic vibrational frequency of the particle in the potential well for motion leading to the saddle point)  is made.  

Tunneling, the rate of which has a very weak dependence on the temperature of the surface, should dominate the diffusion rate at sufficiently low temperatures. However, the tunneling rate depends strongly on the width and height of the energy barrier, as in $$k_\mathrm{d}=\nu_{0}\ \exp{\; \frac{-2a \left(2 m_\mathrm{H} E_\mathrm{d} \right)^{\frac{1}{2}}}{\hbar}},$$assuming a rectangular potential, where $a$ and $E_\mathrm{d}$ are the width and height of the barrier, and $m_\mathrm{H}$ is the mass of the hydrogen atom. In an amorphous solid, there can be a wide distribution of both barrier widths and heights, leading to long residence times of the H atoms in the deep wells of the binding energy landscape. This situation is not unlike the trapping of electrons in amorphous silicon. 

The diffusion of a single atom, by random hopping or tunneling, across the surface is called single particle (or tracer) diffusion, as to distinguish it from the concentration driven diffusion \citep{Tsong:2005} . The direct measurement of tracer diffusion is obtained using visual methods such as field ion microscopy and scanning tunneling microscopy, or with  quasi-elastic particle scattering \citep{MiretArtes:2005}. Due to technical restrictions, and because these techniques are applied to conductive surfaces (typically metals and graphite), they find few applications in probing astrochemically relevant surfaces. Indirect methods, such as the ones used in experiments to study HD and H$_2$ formation (see next Section for details), can provide estimates of the average  mobility of H atoms in astrochemically relevant situations. 

The activation energy for diffusion is empirically related to the binding energy. From experiments probing atoms weakly adsorbed on well-ordered surfaces, we find $E_\mathrm{d}\sim \alpha E_\mathrm{b}$ and $\alpha$=0.3, where $E_\mathrm{d}$ and $E_\mathrm{b}$ are the energy barrier for thermally activated diffusion and the binding energy, respectively \citep{Bruch:2007b}. Analysis of experiments studying  H atoms on dust grain analogs give a wider range of values of $\alpha$, from $\alpha \sim 0.3$ up to $\alpha \sim 0.8$ \citep{Katz:1999,2005ApJ...627..850P,Perets:2007}, depending largely on the morphology of the surface. For the case of H atoms strongly localized on the surface, as in the case of H chemisorbed on graphite or PAHs, the calculated thermal activation energy to migrate out of the adsorption site can be comparable to the energy for desorption. 

Experiments have shown that on the surfaces of a wide variety of solids the formation of H$_2$ occurs via three main mechanisms:  Langmuir-Hinshelwood, Eley-Rideal, and ``hot atom'' \citep{Kolasinski2008}. In the Langmuir-Hinshelwood mechanism, atoms from the gas phase first become accommodated on the surface and then, via diffusion, they encounter each other and react. The resulting molecule might or might not leave the surface, depending on how the energy gained in the reaction is partitioned. 

In the Eley-Rideal reaction, an incoming atom interacts directly with a partner on the surface; the incident atom is not accommodated on the surface. The resulting molecule is likely to leave the surface retaining much of the energy gained in the reaction. Here, the cross-section for the reaction is of the order of atomic dimensions. 
The ``hot atom'' mechanism is similar to the Eley-Rideal one, but here the atom first lands on the surface, without becoming fully accommodated, and proceeds to sample the surface at supra-thermal energy until it finds and reacts with a partner species. 

Examples of detections of Eley-Rideal or abstraction reactions relevant to ISM environments are found in works involving experiments of H atoms interacting with PAHs, hydrogenated amorphous carbon and graphite \citep{1998Natur.391..259S,2002CPL...366..188Z,2006PhRvL..97r6102H,2006PhRvL..96o6104H,2008ApJ...684L..25M,2012ApJ...752....3T,2012ApJ...745L...2M,2016NatSR...619835C}. The various processes involved in the formation of H$_2$ are summarized in Fig.~\ref{figure_processes}.

Theoretical calculations, confirmed by experimental results studying H$_2$ formation on graphite \citep{2008CPL...455..174L,Islam:2007,Islam:2010}, show that  ro-vibrational excitation of H$_2$ leaving the surface upon formation is concentrated around high values of the vibrational quantum number (3-4) and low values of the rotational quantum number. There have been a few attempts to detect a signature of such H$_2$ formation, via observations of  transitions of H$_2$ in appropriate regions of the ISM,  but with no success to date \citep{Tine:2003,Thi:2009}. The  ortho (odd rotational quantum number $J$) to para (even $J$) ratio of H$_2$ can yield information concerning the thermal history of the associated cloud, as well as concerning the conditions associated with H$_2$ formation on the grains of the cloud \citep{Wilgenbus:2000,Lebourlot2000}, although presence of other H$_2$ molecules co-adsorbed on the grain can hinder the release of excited molecules \citep{Congiu:2009}. 

In the laboratory, measurements of the ro-vibrational state of nascent (freshly-formed)  molecules on surfaces have been used to determine whether the ortho  to para  ratio of such newly-synthesized molecules would be different from the statistical ratio \citep{Hama:2013}; for temperatures greater that about 200~K, the statistical ratio is 3. Measurements show that for H$_2$ formed on surfaces of amorphous solid water at low temperature  the nascent molecules possess the appropriate statistical value of the OPR  \citep{2010ApJ...714L.233W,2012ApJ...760...35G}.

Experimental and theoretical results of H$_2$ formation need to be appropriately adapted to the conditions present in the relevant environments of the ISM; this adaption can be performed  by using robust experimental data in computer simulations of processes occurring in ISM environments. Specifically, experiments are performed at much higher H atom fluxes than are present in the ISM, and probe chiefly the kinematics of the reactions.  Hence,  simulations need to translate this laboratory information to reveal its impact under the conditions pertaining in the ISM: such as low fluxes of H atoms impinging on grains and steady-state conditions. For example, \citet{Katz:1999} used rate equations to fit experimental data (temperature programmed desorption traces) of H$_2$ formation on polycrystalline and amorphous silicates and on amorphous carbon. They then used the results to predict the formation of H$_2$ under the conditions of the ISM.  \citet{Cazaux:2005} considered both physisorption and chemisorption interactions in their rate equations and fitted the same data as in \citep{Katz:1999} but with more parameters. They found that only a physisorption interaction between H atoms and the surface was necessary to explain the data. \citet{2005MNRAS.361..565C} instead used continuous-time, random-walk Monte Carlo code to study the effect of surface roughness  on the formation of molecular hydrogen using a model square lattice. These investigators found that roughness increased the grain temperature range over which H$_2$ formation is efficient. Stochastic effects, arising from the fact that the actual size distribution of dust grains in the ISM is skewed to small grains, have also been taken into accounts in models by \citet{Biham:2002,cuppen2006,cazaux:2009,lebourlot2012,Bron14}.

\subsubsection{Silicate surfaces}
The ubiquitous observation of molecular hydrogen in widely varying interstellar environments poses significant challenges in explaining its formation. In diffuse clouds, dust grains are largely bare and the formation of H$_2$ occurs on silicates and amorphous/graphitic carbon (graphite, amorphous carbon, and PAHs). The first experiments studying H$_2$ formation on dust grain analogs involved a polycrystalline silicate \citep{Pirronello:1997a}. In these experiments, the aim was to measure the efficiency of H$_2$ formation under conditions which simulated the ISM.

\begin{figure}
\includegraphics[scale=0.5]{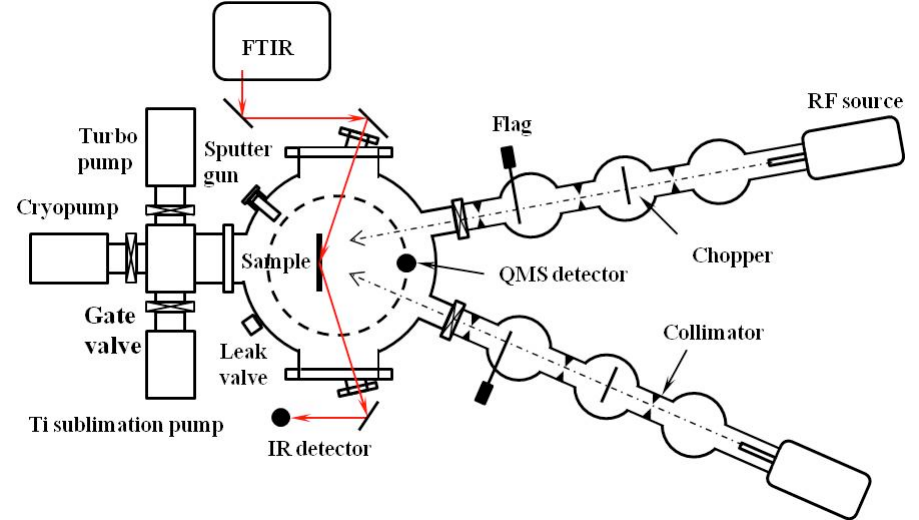}
\caption{Apparatus at Syracuse University used to study H$_2$ formation on silicate surfaces. Two independent beam lines converge on a sample mounted on a rotatable flange. A quadrupole mass spectrometer mounted on a rotatable platform can quantify and identify both the products from the surface and the species in the incident beams.\label{fig1}}
\end{figure}

Quantifying the formation of H$_2$ is particularly challenging. For example, in a typical experiment,  molecular hydrogen is dissociated and the resulting atoms directed onto a sample surface, see Figure~\ref{fig1}. Although it is possible to dissociate up to nearly 90\% of the H$_2$ molecules in such an H atom source, the remaining un-dissociated species will contaminate the sample, making it impossible to determine if  molecules on the surface came from the source or are the product of atomic recombination on the surface. This limitation was lifted in the work of \citet{Pirronello:1997a} by using two beamlines directed at the sample, one dosing H atoms and the other dosing for D atoms. In this situation, under the associated experimental conditions, the formation of HD can only occur on the surface of the sample. Another technical limitation of this class of experiments is associated with contamination. Even in a state-of-the-art ultra-high vacuum apparatus (base pressure 10$^{-10}$ torr), the adsorption of background gas (mostly hydrogen) on the surface of the sample limits the sensitivity and duration of experiments studying H$_2$ formation. Using highly collimated beams, as shown in Figure~\ref{fig1}, allows experimental operating pressures approaching $10^{-10}$ Torr. 

Another technical limitation associated with laboratory experiments is the fact that fluxes of H atoms employed are, for practical reasons, orders of magnitude higher than in the ISM. This mismatch of fluxes cannot be solved directly. However, with careful experimental design and the use of simulations to reproduce the conditions in the ISM, as performed by \citet{Katz:1999} and \citet{Biham:2002}, the efficiency of H$_2$ formation in the ISM can be obtained from experimental kinetic data. Further work by the Biham's group studied the effect of particle size \citep{Lipshtat:2005} and porosity \citep{Perets:2006} on H$_2$ formation in interstellar environments. Diffusion of H atoms was included in the simulations of the formation kinetics \citep{Katz:1999}, revealing that the ratio of the energy barrier for H atom diffusion to the binding energy is considerably higher than the typically assumed value  ($\sim$ 0.3 ) for physisorbed atoms on single crystal surfaces \citep{Bruch:2007b}. Because the H atom binding energy could not be well constrained, only an upper limit of 0.7 could be obtained for this ratio in the fitting of \citet{Katz:1999}. The reason for this unusually high diffusion barrier is likely to be the complex morphology of polycrystalline and amorphous silicates.

Subsequent experiments showed that the efficiency of atomic recombination to form H$_2$ is dependent on the morphology of the surface, the efficiency being larger on amorphous silicates than on crystalline or polycrystalline silicates. The simulations by \citet{Katz:1999} of H$_2$ formation on polycrystalline silicate and on amorphous carbon, and by \citet{Perets:2007} on amorphous silicate, showed that, under the conditions present in diffuse interstellar clouds, the efficiency of atomic recombination to form H$_2$ is high over only a narrow range of temperatures. In the experiments modeled by \citet{Katz:1999} and by \citet{Perets:2007}, most of the molecules that were formed remained on the surface and only a small minority left the surface following their formation. Both the molecules coming off the surface upon formation, and the ones that remained on it, were detected using a quadrupole mass spectrometer whose sensitivity is inversely proportional to the speed of the particles. 

In an experiment studying H$_2$ formation, in which it was possible to detect molecules leaving the surface in superthermal ro-vibrational states, \citet{Lemaire:2010} found that some nascent molecules were formed on, and ejected from, the surface at a temperature as high as 70K. Although the kinetic energy of H$_2$ was not measured, experimental conditions and analogy with the experiments studying H$_2$ or HD formation on graphite, by \citet{2006JChPh.125h4712B,Islam:2007,2008CPL...455..174L,Islam:2010}, suggest that the kinetic energy was of the order of an eV.  Thus, it is possible that earlier experiments underestimated the proportion of nascent molecules immediately leaving the surface following their formation. 

The influence of the morphology of the silicate surface on the kinetics of molecular hydrogen formation was studied by \citet{He:2011}. This work determined the distribution of the binding energy of the species on the surface using TPD. Specifically, the shape of TPD traces, which record the desorption rate (the differential of the  desorption yield), as a function of  temperature, can be fitted using a distribution of binding energies. 

As mentioned before, it is difficult to detect atomic hydrogen in TPD experiments, especially when the surface coverage is less than one layer. However, experiments studying D$_2$ show a dramatic difference between TPD spectra from a single crystal of forsterite (Mg$_2$SiO$_4$) and from an amorphous silicate. The derived binding energy distribution for the amorphous sample is much wider, and centered at a much higher desorption energy, than the one for the single crystal. Experiments studying  H+D $\rightarrow$ HD formation are consistent with the D$_2$ experiments: they show that HD formation on a silicate crystal occurs at lower temperatures than on amorphous silicate, suggesting that thermally activated diffusion plays an important role in the reaction \citep{He:2011}. 

Compared with the significant number of theoretical investigations of the interaction of H atoms with  carbonaceous surfaces, there are few reports of theoretical investigations of H atoms interacting with silicate surfaces. Such calculations have been performed involving the stable (010)  surface of Mg$_2$SiO$_4$ as well as the (001) and (110) surfaces which have higher surface energies.  \citet{Goumans:2009b} used  an embedded cluster approach where part of the surface was described by DFT and part by analytic potentials, while \citet{Garcia-Gil:2013}, \citet{Navarro-Ruiz:2014}, and \citet{Navarro-Ruiz:2015} employed Density Functional Theory.  H$_2$ is formed more readily  on the (010) surface due to the fact that on the other surfaces H atoms are more strongly adsorbed and the barriers to diffusion are thus higher. 	Mg atoms are the most favorable sites for physisorption, while chemisorption is on the oxygen site. However, the physisorption energy of H on crystalline silicates, and the energy barriers to diffusion, are calculated to be considerably higher than the values given by experiments on amorphous silicates. \citet{Goumans:2009b} invoked hydroxilation of the surfaces used in experiments to reconcile the discrepancy between these theoretical and experimental values, while \citet{Navarro-Ruiz:2014} pointed out  the challenge for computational studies in correctly taking into account the large dispersion energies in  weak interactions and open shell systems when using DFT. The calculations show that H$_2$ formation via the Langmuir-Hinshelwood mechanism  is  favored on the (010) surface \citep{Navarro-Ruiz:2014,Navarro-Ruiz:2015}.

\subsubsection{Carbonaceous surfaces}



\paragraph{H$_2$ formation on graphite}

The interaction of atomic hydrogen with graphite surfaces, and the pathways to molecular hydrogen formation on these surfaces, have been studied in considerable detail both theoretically and experimentally. Hydrogen atoms can both physisorb and chemisorb on graphite:

Physisorbed H atoms are weakly bound in a shallow potential well with a depth of 43.3 $\pm$ 0.5 meV resulting in a ground state binding energy of 31.6 +- 0.2 meV, as determined by scattering experiments \citep{1980JChPh..73..556G}. The sticking coefficient has been estimated theoretically to be 5-10 \% for H atoms with translational energies ranging from 0 to 50 meV \citep{2008JChPh.128k4704M,2011JChPh.134k4705L}. 

Once in the physisorbed state, the H atom is highly mobile on the surface with a diffusion barrier predicted by theory to be only 4 meV \citep{Bonfanti:2007}. This high mobility allows H atoms to scan a large area of the surface and recombine with any other H atoms they encounter; atoms which could be physisorbed, chemisorbed or incident  from the gas phase.  This reactivity can occur both at low temperatures, where the atom's high surface mobility is assisted by tunneling, and at higher temperatures, where the high thermally induced mobility may allow a significant surface area of a grain to be explored by the atom, even if the atom's lifetime in the physisorbed state is extremely short \citep{Cuppen2008,2006JChPh.124k4701C}.

Chemisorption of H atoms on the graphite surface is more complex than physisorption. A single H atom can chemisorb above a carbon atom in the graphite surface with a binding energy of 0.7-1.0 eV \citep{Sha:2002b,2006PhRvL..96o6104H,Casolo:2009b,Ivanovskaya:2010a}. However, this binding requires the associated carbon atom to pucker up, out of the surface, by 0.1 \AA . Thus, the chemisorption is associated with a large energy barrier of 0.15-0.2 eV \citep{1999CPL...300..157J,2006PhRvL..96o6104H}. As a consequence of this barrier, the sticking probability for H atoms into the chemisorbed state is highly energy dependent and has mainly been estimated theoretically \citep{2005JChPh.122a4709S,Kerwin2006,Kerwin2008,Morisset2008,Morisset2010,Karlicky:2014,Bonfanti:2015}. 

The diffusion of isolated chemisorbed H atoms on graphite is highly disfavored with calculated barriers of 0.8-1.1 eV \citep{2006PhRvL..97r6102H,2003CPL...368..609F}. The chemisorption of one H atom dramatically changes the reactivity of carbon atoms on specific neighboring graphitic sites, yielding a reduction, or even disappearance, of the barriers for chemisorption of a second incoming H atom nearby  \citep{2006PhRvL..97r6102H, 2006CPL...431..135R}. As a consequence, sticking probabilities of H atoms on sites in the vicinity of previously chemisorbed H atoms are increased. Thus, hydrogen atoms predominantly adsorb into dimer or cluster structures on the graphite surface  \citep{2003CPL...368..609F,2006PhRvL..96o6104H,2006PhRvL..97r6102H}. Diffusion barriers of H atoms within such clusters are reduced and values of 0.2-0.4 eV for diffusion into more energetically favorable adsorption structures have been found \citep{2006PhRvL..96o6104H}. 

Molecular hydrogen formation via reactions between chemisorbed H atoms has a high activation barrier of $\sim 1.4$ eV \citep{2002JChPh.117.8486Z,2006PhRvL..96o6104H}. In contrast, Eley-Rideal reactions between impinging gas-phase H atoms and chemisorbed H atoms have been shown to be barrier-less for the case of abstraction from specific dimer structures \citep{2007CPL...448..223B}, and, in recent calculations, also for abstraction of hydrogen monomers \citep{2011PCCP...1316680B}. Experimental measurements involving 2000 K H atoms impinging on a hydrogenated graphite surface, held at 300 K, show Eley-Rideal (including hot atom) cross-sections of 17 \AA $^2$ at low H coverage decreasing to 4 \AA $^2$ at high coverage \citep{2002CPL...366..188Z}. However, at low collision energies, theory predicts that quantum reflection effects will limit the Eley-Rideal abstraction cross section \citep{2009JPCA..11314545C}.

A few measurements regarding the energy partitioning in molecular hydrogen formation on graphite have been reported. The ro-vibrational distribution has been measured for molecular hydrogen formed via reactions involving physisorbed H atoms on a graphite substrate held at 15-50 K. The measurements show high vibrational excitation energies with the vibrational distribution peaking at $v$=4 \citep{2008CPL...455..174L}, while only minimal rotational excitation, corresponding to an excitation temperature of 300 K, was observed \citep{2006JChPh.124k4701C,2008CPL...455..174L}. Theoretical calculations predict significantly higher vibrational excitation \citep{Morisset:2005}. 

The kinetic energy for hydrogen molecules formed from reactions between chemisorbed H atoms on graphite has been measured experimentally and yielded an average of 1.4 eV \citep{2006JChPh.125h4712B}. No measurements exist of the energy partitioning for molecular hydrogen formed by Eley-Rideal abstraction reactions with chemisorbed species. However, using relaxed surface, theory predicts that the majority of the released energy goes into vibrational excitation \citep{Martinazzo:2006a,Martinazzo:2006b,2009PCCP...11.2715B,2010CPL...498...32S,2003PCCP....5..506M,morisset2004role} . In these calculations geometrical constraints limit information on the partitioning into rotational excitation. However,  approximating the surface as rigid, theoretical results \citep{Farebrother:2000,Meijer:2001} reveal H$_2$ forms with low vibrational excitation and  high rotational excitation.

\paragraph{H$_2$ formation on amorphous carbon surfaces}
Molecular hydrogen formation was studied experimentally on low temperature (5-20 K) compact amorphous carbon surfaces upon irradiation with 200 K H and D atoms. Formation efficiencies above 50 \% were observed at 5 K, followed by a rapid fall off in efficiency with increasing surface temperature to below 10 \% at 18 K \citep{1999A&A...344..681P}. This fall off in efficiency was ascribed to the short lifetime of weakly bound physisorbed H/D atoms at increased temperatures, and may also indicate a reduced mobility of the adsorbed atoms, compared with the case of graphite. The height of the diffusion barrier in this situation is expected to be very strongly dependent on the exact nature of the carbonaceous surface.

Molecular hydrogen formation on hydrogenated porous, defective, aliphatic carbon surfaces has also been studied experimentally. H atoms chemisorb strongly to carbon defects with low activation barriers, an experimentally determined activation barrier of 70 K has been reported \citep{Mennella2006}, and typical binding energies in the range of $3-6$ eV. Hence, molecular hydrogen formation from two H atoms chemisorbed on such defective carbon surfaces is generally not energetically favorable. However, once the high energy binding sites have been saturated with H atoms, less tightly bound H species may then adsorb and be available for reaction. Furthermore, molecular hydrogen formation involving chemisorbed H atoms on these surfaces can still proceed via Eley-Rideal or hot atom abstraction reactions. The cross-section for such reactions has been determined for a hydrogenated porous, defective, aliphatic carbon surface over a substrate temperature range $13-300$ K and for H atom temperatures ranging from 80 to 300~K. The results show that the reaction is barrierless at low surface temperatures, while a small activation barrier of 130 K was found at surface temperatures above $\sim 100$ K \citep{2008ApJ...684L..25M}. An abstraction cross-section of 0.03 \AA $^2$, roughly $1/100$ of the value found on graphite, was reported for 300 K H atoms impinging on a 300 K sample \citep{2008ApJ...684L..25M}. The energy partitioning in molecular hydrogen formation on these surfaces is determined by the exoergicity of the reaction, as a C-H bond has to be broken, and by the finding that the majority of the synthesized  molecules are retained in the porous structure following their formation. As a result, molecular hydrogen formed on these surfaces is expected to desorb with low ro-vibrational excitation and low kinetic energy, that is with a temperature, effectively, that of the surface.

An alternative molecular hydrogen formation pathway on hydrogenated amorphous carbon is via irradiation with VUV photons. Experimental investigations show that irradiation of hydrogenated amorphous carbon with 6.8-10.5 eV photons leads to breaking of one C-H bond for every 70 incident photons. The majority of the  H atoms released by this process ($\sim 95 \%$) were observed to react to form molecular hydrogen \citep{2014A&A...569A.119A}. The H$_2$ molecule synthesized in this manner was observed to be retained in the surface structure.  These dynamics again indicate that when these H$_2$ molecules eventually desorb, they will do so with a ro-vibrational distribution and kinetic energies determined by thermal equilibrium with the substrate.

\paragraph{H$_2$ formation on PAHs}
The ability of PAH molecules to catalyze molecular hydrogen formation has been investigated by several authors \citep{1998Natur.391..259S,2008ApJ...679..531R,2012ApJ...752....3T,2012ApJ...745L...2M,2016NatSR...619835C}. Simple abstraction of H atoms from the PAH, by incident H atoms to generate H$_2$, is not energetically favorable on un-functionalized PAH molecules, due to the high C-H bond energies of $\sim 4.8$ eV. However, such reactions can be activated by the initial addition of excess H atoms to the PAH molecule, which then becomes superhydrogenated. Both experimental and theoretical calculations have demonstrated that this superhydrogenation via hydrogen atom addition is possible \citep{2008ApJ...679..531R,2012ApJ...752....3T,2012ApJ...745L...2M,2016NatSR...619835C}. Addition barriers for PAHs vary, depending on charge state and degree of superhydrogenation \citep{2008ApJ...679..531R,2016NatSR...619835C}. 

On the cation of coronene (a prototypical PAH), the first H addition to  a C atom on the periphery of the molecule, an ``edge'' carbon,  has a very small barrier (10 meV); while a second H atom addition has been calculated to have  a barrier of 30 meV \citep{2016NatSR...619835C}. Subsequent H addition reactions have barriers which alternate in magnitude. Some of the barriers are much higher (of the order of 0.1 eV), leading to the predominance of coronene cations with a magic number of H atoms attached (+5, +11 and + 17 extra hydrogens). The barriers for H addition to a coronene cation with an even number of extra H atoms is large, while hydrogenation of a coronene cations with an odd number of extra H atoms is small, and this alternation occurs until full hydrogenation is reached. 

On neutral coronene, a barrier of 60 meV for the addition of the first H atom to an edge site has been calculated. Subsequent H addition reactions have lower barriers or are even barrierless \citep{2008ApJ...679..531R}. Comparable experiments on coronene show that the first H atom addition, for atoms with energies of 1400-2000~K, has a cross-section of 0.7 $\pm$ 0.4 \AA $^2$ \citep{2012ApJ...752....3T}. Experiments on both cationic and neutral PAHs demonstrate that, for H atom beams with energies ranging from 300~K - 2000~K, high degrees of superhydrogenation (in many cases complete superhydrogenation) with one excess H atom per C atom, are attainable \citep{C3FD00151B}.

For the neutral coronene molecule, hydrogen addition measurements can be simulated using addition cross-sections ranging from 0.55-2.0 \AA $^2$ \citep{2012ApJ...752....3T,C3FD00151B,2016NatSR...619835C}. However, once the molecule is super-hydrogenated, abstraction reactions to form H$_2$ compete with H atom addition \citep{2008ApJ...679..531R}; such abstraction reactions are often barrierless. These abstraction reactions have been directly observed, via H-D exchange, using the coronene molecule \citep{2012ApJ...745L...2M,2012ApJ...752....3T}. For incident H atoms at 300~K,  an abstraction cross-section of 0.06 \AA $^2$ per excess H atom was measured \citep{2012ApJ...745L...2M}.  For more energetic H atoms (2000~K) a lower cross-section of 0.01 \AA $^2$ per excess H atom was observed to yield better agreement between simulations and measurements \citep{C3FD00151B}.

\subsubsection{Water ice surfaces}\label{water_section}
In cold molecular clouds where the visual extinction is larger than 3, cosmic dust grains are covered by water ice. The abundance of the H$_2$ molecule in molecular clouds is determined by the balance between destruction in the gas phase and formation of H$_2$ on the icy dust surfaces. In this section, we briefly overview the experimental work studying the physico-chemical processes involving hydrogen, which lead to H$_2$ formation on these icy mantles covering the  dust grains. As discussed above, H$_2$ formation generally occurs via a sequence of elementary processes: sticking, surface diffusion, and recombination reactions.  Each process has been targeted by various classes of experiments which will be described in the following subsections. 

Upon H$_2$ formation, dissipation of the heat of reaction must occur. Following the recombination event, the excess energy of about 4.5~eV has to be partitioned between the kinetic and internal energy of the nascent H$_2$ and the grain. The kinetic energy of the molecule is redistributed by collisions and remains within the cloud, while internal excitation followed by deexcitation generates IR photons that may escape from the cloud. Special attention should be paid to  the OPR of the nascent H$_2$ molecules. The energy difference of $\sim$14.7 meV (corresponding to 170 K) between the ground states of the two nuclear spin isomers (ortho and para states) is large enough to affect the gas phase chemistry of H$_2$ in molecular clouds at around 10 K. Furthermore, since the conversion between the nuclear spin states is forbidden in the gas phase, ortho-para interconversion on the ice surface plays a crucial role in determining the OPR of H$_2$.

\paragraph{Interstellar Ices}
Astronomical observations show that the 3 $\mu$m-bands seen in the infrared absorption spectra of dust are well explained by the existence of water ice as a mantle in polycrystalline and/or amorphous forms \citep{1988ApJ...334..209S,1993dca..book....9W}. Amorphous ice (hereafter, Amorphous Solid Water: ASW) dominates at low temperatures over the polycrystalline (PCI) phase. Therefore, experimental studies began by preparing and characterizing ASW as an analogue of interstellar ice mantles. The quantity of ice observed in such mantles cannot be due to simple freezing out (accretion) of H$_2$O molecules synthesized in the gas phase; thus, reactions on the grain surface between hydrogen and oxygen  are required. Nevertheless, it is generally believed that ASW produced experimentally by the deposition of water vapor onto cold substrates reproduces interstellar ice-mantles fairly well. However, it should be noted that the porosity of ASW strongly depends on the substrate temperature and the deposition method of H$_2$O vapor \citep{1999Sci...283.1505S}. Furthermore, it is known that additional processing due to D-atom exposure on pre-deposited ASW \citep{2011PCCP...13.8037A} and also synthesis of ASW through D + O$_2$ reaction on cold surfaces \citep{Oba:2009} result in formation of compact (less-porous) amorphous ice. 
For the experiments relevant to H$_2$ formation on icy dust grains, the ASW samples were typically prepared at around 10 K by introducing H$_2$O vapor into vacuum chambers, sometimes through a capillary plate. 

The chemical and physical properties of ASW are an important research target not only for astronomers but also for physicists and chemists \citep[for a review, see][]{2008PrSS...83..439W}. Briefly, the main remarkable features of ASW produced at $\sim$10 K are: the porous structure with its large surface area and, because of its irregular morphology, the wide variety of different adsorption sites for hydrogen atoms and molecules. The surface area of ASW was found to be 10 times larger than that of polycrystalline ice when ASW was deposited with the amount of $10^{17}$ molecules cm$^{-2}$ at 10 K  \citep{2008CPL...456...36H}. These features significantly influence the associated physico-chemical processes of hydrogen atoms and molecules, such as adsorption, desorption and diffusion, when they interact with the ice surface. 

\paragraph{Sticking and adsorption energy}
H$_2$ formation by H-H recombination first requires adsorption of H atoms onto the icy mantle of a dust grain. Therefore, the sticking coefficients and adsorption energies of H atoms on these surfaces are key parameters. Because of significant  experimental difficulties in quantifying these processes, these quantities have often been obtained theoretically. In such calculations, both semi-classical and quantum approaches have been employed \citep{1970JChPh..53...79H,hollenbach71,1991ApJ...379..647B,1991JChPh..95.6026B,1998A&A...330..773M,1999MNRAS.306...22T,2007MNRAS.382.1648A}. Here adsorption energies of H atoms on crystalline water ice and ASW were determined to be in the range of 300-600 K. \citet{2007MNRAS.382.1648A} showed that the adsorption energy for H atoms on ASW is higher than that for crystalline ice by approximately 200 K. 

 Molecular Dynamics calculations (MD) showed that the sticking coefficients of H atoms on ASW are near unity at around 10 K, when the kinetic energy of the impinging atoms is less than about 100 K. In the MD calculations \citep{1991ApJ...379..647B,2004A&A...422..777A,2007MNRAS.382.1648A}, adsorption is defined as trajectories having stabilized at a total H-atom energy of $\sim$100~K for duration of $<$10-15 ps, where the total energy consists 
of the adsorption and kinetic energies of H atom. The sticking coefficient was found to decrease steeply with the increasing incident energy of atoms. However, it should be noted that MD calculations can only follow a very short time period of the H-atom dynamics.  Thus, the calculated sticking coefficients may differ from those derived under realistic conditions in experiments and astrophysical environments \citep{2004A&A...422..777A}.
\citet{1991ApJ...379..647B} evaluated the temperature dependence of the sticking coefficient. A very good match between the experimental values of the sticking coefficient for molecular hydrogen and the theoretical value has been found \citep{Matar:2010}. The sticking coefficients for H atoms calculated to date are summarized in \citet{2014ApJ...790....4V}. However, direct measurements of the sticking coefficients and
adsorption energies of H atoms are not easy to perform and thus have not yet 
been made.

\paragraph{Diffusion}
In molecular interstellar clouds, the accretion of an H atom onto a dust grain occurs about once a day and the coverage of H atoms on the dust grains is considered to be very low. In this situation, the LH mechanism should dominate the recombination process. Since the recombination itself is a radical-radical barrier-less reaction, the surface diffusion of H atoms across the dust surface is the rate-limiting process for H$_2$ formation. 

The diffusion rate, via thermal hopping over the appropriate energy barriers, can be simply derived from the Arrhenius equation. For example, assuming activation energies of 200 and 500 K with a frequency factor of $10^{12}$~s$^{-1}$, the hopping rates become $10^3$ and $10^{-10}$~s$^{-1}$, respectively, at a surface temperature of 10~K. However, since in reality the surfaces of ASW and even those of PCI consist of a variety of adsorption sites accessed by different diffusion barriers, it is not easy to predict the actual diffusion rate. Furthermore, diffusion by tunneling should also be considered at low surface temperatures. 

The tunneling-diffusion rate is much more sensitive to the shape, especially width, of barriers than thermally activated hopping. Although a rectangular potential is often used for estimating a tunneling rate, this is a very simplified model. The tunneling diffusion of H atoms on ice was first studied theoretically by \citet{1983JPhCh..87.4229S}. He evaluated tunneling diffusion for a non-periodic amorphous structure and indicated that the diffusion on ASW is significantly slower than on PCI, leading to a corresponding recombination rate at least three orders of magnitude lower on ASW. 

In recent years, assuming a thermal hopping mechanism, activation energies for H atom surface diffusion have been derived from TPD experiments. Analyzing the TPD spectra obtained experimentally after H or D atoms have been deposited on ASW, \citet{2005ApJ...627..850P} reported the activation energy to be in a range of 41-55 meV. Similarly, applying a simple rate equation model to the results of TPD experiments, \citet{2008A&A...492L..17M} determined the activation energy of D atom diffusion to be about 22 meV. This discrepancy in the results between the two groups could partly originate from different surface coverages of atoms, which require different diffusion lengths to be covered for recombination. 
Since the  energy depths of  adsorption sites have a distribution, long-distance diffusion tends to be limited by higher activation barriers than short-distance diffusion. Also, for an irregular surface structure, like that of ASW, TPD experiments, and their analysis by rate equation models with multiple parameters, result in some ambiguity in determining the value of the diffusion barrier. 

Recently, using a combination of laser-induced photo-desorption and resonance enhanced multi-photon ionization (REMPI) methods, Watanabe and coworkers clarified how the diffusion mechanisms, either tunneling or thermal hopping, depend on the diffusion distance on ices \citep{2015PhRvL.115m3201K}, and measured the activation barriers for thermal hopping of H and D atoms on ASW \citep{2010ApJ...714L.233W,2012ApJ...757..185H}. Watanabe and coworkers demonstrated that the ASW surface contains various sites that can be categorized into at least three groups: very shallow-, middle-, and deep-potential sites, with associated diffusion activation energies of $<$18, 22 (23 meV for D atoms), and $>$30 meV, respectively. These values of barriers cover those obtained by TPD experiments, and have been confirmed by recent quantum  calculations \citep{Senevirathne:2017}.

\paragraph{H$_2$ formation on ices}
\subparagraph{Formation by recombination}
H$_2$ formation by atomic recombination has most commonly been  investigated in TPD experiments where the yields of H$_2$, HD, and  D$_2$  were detected as a result of the recombination of H and/or D atoms which were deposited on ASW \citep{2001ApJ...548L.253M,2002ApJ...581..276R,2003Sci...302.1943H,2005ApJ...627..850P,Amiaud:2007}. In these experiments, recombination efficiencies, as a function of the surface temperature, were determined together with several parameters characterizing elementary processes such as  diffusion. These results are well-summarized elsewhere \citep{2013ChRv..113.8762V}. Another class of experiment was performed by monitoring the yield of H$_2$ or HD species, formed by recombination, which were then photodesorbed from the ASW surface; no formal heating of the surface was performed \citep{2003Sci...302.1943H,2010ApJ...714L.233W}. 

As noted above, the formation of H$_2$ by atomic recombination releases an excess energy of about 4.5 eV. The partitioning of this energy resulting from H$_2$ formation is an important issue in astronomy.  This importance arises because ro-vibrational emission observed from H$_2$  may  partly be due to excitation by the energy released by recombination \citep[][and references therein]{2001ApJ...561..843T}. The number of studies probing the energy partitioning upon H atom recombination on ASW is limited. MD calculations showed that H$_2$ synthesized from H atoms on ASW can be vibrationally excited in $v$=7-8 \citep{1999ApJ...520..724T}. \citet{2003ApJ...596L..55R} demonstrated experimentally that the kinetic energy of the desorbing H$_2$ molecules, formed by recombination on ASW, is almost equivalent to the surface temperature. Considering the vibrational energy of the nascent H$_2$, no highly excited species have been detected following atomic recombination on ice \citep{2003Sci...302.1943H,2012ApJ...757..185H}. This lack of vibrational excitation may well be because most of the freshly synthesized H$_2$ molecules are re-trapped within the  ASW layer and thermalized there \citep{2010ApJ...714L.233W}.

\subparagraph{Formation by energetic processes}
  Energetic agents (ions, electrons and photons) irradiating ices and ice mixtures can dissociate molecules. The fragments of such dissociative processes can recombine to generate new molecules. 

\begin{itemize}
 \item [-] Ion bombardment: \citet{1982NIMPA.198....1B} performed the first quantitative measurements of H$_2$ and O$_2$ formation following ion irradiation by sending 1.5 MeV He$^+$ onto layers of ASW a few hundred Angstrom thick, with the surface held at 10 K. The radicals generated, that are mobile in the energized region around the ion track, recombine in a variety of different ways and diffuse out of the ice layer during the irradiation. When investigating effects induced by energetic agents, it is mandatory to use dose rates in the range in which the effects (molecule formation, sputtering and so on) are proportional to fluxes. Only under such conditions single particle effects, not cumulative ones, are measured.  Under these conditions experimental results may be applied to interstellar environments, where particle fluxes are orders of magnitude lower than those used in the laboratory. With this "caveat", the yields (i.e. the average number of molecules released by the ice per each impinging ion) were measured as function of the ice temperature \citep[see Fig. 7 of][]{1982NIMPA.198....1B}. Values measured in the laboratory, for a narrow range of energies of the impinging ions, may be extrapolated to all other energies encountered in space by taking into account  that the formation and release of product species scale with the stopping power d$E$/d$x$ (the energy lost per unit path length in the solid) of swift ions. The yield then reaches its maximum for kinetic energies of the fast ions around roughly 100 keV per nucleon, in the case of light ions \citep{1982NIMPA.198....1B}. 
\citet{1988A&A...196..201P}  and \citet{1991A&A...245..239A} used these laboratory results in a Monte Carlo simulation to asses the importance of the production of H$_2$ by cosmic ray bombardment of icy grain mantles inside, and at the edge of, dense clouds. Their evaluation should be revised due to the fact that the ionization rates involving cosmic rays in dense clouds have been increased by about an order of magnitude since their papers where published \citep[see e.g.][]{1998ApJ...499..234C}.


    \item [-] UV Irradiation:
\citet{1995P&SS...43.1311W,1995Natur.373..405W} observed the H$_2$ yields desorbed from 50 to 100~K ice during irradiation with Ly-$\alpha$ photons. \citet{1991ASSL..173...87K} measured the thermal desorption spectra of H$_2$ molecules from  UV-irradiated pure water ice. These were both rather qualitative investigations. Later, UV irradiation of water ice \citep{2000ApJ...541..772W} showed that the fraction of D$_2$ molecules formed in D$_2$O ice, at 12 K, after a dose of $10^{18}$ cm$^{-2}$ UV photons ($\lambda\sim$ 126, 172 nm) was around 2\% of the number of irradiated D$_2$O molecules. Most of the D$_2$  molecules produced, in contrast to the case of ion irradiation, remained trapped in the ice layer. The estimated cross section for this molecular synthesis was about $2.4\times 10^{-18}$ cm$^2$, a value that may render such a process non-negligible under the UV conditions pertaining in certain molecular clouds.
\citet{2008ApJ...682L..69Y} detected the formation and release of H$_2$ molecules in highly excited states (v=0-5; J=0-17) following UV (157 nm) irradiation of ASW at 100 K. Using the REMPI technique, they were able to recognize two distinct formation mechanisms of H$_2$: a hot H atom abstracts an H atom from an H$_2$O molecule, yielding cold H$_2$ molecules, and recombination of two H atoms that yields ro-vibrationally hot H$_2$.

    \item [-] Electron Irradiation:
In H$_2$ formation induced by electron irradiation of water ice it is important to understand the process of secondary electron generation by ion bombardment of ice. The process of direct injection of free electron from the gas phase is not very relevant to interstellar environments. \citet{1996PhRvL..77.3983K} and \citet{1994JChPh.101.3282K} studied the production of molecular H(D)$_2$ by irradiating ASW with low energy (5-100 eV) electrons. The D$_2$ yield increased monotonically with the temperature of the substrate in the range 88 - 145~K, doubling at the highest temperatures. The D$_2$ molecules showed very little translational energy (20 - 50 meV) but were vibrationally ($v$ = 0 - 4) and rotationally ($J$ = 0 - 12) excited. The authors suggested that the dominant mechanisms for production of D$_2$ at 100 eV incident electron energy are dissociative recombination of holes (D$_2$O$^+$ or D$_3$O$^+$) with quasi-free or trapped electrons and dissociation of excitons at the vacuum-surface interface. 
\end{itemize}

\paragraph{Ortho to para conversion of H$_2$ on ice}
The ortho-to-para ratio (OPR) of H$_2$ released into the gas phase  affects the chemical evolution of a cloud, including deuterium fractionation, \citep{2011ApJ...739L..35P,2016A&A...588A..27B,2017MNRAS.466.4470M}.  As noted above this affect of the OPR arises because the energy difference between the ortho and para rotational ground states is $\sim$14.7~meV; this energy  corresponds to a temperature of 170~K, which is significantly higher than the local temperature of molecular clouds. 

The radiative transformation between ortho and para states is forbidden in the gas phase. Therefore, it is highly desirable to understand how the OPR behaves when the molecules are adsorbed on icy dust grains. \citet{2010ApJ...714L.233W} demonstrated experimentally that the OPR of nascent H$_2$ formed on ASW, at a surface temperature of around 10 K, is at the statistical value of 3, but diminishes rather quickly when H$_2$ is adsorbed on ice at 10 K. Similarly, \citet{2012ApJ...760...35G} reported that the OPR of D$_2$ formed on ASW showed a statistical OPR. 

In recent years, using the REMPI method, ortho-to-para conversion rates have been measured on ASW. \citet{2011NatPh...7..307S} determined a conversion rate of $\sim2.7\times 10^{-3}$~s$^{-1}$ for H$_2$ and found that the conversion rate for D$_2$ is much slower than that for H$_2$; a fact later confirmed by another group \citep{2012ApJ...757..185H}. \citet{2011PCCP...13.2172C} reported that trace amounts of O$_2$  on ASW accelerates the nuclear spin conversion for D$_2$. The conversion mechanism on ASW was first proposed by \citet{2011NatPh...7..307S}. However, the role of surface temperature is not considered in their model. Very recently, the conversion rate for H$_2$ was found to increase steeply, from $\sim2.4\times 10^{-4}$ to $\sim1.5\times 10^{-3}$~s$^{-1}$, over the temperature range of 9.2$-$16~K \citep{2016PhRvL.116y3201U}. This temperature dependence of the conversion rate can be explained by an energy dissipation process via phonons.

\subsection{Astrophysical models}   

In the domain of astrophysics, numerical models are of tremendous importance. Astrophysical objects are complex systems coupling a wide range of physical and chemical processes and of spatial and temporal scales where physical conditions (density, UV flux, etc...) vary by orders of magnitude. In contrast, laboratory experiments isolate one or a few processes, which they study over a limited range of physical conditions. Numerical models then constitute the indispensable link between our experimental and theoretical knowledge of the elemental processes and astronomical observations, transforming experimental results into observable predictions, and tracing back unexpected observations to specific processes that need to be investigated in the laboratory.

Due to computational limitations, models cannot treat all aspects at the most detailed level. Astrophysical models differ in the way they focus on some parts of the physics and chemistry, which they treat in great detail because they are the most relevant in a given astrophysical environment, while neglecting, or approximating more crudely, other aspects that are less relevant. 
We first present here the main approaches to modeling the H$_2$ formation rate itsef, before discussing the broad families of astrophysical models where these approaches are used, defined in term of the areas of physics and chemistry where they place their focus.

%
%

\subsubsection{Calculating H$_2$ formation efficiency in models}
In order to calculate the H$_2$ formation rate in models based on laboratory experiments and/or astrophysical observations, different methods (corresponding to different levels of approximation) have been used.

Until relatively recently, most types of models used an approach based entirely on observational measurements of the H$_2$ formation rate in diffuse molecular clouds \citep{1974ApJ...191..375J,Gry2002}. This approach simply parameterizes the formation rate as the product of the atomic hydrogen abundance $n(\mathrm{H})$, the total gas density $n_\mathrm{H}$ (as a tracer of the dust abundance), and an efficiency parameter $R_f$, determined to be roughly $3\times10^{-17}\,\mathrm{cm}^3 \mathrm{s}^{-1}$. Additional dependencies, like metallicity and gas temperature, were often included, sometimes also taking the experimentally measured sticking efficiency into account.

The simplest approach based only on laboratory data is known as the \emph{rate equation approach}, treating the surface abundances (here of H and H$_2$) as continuous variables obeying differential equations (similar to gas phase chemistry), using the experimentally constrained rates for each process (desorption, migration, etc...). One such widely used rate equation model of H$_2$ formation is \cite{2002ApJ...575L..29C,Cazaux2004}. Despite its advantages of simplicity and low computational cost, several approximations in this approach can be problematic in some regimes so that other numerical techniques have been developed \citep{2003Ap&SS.285..725H}. The first main problem is the fact that surface populations on dust grains can be of the order of unity, which make their discrete nature crucial (e.g. Langmuir-Hinshelwood formation goes from impossible to possible when going from a single adsorbed atom on a grain to two).
In addition, this approach misses any spatial effects on the surface (preferential chemisorption on neighboring sites to a chemisorbed atom, different adsorption sites have different binding energies on amorphous surfaces, etc...). The methodology also assumes a fixed dust temperature while the temperature of small grains fluctuates due to stochastic heating by UV photons.

The discrete nature of surface population results in random fluctuations (due to, for example, the stochastic nature of atom adsorption and desorption events). This can be accounted for by describing the surface populations. in statistical terms, by a probability distribution function obeying a master equation (again based on the experimentally constrained rate for the elemental processes) and solved to compute the average formation efficiency \citep{Biham:2001,Green:2001}. This approach is thus commonly called the \emph{master equation approach}. Analytical solutions are sometimes possible \citep{Biham:2002}, otherwise, numerical solution of the master equation incurs a significantly higher numerical cost than rate equation approaches. Approximations allowing faster solution have been proposed, such as the moment equation approach \citep{Lipshtat:2003,LePetit:2009}. A careful comparison of the master equation methodology to various approximations of this approach, and to the simpler rate equation formalism, was presented in \cite{Rae2003}. The master equation approach has also been extended to simultaneously account for dust temperature fluctuations of small grains \citep{Bron14}. 

Finally, \emph{Monte Carlo simulations} can, in principle, take any level of detail into account, by simulating realizations of the random evolution of the system (again using experimentally constrained rates), over which averages and other statistics can be computed \citep{Charnley2001}. This approach is, however, very costly in term of computational time, and is thus rarely used inside complete astrophysical models. It has been used to study spatial effects neglected by other approaches, such as variations of binding energies across the surface \citep{Chang:2005,2005MNRAS.361..565C}, or the effect of dust temperature fluctuations \citep{cuppen2006}. \citet{2009ApJ...691.1459V} and \citet{2014ApJ...787..135C} have presented a new generation of Monte Carlo models computing both the gas-phase and surface chemistries for large networks (similar to the ones used by rate equation models). These models are called unified Monte Carlo models. \citet{2014ApJ...787..135C} however use a microscopic approach for the surface chemistry and a macroscopic one for the gas whereas \citet{2014ApJ...787..135C} use a macroscopic approach for both these environments.

We also note that modifications to the simple rate equation approach have been proposed to approximate effects such as discrete population fluctuations \citep{Garrod:2008a,Cuppen2011}, diffusion on a spherical rather than plane surface \citep{Lohmar:2006,Lohmar:2009}, or surface inhomogeneity \citep{2006A&A...458..497C}.


\subsubsection{Hydrodynamical and MHD Simulations}


Hydrodynamical and Magneto-Hydro-Dynamical(MHD) simulations aiming at reproducing astrophysical systems, such as galaxy formation and star formation within molecular clouds, solve the equations for the hydrodynamics (or the magnetohydrodynamics, when fluids are magnetized) in order to describe the evolution in time of the system. However, the coupling of detailed chemistry to these high-resolution, multi-dimensional time-dependent simulations is still challenging.

While most hydrodynamical and MHD simulations only treat gas dynamics (possibly including thermal balance and gravity), it is only in recent years that computational capabilities have allowed us to include the formation of molecular hydrogen. However, since in astrophysical simulations the scales at which the H$_2$ molecules form are never resolved, the formation and destruction processes are included as sub-grid models, described by reaction rates derived from observational constraints \citep{1974ApJ...191..375J, Gry2002}. 

Approximate, less computationally demanding, approaches to compute H$_2$ abundances are usually stationary 
\citep{robertson2008, krumholz2009, kuhlen2012}. In contrast, a time-dependent and self consistent description of the chemical evolution of the system requires much greater computational resources \citep{gnedin2009, pelupessy2006, bekki2013, henderson2016, micic2012, micic2013, dobbs2008, glover2010, glover2008, glover2007a, glover2007b, maclow2012, smith2014, koyama2002,walch2015,hocuk2016,valdivia2016}. 

The treatments used can involve different degrees of detail in representing the chemical networks.  Approaches range from following the evolution of the H$_2$ abundance only \citep{valdivia2016}, or that of a few more species in reduced networks \citep{lim1999, lim2001, glover2010}, to much more complex chemical networks containing hundreds of chemical species and thousands of coupled chemical reactions \citep{grassi2013, ziegler2016}.
The elemental processes of H$_2$ formation (diffusion, desorption,...) are usually not treated in these models, and simplified formulae are used, based observational determinations of H$_2$ formation efficiency \citep{1974ApJ...191..375J, Gry2002} and scaled by the sticking coefficient of hydrogen atoms and the total dust cross-section per hydrogen nucleon \citep{bekki2013}. 



The dominant destruction of H$_2$ by photo-dissociation brings a complication, as it requires some calculation of radiative transfer. Dust extinction and self-shielding by H$_2$ (in the approximation of \citealt{draine1996}) are usually taken into account, and are crucial in order to obtain the right photo-dissociation rates \citep{hartwig2015}. Since the computational cost of calculating column densities is high \citep{clark2012,valdivia2014}, simulations use simplified treatments, such as ray-tracing schemes \citep{inoue2012}, logarithmic approximations \citep{clark2012, valdivia2014}, and even constant shielding parameters to avoid modeling radiative transfer in the Lyman and Werner bands.

Hydrodynamical and MHD simulations have shown that the formation of H$_2$ begins in the initial phases of the molecular cloud formation
\citep{glover2008, clark2012}, and that compressive motions within the gas (either driven by the galaxy dynamics or the turbulence) accelerate the formation of molecular hydrogen on galactic scales \citep{dobbs2006}, as well as on the scale of interstellar clouds \citep{micic2012, glover2007a, glover2007b, valdivia2016}. 
\citet{clark2012} and \citet{smith2014} have shown that the formation of CO, which is the main tracer of molecular gas, is significantly delayed with respect to the formation of H$_2$.  Theoretical \citep{wolfire2010,smith2014} and observational studies \citep{grenier2005,paradis2012, langer2014} point out that there is a significant amount of H$_2$ (between $30$ and $70$ percent) which is not traced by CO. 

Simulations have shown that the multiphase nature
of molecular clouds (where a cold and dense phase coexists with a diffuse and warm one), and their turbulent motions, have a significant effect on the abundance of H$_2$. Turbulent motions mix the cold and warm phases, while the multiphase structure allows the survival of molecules in shielded pockets of warm gas \citep{valdivia2016}. The presence of a hot phase provides a confining pressure that increases the abundance of H$_2$ in cold gas clumps \citep{dobbs2007}. Such a conclusion is consistent with the results of \citet{henderson2016}, who find that in disk galaxies a strong ram pressure enhances the formation of H$_2$.

Since molecular hydrogen also contributes to the cooling of the gas, it is perhaps expected that abundances of H$_2$ should affect the star formation rate and the spatial distribution of star forming regions \citep{christensen2012}.  However, the recent works of \citet{glover2014} and \citet{richings2016} have shown that H$_2$ has little influence on the star formation process. On the other hand, the abundance of H$_2$ is important for molecular outflows \citep{richings2016}, and it plays an important role on the warm chemistry of the ISM \citep{valdivia2017}.

\subsubsection{Kinetic models for complex chemistry}

Astrochemical models aiming at studying the molecular complexity in dense and shielded interstellar regions (cold cores for instance) focus on a detailed description of the complex chemical network at play, and on the time-dependent aspect of this chemistry (as some chemical timescales become comparable to the dynamical lifetime of these clouds). Such models thus solve kinetic differential equations in which each individual chemical and physical process is assigned a rate \citep{Wakelam2013}.

 The first kinetic models \citep{1980ApJS...43....1P,1982ApJS...48..321G,1984ApJS...56..231L} were based solely on gas-phase chemistry (bimolecular reactions, interactions with UV photons or cosmic-ray particles). In these models, the H$_2$ formation rate was computed from the collision rate of H atoms with dust grains \citep{hollenbach71,Hollenbach:1971b}. In the 90's \citep{1992ApJS...82..167H,1993MNRAS.261...83H,1993MNRAS.263..589H}, models including gas-grain interactions and grain surface chemistry began to be developed. Models of the surface processes were based on the Langmuir-Hinshelwood mechanism. 
The H accretion rate depends on the grain's geometrical cross section, the concentration of grains, the gas-phase thermal velocity of the H atoms and a sticking probability. The model then computes the time taken for a hydrogen atom to explore the entire grain via diffusion ($t_\mathrm{diff} = \frac{N_\mathrm{s}}{\nu_0 \exp(-E_\mathrm{diffus}/(k_\mathrm{B}\,T_\mathrm{dust}))}$ with $N_\mathrm{s}$ the total number of surface sites on a grain). The rate of H$_2$ formation is $2t^{-1}_\mathrm{diff,H}N_\mathrm{H}^2n_\mathrm{d}$ with $N_\mathrm{H}$ the number of H atoms on the grains and $n_\mathrm{d}$ the number density of grains \citep[see][]{1992ApJS...82..167H}. This approximation assumes that there are at least two hydrogen atoms on the grain at each time and therefore overestimates the H$_2$ production at the accretion limit (at low density) \citep{1998ApJ...495..309C}.

In these models, the precise nature of the grain surface is usually ignored and it is assumed that the grains are covered by water ice; hence, only the H$_2$ binding energy on H$_2$O is used. The parameters required are associated with the sticking, diffusion and desorption processes in the physisorption regime. Due to their connection with deuteration, ortho-to-para conversion on grains may also be an important phenomenon to model \citep{Pagani:1992,2017MNRAS.466.4470M}. H$_2$ formation on these icy surfaces may induce some secondary effects due to the significant amount of energy released in the formation of the H-H bond. Some experimental observations, such as the reduction of the porosity of the ice \citep{2011PCCP...13.8037A} associated with H$_2$ formation, are not included in models. Conversely, some proposed phenomena, which have not been observed experimentally \citep{Minissale:2016,Amiaud:2007}, such as molecular desorption induced by H$_2$ formation \citep{Takahashi:2000} or the H rejection mechanism, are sometimes included in models \citep{LePetit:2009}.

\subsubsection{PDR models}
PDR models focus on the interaction of UV radiation with molecular clouds. They involve simultaneous computation of radiative transfer and the chemical and thermal state of the gas, with a detailed treatment of the micro-physics of both gas and grains. Current major examples include the Meudon PDR code \citep{2006ApJS..164..506L,lebourlot2012}, the KOSMA-$\tau$ code \citep{Storzer1996,Rollig2013}, CLOUDY \citep{Ferland2013}, and the PDR codes of \cite{Kaufman1999,Kaufman2006} or of \cite{Meijerink2005}. Most of these codes are one dimensional and stationary, although some higher dimensional and/or time-dependent codes have been developed recently \citep[e.g.,][]{Bisbas2012,Motoyama2015,Hosokawa2006}, often at the cost of a simplified micro-physics and chemistry.

In PDR models H$_2$ formation was initially described empirically by a constant formation efficiency (e.g. using \citealt{1974ApJ...191..375J}), but some codes have progressively included a more detailed calculation over a full distribution of grain sizes \citep[e.g.,][]{lebourlot2012,Rollig2013}, using a rate equation treatment of the adsorption, migration, reaction and evaporation of physisorbed and chemisorbed hydrogen atoms.
Grains are heated by the UV photons and H$_2$ formation thus occurs on bare silicate and carbonaceous surfaces (in the warm surface layer of the region, where UV photons are not yet absorbed). On these warm bare grains, physisorbed atoms can evaporate quickly and chemisorption can play an important role in allowing an efficient H$_2$ formation (e.g. \citealt{lebourlot2012}) as observed in PDRs (e.g. \citealt{Habart04}). 

Very small grains and PAHs represent the majority of the dust surfaces on which H$_2$ formation can occur. H$_2$ formation on PAHs could be an important contribution to the total formation rate \citep{boschman2015}. In addition, small grains undergo large temperature fluctuations on short time scales, as their low heat capacities make their temperatures sensitive to single UV photon absorption events. These fluctuations can significantly increase the efficiency of physisorption-based, Langmuir-Hinshelwood, H$_2$ formation in the UV-rich environments of PDRs, as was shown by \cite{Bron14} by coupling a master equation approach to the Meudon PDR Code. In addition, the sticking coefficients at the warm gas temperatures of PDRs (several 100 K) are also  important parameters \citep{Cazaux2011}.
Finally, the role of surface processes in H$_2$ ro-vibrational excitation (in the gas) has also been investigated. Ortho-para conversion on the surface was shown to have an important impact on the gas-phase OPR of H$_2$ in PDRs \citep{Lebourlot2000,2016A&A...588A..27B}. The impact of the internal energy content of the H$_2$ molecule following its formation  has been less explored in models, in part due to the fact that there is little associated observational data to reproduce.


\subsubsection{Shock models}

Ubiquitous in the Universe, shocks are produced by violent pressure disturbances such as star-driven (or AGN-driven) jets and winds, supernovae explosions, and collisions between molecular clouds or, at larger scales, collisions between galaxies and between accretion filaments and halos of galaxies. 

 Due to the coupling between the chemistry and the dynamics in shocked regions, the formation of H$_2$, and emission from H$_2$, are among the most difficult astrophysical processes to model \citep[see reviews addressing the physics and chemistry of interstellar shocks by][]{Draine1980, McKee1980, Chernoff1987, Hollenbach1989, Draine1993, Hartigan2003}. Since H$_2$ is one of the main coolants in shocks,  its energetics have always been considered in molecular shock models \citep{Flower1985, 2003MNRAS.341...70F, Guillard2009, Flower2010}. Magneto-hydrodynamic (MHD) shock models have been updated with a benchmark of experimental results associated with H$_2$ formation on carbonaceous material -- see e.g. models by \citet{Guillet2009, Flower2015} with data from \citet{Cuppen2008, Cuppen2010}.
In this context, the role of chemisorption and the transfer from chemisorbed states to physisorbed states \citep{Cazaux2004} is a key point, which allows models to bridge the low and the high temperature regimes in these environments.  Due to their intrinsic high temperature gradients, reproducing shocked regions without taking into account both the chemisorption and physisorption of hydrogen, and their interplay, seems to be an oversimplification.

In the presence of a magnetic field, ubiquitous in the ISM, the shock wave satisfies not only the equations representing the fluid dynamics, but also Maxwell's equations. 
The magnetic field ($B$) does not interact the same way with charged and neutral particles, which can result in a decoupling of the charged (ions and electrons) and neutral fluids. Depending on the intensity of $B$ and the ionization fraction of the gas, we distinguish between two types of MHD shocks, $C$- and $J$-types:
\begin{description}
\item[$J$-shocks:] if $B$ is weak or absent, or if the ionization fraction of the gas is high, the collisional coupling between charged and neutral particles is strong. These particles behave like a \textit{single fluid}, coupled to the magnetic field. The properties of these shocks are similar to that of the hydrodynamical shocks. Across the shock front, the variables (pressure, density, velocity, etc.) change abruptly (hence $J$ for {\it Jump}. The transition region, which has a scale of the order of the mean free path, is treated as a discontinuity. The pre-shock and post-shock  values of the gas density, pressure and temperature are related by the Rankine-Hugoniot jump conditions.

\item[$C$-shocks:] if the magnetic field is present and the ionization fraction low (typically $x_\mathrm{e} = 10^{-7} - 10^{-8}$ in dense $n_{\rm H} = 10^{3} - 10^{4}$~cm$^{-3}$ molecular clouds), the neutral and charged fluids are decoupled from each other. These shocks are named \textit{multi-fluids} or $C-$shocks ($C$ for \textit{Continuous}), because in this case the discontinuity is smoothed and the gas parameters vary continuously across the shock front.
\end{description}

In the shock wave, H$_2$ is collisionally excited (or destroyed) and chemically destroyed, but reforms on grain surfaces in the post-shock gas. In state-of-the art MHD shock codes \citep[e.g.][]{Flower2015}, the rate of H$_2$ formation is simply parametrized as a function of the gas hydrogen density $n_{\rm H}$, the grain density $n_\mathrm{gr}$ and grain size $r_\mathrm{gr}$, and a coefficient that incorporates the probability that an H atom sticks to a grain, $1/(1+ T_\mathrm{eff}/T_\mathrm{cr})^{0.5}$, with $T_\mathrm{eff}$ being the effective grain temperature \citep{LeBourlot2002} and $T_\mathrm{cr} \approx 30-100~$K a critical grain temperature below which the sticking probability tends to 1. In the shock codes, the H$_2$ formation rate [cm$^{-3}$~s$^{-1}$] is thus often written as 
 \[ n_{\rm H} n_\mathrm{gr} \pi r_\mathrm{gr} ^2 \left( \frac{8 \pi T_\mathrm{eff}}{\pi m_\mathrm{H} (1+ T_\mathrm{eff}/T_\mathrm{cr})} \right) ^{0.5} . \] 

\begin{figure}
    \includegraphics[width=0.45\textwidth,angle=90]{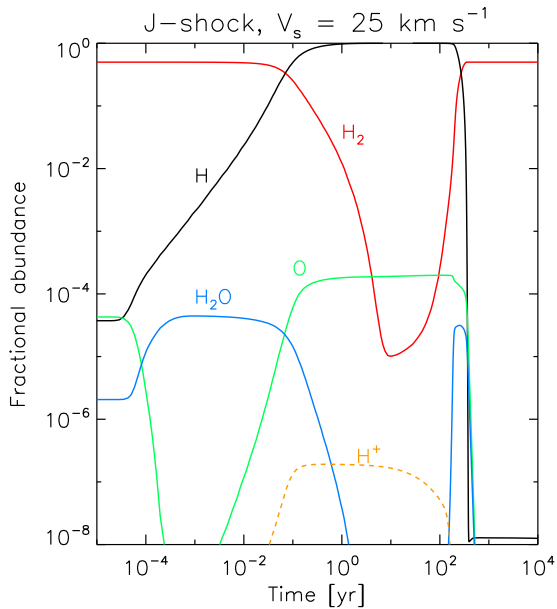}
      \caption{Fractional abundance profiles in a $J$-shock computed with the \citep{Flower2015} MHD shock code. The shock velocity is 25~km~s$^{-1}$ and the pre-shock gas density $n_{\rm H} = 10^4$~cm$^{-3}$.  The pre-shock magnetic field is $B_0=0\,\mu$G. Abundances are relative to H ($n(X) / n_{\rm H}$).} 
       \label{fig:shocks_J}
\end{figure}

\begin{figure}
    \includegraphics[width=0.45\textwidth]{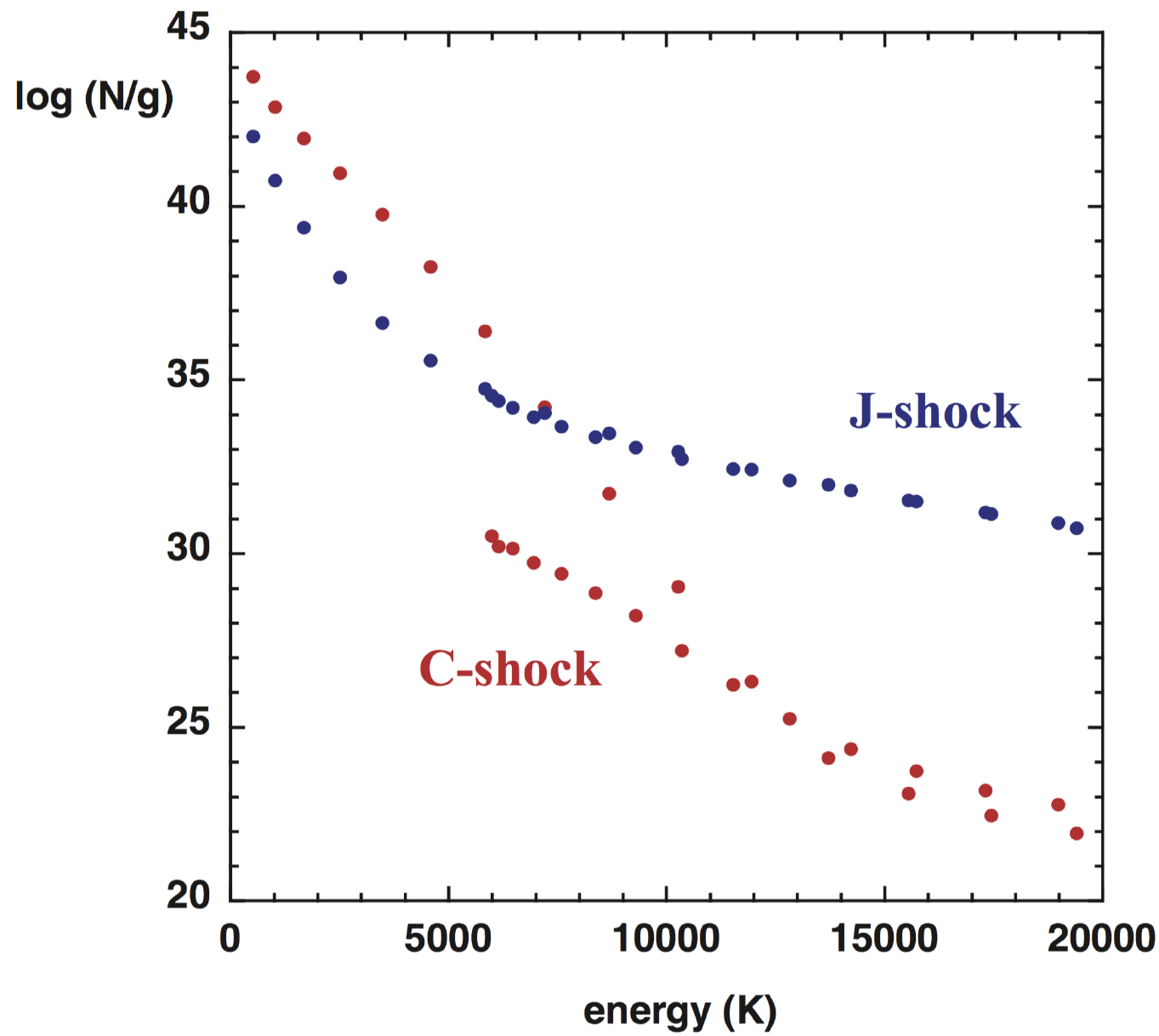}
      \caption{Comparison of the H$_2$ excitation diagrams of $C$- and $J$-type shock waves from the MHD code of \citet{Flower2010}. For both types of models, the shock velocity is $20~\mathrm{km~s^{-1}}$ and the pre-shock density is $n_\mathrm{H} = 2 \times  10^4$~cm$^{-3}$. For the the $C$- (resp. $J$) type model, the initial transverse magnetic field intensity is $B = 140\, \mu$G (resp. $B=14\, \mu$G). } 
       \label{fig:h2excit_C-J_shocks}
\end{figure}

Figure~\ref{fig:shocks_J} illustrates the impact of a dissociative $J$-shock wave in a molecular cloud. 
The passage of a shock wave impacts the physical and chemical quantities upon which the H$_2$ formation rate depends: the hydrogen density and grain properties (see \S~\ref{subsubsec:shocks_dust} for a discussion about the effect of shocks on grains). Rapid dissociation occurs, followed by H$_2$ re-formation in the post-shock gas because a significant number of grains survive the passage of the shock. Figure~\ref{fig:h2excit_C-J_shocks} illustrates the impact of the nature of the shock (i.e. $B$ field and ionization fraction) on the H$_2$  excitation. For high energy levels, the diagram is flatter for $J$-shocks than for $C$-shocks because the maximum temperature reached in the shock is higher.

MHD shock codes have been used to fit the observed H$_2$ diagrams, which provide clues on the dominant excitation mechanism, and physical conditions (pre-shock densities, shock velocities, magnetic field intensity), both in Galactic \citep[e.g.][]{Gusdorf2012} and extragalactic environments \citep{Guillard2009, Guillard2012, Lesaffre2013, Appleton2017}. 


\section{Recommended parameters}  
The aim of this section is to provide a table of reviewed physical parameters as inputs for models. In the previous parts, we have revealed the heterogeneity of results associated with interstellar H$_2$ formation and the large variety of systems involved with this process.
	Even though a physical parameter can be precisely defined and evaluated, it is often  not directly usable in models in such a fundamental form. For example, experiments and calculations have shown that for a correct description of H atom diffusion, a single height (and width) of a diffusion barrier is not sufficient; a distribution of barrier heights is a more realistic approach. However, in astrochemical models the use of such distributions is not implemented and therefore one has to choose a reasonable parameter that approximates the real complexity.
	We are well aware of the imperfections of this approach to parametrization, but at present it is the only approach that is viable. Therefore, each value in the table of physical parameters is the best approximation in the view  of the authors. 
	The parameters we present in the table are those commonly used in existing models.
    
	We propose 8 prototypical classes of surface, each class, of course,  often representative of a large variety of substrates: (1) amorphous and (2) crystalline silicates, (3) amorphous carbon (including a very large variety of compounds, from HAC to onion rings), (4) graphite, (5) PAHs and (6) PAH cations, and two morphologies of ice, one geometrically flat such as (7) polycrystalline ice, and one with a high degree of molecular disorder, such as (8) porous amorphous ice (ASW) grown at low ($<$30 K) temperature.
	
	\subsection{Sticking coefficients}
	 For physisorption, the sticking coefficient is predominantly a function of the thermal energy of the incident atoms. Of course, in principle, the surface temperature is also a parameter. However, there is a distinct lack of measurements of sticking as a function of relevant surface temperatures, and we have to neglect this parameter. This simplification is supported by the fact that physisorption plays a major role in environments where the surface temperature of grains is low and will not strongly affect sticking.
     We use the parametrization of \citet{Matar:2010}, of the sticking probability  as a function of $T_\mathrm{gas}=T$ which depends on 3 parameters ($S_0$, $T_0$, $\beta$) such that 
	 $$S_\mathrm{phys}(T) = S_0  \frac{1+T/T_0}{(1+T/T_0)^\beta}  $$
	 
	  where $\beta$ is a geometrical factor which is equal to 2.5 for the case of isotropic deposition, which occurs in the ISM. Thus, this leaves the parameters $S_0$ and $T_0$. When no information exists we use unity for $S_0$ as it is usually close to one. $T_0$ is the only scaling parameter of the sticking coefficient for physisorption.  

	For sticking involving chemisorption we use the formalism of \citet{2005JChPh.122a4709S} referred to in \citet{Cuppen2010} for the sticking probability. This probability also depends on the surface temperature  $T_s$ of the grain;
	
	 $$S_\mathrm{chem}(T, T_\mathrm{s}) =  \frac{\exp(-E_\mathrm{chem}/T)}{(1+5\times 10^{-2}\sqrt{T+T_\mathrm{s}}+ 1\times 10^{-14} T^4}$$

'Therefore $S_\mathrm{chem}$ is only governed by one physical parameter $E_\mathrm{chem}$ which is the entrance barrier to the chemisorption sites. However, in case of large barriers, this expression cannot be used because H atoms can tunnel through the barrier and transmission coefficients across this barrier should be accounted for, especially at low H atom energies \citep{Cazaux2011}. In this latter study, a mixed classical-quantum dynamics calculation showed that by considering phonons, the resulting sticking probabilities are different than the transmission probabilities of the chemisorption barrier, which are typically used to model sticking probabilities. However, the resulting H$_2$ formation rate is only slightly influenced by the effect of phonons at low temperatures.

	\subsection{Energy barriers}

In the table of parameters below, we let $E_\mathrm{diffus}$ correspond to the height of the diffusion barrier between two physisorption sites. A lower value means faster diffusion.  $E_\mathrm{binding}$ is the energy required for an H atom to desorb from a physisorption site. $E_\mathrm{diffPh-Chem}$, is the height of the diffusion barrier for going from a physisorption site to a chemisorption site. 
For the specific case of aromatic carbon rings, the barrier to chemisorption is dependent on whether one or more H atoms are already chemisorbed. In this case we indicate the values for the first and second hydrogen atom adsorption reaction. For graphite and neutral PAHs  adsorption events beyond the first H atom often have a significantly lower or even a vanishing barrier. 


We have not listed the diffusion barrier between two chemisorption sites because under astrophysical conditions this slow diffusion rarely contributes efficiently to H$_2$ formation. However calculated values are available for some situations (\cite{2003CPL...368..609F,2006PhRvL..96o6104H}).   

$E_\mathrm{Reac Ch-Ch}$ is the barrier to the reaction of two adjacent chemisorbed H atoms. 
$E_\mathrm{Bind-Ch}$ is the binding energy of an isolated H atom on the listed surface.

$\sigma_{ER}$ is the cross-section in \AA$^2$ pr. chemisorbed H atom for H$_2$ production via Eley-Rideal abstraction reactions with gas phase H atoms

\subsection{Table of recommended values}
In Table~\ref{table_parameters} we typeset values which come directly from a value already described above, or in a paper cited earlier, in regular roman font.  Sometimes a range of existing values is indicated.
For some other cases, the adopted value is an educated choice and we indicate such values in italic. When several determinations of a given parameter exist, we use bold-face in order to indicate that these values have been confirmed by different methods (experiments or calculations). 

{
\footnotesize
	\begin{table*}
\caption{Selected values characterizing processes of hydrogen interaction with surfaces of dust grain analogs. $S_\mathrm{H}(T)$ is the sticking coefficient of H at the temperature T, E$_{diffus.}$ is the activation energy of H diffusion, E$_{binding}$ is the binding energy of H to surfaces; energy values are in meV; 1 meV = 11.6 K/k$_B$.
\label{table_parameters}}


	\begin{tabular}{|*{9}{p{1.7cm}|}}
		
		\hline
		\multicolumn{9}{|c|}{PHYSISORPTION	}\\
		\hline
	
		\hline &1: Amorphous Silicate &2 : Crystalline Silicate  &3: Amorphous Carbon  &4: Graphite  & 5: PAH & 6: PAH$^+$ & 7: Crystalline Ice &8: Porous Amorphous Solid Water\\ 
	
		\hline A:  $S_\mathrm{H}$(T$_{gas}$)&S$_0$=1.0, T$_0$=25 K&  \em{S$_0$=1.0, T$_0$=25 K } & \em{S$_0$=1.0, T$_0$=125K}& S$_0$=1.0, T$_0$=24K
	& \em{S$_0$=1.0, T$_0$=24K} & \em{S$_0$=1.0, T$_0$=24K}& S$_0$=1.0, T$_0$=52K  &\em{S$_0$=1.0, T$_0$=52K }  \\ 
	
		\hline B:  E$_{diffus.}$ &35   &8.5  &  44& 4 &\em{4}  & \em{4}  & $< 18$, \em{16} & \bfseries{22} \\ 
		\hline C: E$_{binding}$ & 44&  25 &  57&40 &\em{40}  &\em{40}  & 30  &\bfseries{50}  \\ 
		\hline \hline
			\multicolumn{9}{|c|}{CHEMISORPTION}\\
		
		\hline D: S$_{chem}$(T,T$_g$) & E$_{chem}$=300&  E$_{chem}$=260 & E$_{chem}$=6 &E$_{chem}$= 150 & E$_{chem}$=60 &E$_{chem}$=10 &n.a  &n.a  \\ 
		\hline E: E$_{dif Ph-Chem}$   & 332 &  278& \em{6} & 190,0 & 100,0 &50,70  & n.a & n.a \\ 
	
		\hline F: E$_{Reac Ch-Ch}$  &  48&  1510 (74)& $>$ \em{1400} &  1400  & n.a & n.a  & n.a &n.a  \\ 
		\hline G: E$_{Bind-Ch}$  &1786  & 398 & $>$\em{870},4000 &  870& 1450 & 2810  &  n.a&n.a  \\ 
        \hline H: $\sigma_{ER}$  & n.a  &  n.a & 0.03 & 4-17 & 0.06 & {\em 0.06} &  n.a&n.a \\ 
		\hline \hline
		\multicolumn{9}{|c|}{DYNAMICS AND OTHER PROCESSES}\\
		\hline I: Nascent OPR  & - & - & - & - &  -& - & - &3 to 1  \\ 
		\hline J: NSC  & - & - & - & Yes & - & - & - & Varies with T  \\ 
		\hline K: Int. Energy  & medium $v$, low $j$  &-&   No  & $v$ 2-5, low $j$ &  -&  -&  No & No   \\ 
		\hline L: Kin. Energy  &- &  - & {\em No/Low}  & 1.3 eV   & - & - & - & \em{No}  \\ 
		\hline M: Energetic
		formation & -&  - & Yes  & - &  -& yes &  -& Yes  \\ 
		\hline
	\end{tabular}

\end{table*}	
}

The lines of the table are labeled by a letter, and rows by a number: number 1 for the amorphous silicate, number 2 for crystalline silicate, number 3 for amorphous carbon, number 4 for graphite, number 5 for PAH, number 6 for PAH cation, number 7 for crystalline ice and number 8 for porous amorphous solid water. 
\\
Letter A refers to the sticking to physisorption sites. 
\\
A1 : From reference \cite{Chaabouni:2012}.
\\
A2 : No data, we propose the values of A1.
\\
A3 : No clear data, however due to the possibly high porosity and roughness, and better mass matching of aliphatic groups, the sticking should be higher than on other substrates. The adopted values corresponds to a sticking of 50$\%$ at 300 K.
\\
A4 : Best fit from \citet{PhysRevLett.107.236102}. We point out that values for silicates and graphite surfaces seem to be very similar despite the fact that the adopted methods are independent (experiments vs calculations).
\\ 
A5, A6: No data. We propose the values of A4.
\\
A7 : From reference \citet{Matar:2010}, although the measurements were performed on compact ASW. This choice was made because compact ASW has the same flat topology as the crystalline ice and differs from the crystalline structure at meso-scale (tens of molecules). Therefore, the sticking properties of this type of ice are closer to crystalline ice than the very disordered amorphous solid water ice.
\\
A8 : We take the values of A7. They are lower limits because it is known that the sticking of light species is enhanced by the porous structures of ice (e.g. \cite{Hornekaer:2005}). We do not adopt different and higher values because the degree of porosity varies too much.
\\
\\
Letter B refers to the energy of diffusion E$_{diffus.}$ between two physisorption sites
\\
B1 : From reference \citet{Perets:2007}
\\
B2 : From reference \citet{He:2011}. Note that in the case of polycrystalline silicates there are deeper adsorption sites and that the value of the diffusion barrier has been estimated to be 24 meV  \citep{Katz:1999}. Furthermore, \citet{Navarro-Ruiz:2014} have computed a value of 60 meV using a molecular cluster representative of crystalline forsterite. We keep the low value of barriers from the reference \citet{He:2011} because it compares different morphologies of silicates. The set of adopted binding and diffusion barriers may be less than perfect, but it is at least self-consistent.
\\
B3 : From reference \citet{Katz:1999}. We point out that this value is the highest value for the diffusion barrier of all surfaces. We propose only this one existing value although amorphous carbon can adopt a large variety of morphologies wiht potentially different physical properties. 
\\
B4 : From calculations of \citet{Bonfanti:2007}
\\
B5, B6 : No data. We propose using the value of B4
\\
B7: Diffusion on crystalline ice is known to be faster than the capabilities of experimental measurements at 8-15 K \citep{2012ApJ...757..185H} which implies that the diffusion barrier is less than 18 meV. We choose a conservative values of 16 meV, although calculations propose even lower values \citep{Senevirathne:2017}.
\\
B8 : There is different experimental and theoretical work relatively convergent to establish a distribution of diffusion barriers (see previous section). The value $22~\mathrm{meV}$ that we finally adopt is an effective value which is a consensus between the authors. However, note the reference \citep{2005ApJ...627..850P} which gives a higher value (55 meV), the signature of a very deep physisorption site.
\\
\\
Letter C refers to the binding energy E$_{binding}$ of the H atom physisorbed on the surface.
\\
C1 : From reference \citet{Perets:2007}
\\
C2: From reference \citet{He:2011}. See comment in B2
\\
C3: From reference \citet{Katz:1999}. See comment in B3.
\\
C4: From \citet{1980JChPh..73..556G} and \citet{Bonfanti:2007}
\\
C5,C6: No data. We propose using the value of C4
\\
C7 and C8: Consensus values adopted from the work in four different groups.
\\
\\
Letter D refers to the sticking to chemisorption sites. 
\\
D1 : From reference \citet{Navarro-Ruiz:2015}. 
\\
D2 : From reference \citet{Navarro-Ruiz:2014}. The presence of an adsorbed H on the surface changes the barriers significantly: the reaction between a chemisorbed atom and a physisorbed atom can have a barrier as low as 74 meV.
\\
D3: From reference \citet{Mennella2006}.
\\
D4, D5, D6: From references \citet{2006PhRvL..96o6104H,2008ApJ...679..531R,2016NatSR...619835C}.
\\
\\
Letter E refers to diffusion barrier of H atom going from the physisorption site to a chemisorption site E$_{dif Ph-Chem}$
\\
E1: From reference \citet{Navarro-Ruiz:2015}.
\\
E2: From reference \citet{Navarro-Ruiz:2014}. See comment in D2.
\\
E3: Large spread in values expected. We propose using the value from D3.
\\
E4, E5, E6: These values are obtained from the corresponding D values by adding the binding energy of the physisorption state.
\\
\\
Letter F refers to the barrier to the reaction of two adjacent chemisorbed H atoms E$_{Reac Ch-Ch}$.
\\
F1: From reference \citet{Navarro-Ruiz:2015}.
\\
F2: From reference \citet{Navarro-Ruiz:2014}. See comment in D2.
\\
F3: Large spread in values expected depending on e.g. aliphatic vs. aromatic nature of the amorphous film. We suggest using the value of F4 as a lower limit, since aliphatic structures are expected to offer even higher energy binding sites for H atoms with associated higher recombination barriers. The main reactivity for chemisorbed H atoms is expected to be via Eley-Rideal or hot-atom reactions or via reactions with physisorbed H atoms.
\\
F4: From reference \citet{2006PhRvL..96o6104H}
\\
\\
Letter G refers to the binding energy of an isolated H atom in chemisorption site E$_{Bind-Ch}$. 
\\
G1: From reference \citet{Navarro-Ruiz:2015}.
\\
G2: From reference \citet{Navarro-Ruiz:2014}. See comment in D2.
\\
G3: Large spread in values expected depending on e.g. aliphatic vs. aromatic nature of the amorphous film. For predominantly aromatic films we suggest using the value of G4 as a lower limit, while a value of 4 eV is suggested for mainly aliphatic structures following reference \citet{2008ApJ...684L..25M}. It is to note that any model using these combined values should also be coherent with the value of the cross section experimentally measured.
\\
G4 : From references \citet{2006PhRvL..97r6102H,PhysRevLett.93.187202}
\\
G5: Listed value is for the most stable adsorption site of \citet{2008ApJ...679..531R}: the energies of the different adsorption sites are 700, 700, 1450 meV for central, edge and outer-edge sites, respectively.
\\
G6: Listed value if for the most stable absorption site of \citet{Cazaux2011}: the energies of the different adsorption sites are 1900, 2140, 2810 meV for central, edge and outer-edge sites, respectively.  
\\
\\
Letter H refers to the cross-section $\sigma_{ER}$ in \AA$^2$ pr. chemisorbed H atom for H$_2$ production via Eley-Rideal abstraction reactions with gas phase H atoms
\\
H3: From reference \citet{2008ApJ...684L..25M}. 
\\
H4: From reference \citet{2002CPL...366..188Z}. 
\\
H5: From reference \citet{2012ApJ...745L...2M}.
\\
H6: No data. We propose using the value of H5. 
\\
\\
The Letters I to M refer to the dynamics and other processes about the H$_2$ formation on the listed surface. The blank spaces in this part of the table remind us that there are some important issues related to the formation of H$_2$, like the internal energy of the molecule released, and the capability of the surface to induce nuclear spin conversion, which are still unknown for many of our prototypical surfaces.
\\
Letter I and J refer to the nascent Ortho-Para Ratio (OPR) and to the Nuclear Spin Conversion (NSC), respectively.  
\\
I8: from references \citet{2010ApJ...714L.233W,2012ApJ...760...35G}
\\
J4: from \citet{Palmer1987}
\\
J8: from \citet{2016PhRvL.116y3201U}
\\
\\
Letter K refers to the repartition of the internal energy of the H$_2$ molecule formed.
\\
K1: from reference \citet{Lemaire:2010}
\\
K3: from reference \citet{2008ApJ...684L..25M}
\\
K4:  Experiments from \citet{2006JChPh.124k4701C,Islam:2007,2008CPL...455..174L}. Many different calculations (e.g. \citet{Morisset:2005})
\\
K7: from reference \citet{Congiu:2009} when some H$_2$ is covering the surface, unless probably medium $v$.
\\
K8: from reference \citet{Congiu:2009}
\\
\\
Letter L refers to the informations about kinetic energy of the  H$_2$ molecule formed.
\\
L3:Expected from porous nature of the grains.
\\
L4: from reference \citet{2006JChPh.125h4712B}
\\
L8: Expected from the porous nature of the ice 
\\
\\
Letter M refers to the informations about the energetic formation, such as UV, electrons or ions bombardments, that would lead to H$_2$ formation and release in the gas phase.
\\
M3: from reference \citet{2014A&A...569A.119A}
\\
M6: calculations \citet{Allain1996,lepage2001}
\\
M8: Many references with ions, UV and electrons (see text above on water ice section)

\section{H$_2$ formation in different astrophysical environments}\label{sect:environments}

The interstellar medium involves a wide range of physical (temperature and density) and irradiation (UV, X-ray and cosmic-ray particles) conditions. In this section of the paper, we focus on some specific regions of the ISM were H$_2$ is formed, observed or has a particular role in the chemistry of more complex species.  Specifically, we discuss: diffuse clouds, dense photo-dissociation regions (PDRs, for which we distinguish high illumination PDRs, abbreviated HI, and low illumination PDRs, abbreviated LI), cold shielded regions, hot shielded regions and molecular shocks inside the Galaxy and in other galaxies. 
 Table~\ref{physical_parameters} summarizes some key physical parameters of these sources. The first column gives the UV illumination measured by the $G_0$ parameter, which measures the local energy density in far UV photons (between 912 {\AA} and 2400 \AA), normalized to the energy density of the average interstellar UV field $u_{\mathrm{Habing}}=5.3\times10^{-14}\,\mathrm{erg}\cdot\mathrm{cm}^{-3}$ \citep{1968BAN....19..421H}:  $$G_0 = \frac{\int_{912\,A}^{2400\,A}\,u_{\lambda}(\lambda)\,d\lambda}{u_{\mathrm{Habing}}}.$$
 The second column gives the proton density $n_\mathrm{H}$ of the medium (in cm$^{-3}$). The third and fourth columns list the gas and dust temperatures. The fifth column gives the fraction of hydrogen in the molecular form $f_{\rm H_2}$ (when this parameter is unity, all the hydrogen is molecular). The sixth column is the observed H$_2$ ortho-para ratio (OPR). Finally, the last column gives the H$_2$ formation efficiency $R_{\rm H_2}$ derived from observations (in cm$^{3}$~s$^{-1}$). 
In the following, we present these environments and discuss the relevant physical-chemical processes involved in the formation and destruction of H$_2$ in each of these regions of the ISM.

\begin{figure*}
   \centering
    \includegraphics[width=0.80\textwidth]{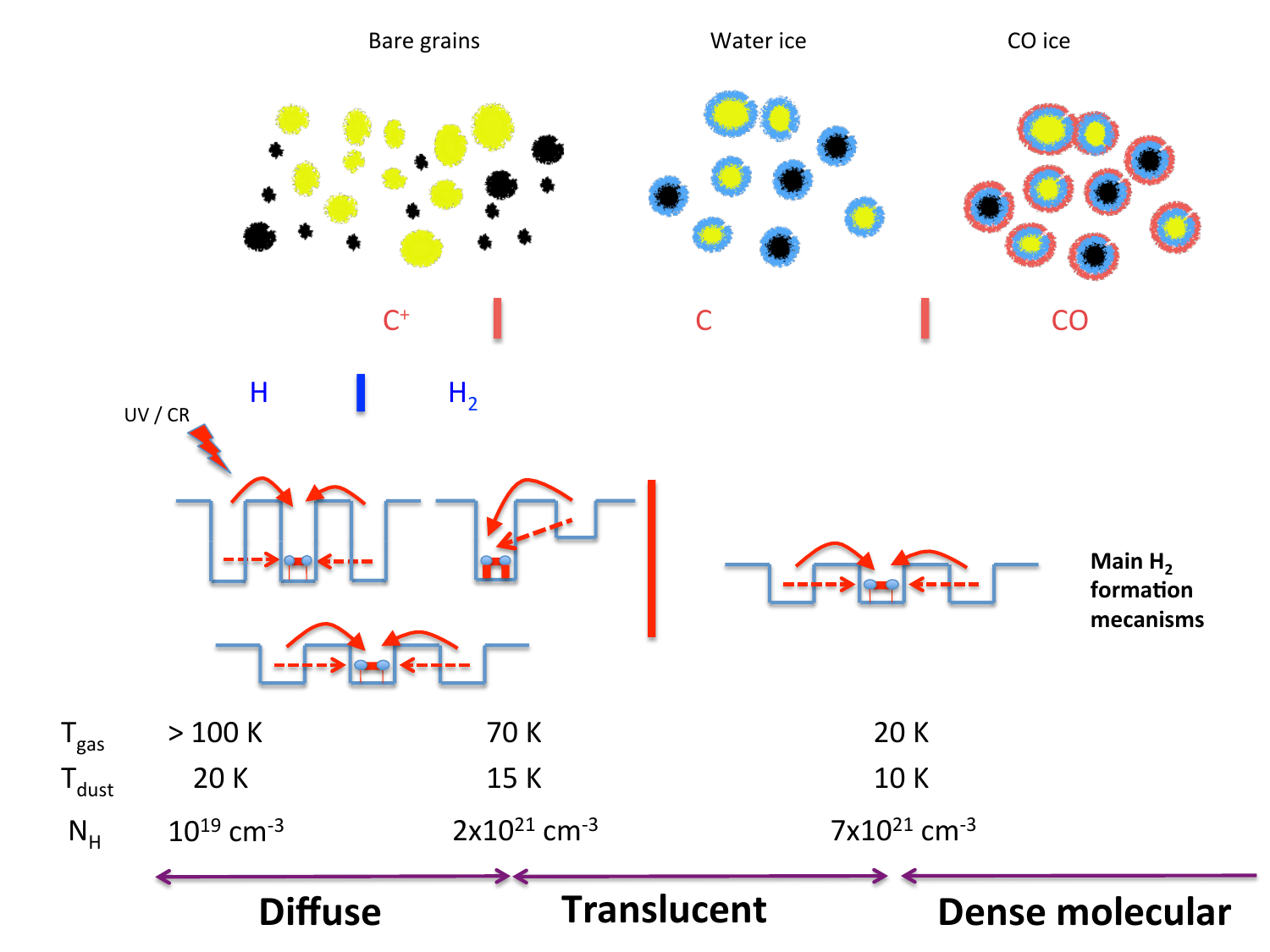}
      \caption{Sketch of the main H$_2$ formation processes in diffuse to dense interstellar medium. }
       \label{main_H2formation}
\end{figure*}
%

\subsection{Diffuse and translucent interstellar medium}\label{diffuse_medium} 

Diffuse clouds represent the regime in the ISM where moderate density clouds are exposed to the background UV field of the galaxy (the interstellar radiation field), and consequently where nearly all molecules are rapidly destroyed by photo-dissociation. In diffuse clouds, molecular hydrogen has been observed through its FUV absorption lines (\citealt{Savage1977} with Copernicus, \citealt{Snow2000,shull2000,rachford2001,Rachford2002,Gry2002} with FUSE). In such clouds, from 10$^{-6}$ to about half of the total number of hydrogen nuclei are bound in hydrogen molecules \citep[e.g.,][]{1974ApJS...28..373S,Shull:1982}. From their H$_2$ content, these clouds can be categorized as 
\begin{itemize}
\item (1) {\em Diffuse atomic clouds.} Here the molecular fraction  $f_{\rm{H}_2}$=$\frac{2\rm{nH}_2}{2\rm{nH}_2+\rm{nH}}$ $\le$ 0.1. The density ranges from 10 to 100 cm$^{-3}$ and the gas and dust temperatures are around 30-100 K and 15-20 K respectively \citep[][for a review]{2006ARA&A..44..367S}. 
\item (2) {\em Diffuse molecular clouds.} Here the molecular fraction  $f_{H_2}$ $\ge$0.1, but carbon is still mostly ionized $f_{C^+}$=$\frac{\rm{nC^+}}{\rm{nC^+}+\rm{nC}+\rm{nCO}}$ $\ge$0.5. They exist from an extinction of $\sim$ 0.2, and have densities ranging from 100-500 cm$^{-3}$ and gas and dust temperatures around 30-100 K and 15-20 K respectively. 
\item (3) {\em Translucent clouds.} With sufficient protection from interstellar radiation (from extinctions of 1-2) carbon begins its transition from an ionized atomic state into a neutral atomic (C) or molecular (CO) form. Clouds in this transition phase have been defined as ``translucent'' \citep{1989ApJ...340..273V}. Here, until Av~$\sim 2$, hydrogen is mostly molecular with densities from 500-5000 cm$^{-3}$. The gas and dust temperatures are about 15-40 K \citep{1989ApJ...340..273V} and 13-19 K  \citep{2012A&A...547A..11N,2013A&A...551A..98L,2014A&A...568A..27F,2014A&A...562A.138R} respectively 
\end{itemize}


\subsubsection{Dust properties in the diffuse and translucent ISM}

The dust properties in the diffuse ISM have been the subject of numerous studies; including, for example, the \citealt{Draine07} and \citealt{Compiegne11} models.
More recently, theoretical modeling based on laboratory experiments (THEMIS\footnote{The composition of dust grains of \citet{jones2013} and \citet{koehler2014b}, for sizes up to 20 nm diameter, consists purely of aromatic-rich H-poor amorphous carbon (a-C), whereas bigger grains have a core/mantle structure, where the core consists either of amorphous silicate (forsterite and enstatite-normative compositions, Mg-rich) or aliphatic-rich H-rich amorphous carbon, a-C:H. For both core types, the mantle consists of aromatic-rich H-poor amorphous carbon.}, \citealt{jones2013}) was applied with great success in the analysis and interpretation of Planck, Herschel and Spitzer observations, even in the most diffuse regions. 
\citet{2015A&A...577A.110Y} are able, with small variations in the dust properties, to explain most of the variations in the dust emission observed by Planck-HFI in the diffuse ISM. 

The dust size distribution in the diffuse ISM, derived by \citet{2015A&A...577A.110Y}, can be described as in the following. Small grains ($\le 20$ nm) follow a power-law size distribution\footnote{The built-in size distributions $dn/da$ (the number of grains of radius between $a$ and $a+da$) are either power-law ($dn/da\propto a^\alpha$) or log-normal ($dn/dlog (a) \propto exp(-(log(a/a_0)/\sigma)^2$) with $a_0$ the centroid and $\sigma$ the width).} with $\alpha$=-5, with a minimum size of $a_{min}= 40$ \AA. 
Large grains follow a log-normal size distribution with a peak at a size of~0.15 $\mu$m. 
Using this dust size distribution and the radiation field and the gas density distribution found in the diffuse ISM, the (mass-weighted) mean temperature of dust grains derived  as being  $\sim$19 K for small and big carbon grains, and $\sim$16 K for big silicate grains. The maximum temperature has been derived as being  $\sim$160 K for small carbon grains, $\sim$50 K for big carbon grains, and $\sim$16 K for big silicate grains.
The  temperature of big dust grains has also been estimated observationally from modified blackbody fits as $T\sim$20 K (e.g., Planck collaboration XI, 2014), but this value results from a mix of dust at different temperatures along the line of sight. 

\subsubsection{Main chemical processes for the formation of H$_2$ depending on the environment }

\paragraph{Diffuse clouds with low radiation field}
In diffuse clouds, H$_2$ forms with a rate of $1-3\times 10^{-17}n_\mathrm{H} n(\mathrm{H})~\mathrm{cm^{-3}s^{-1}}$ where $n_\mathrm{H}$ is the total proton density and $n(\mathrm{H})$ is the density of H-atoms \citep{1974ApJ...191..375J,Hollenbach:1971b,Gry2002}. At grain temperatures typical of diffuse cloud conditions \citep{1983A&A...128..212M,Gry2002,hocuk2016}, which is 15-20 K, physisorbed H atoms can still remain attached to dust grains and H$_2$ can form efficiently through the reaction of 2 physisorbed H atoms (Langmuir-Hinshelwood mechanism). This formation process is very efficient and depends on the binding energies of the physisorbed H atoms, which have been derived experimentally and are reported in Table~\ref{table_parameters}. Rate equations and Monte Carlo simulations for the formation of H$_2$ on such surfaces, where both physisorption and chemisorption are considered \citep{2002ApJ...575L..29C,Cazaux2004,cazaux:2009}, or where the roughness of the surface is taken into account \citep{2005MNRAS.361..565C,2006A&A...458..497C}, show that H$_2$ formation efficiency involving physisorbed H atoms can reach 100\% for a surface temperature of 20~K. Therefore, in diffuse environments with low radiation field (i.e. with dust temperatures lower than 20~K), the formation of H$_2$ is mainly due to the involvement of physisorbed H atoms on dust grains. In figure \ref{main_H2formation}, the main mechanisms responsible for the formation of H$_2$ in diffuse, translucent and molecular clouds are presented. For a diffuse cloud with low $G_0$, the main process is the encounter of two H physisorbed H atoms: the Langmuir-Hinshelwood mechanism. This mechanism dominates for low dust temperatures ($T_\mathrm{dust}\le 20~\mathrm{K}$).

\paragraph{Diffuse clouds with high radiation field} In environments where the radiation field is significant, but not strong enough to dissociate PAHs, which corresponds to $n_\mathrm{H}/ G_0$ $\ge$ 3 10$^{-2}$ \citep{2009ApJ...704..274L}, where $G_0$ is the radiation field in Draine's units, PAHs could also play a role in the formation of H$_2$. Under such conditions, PAH cations can contribute to the formation of H$_2$ with rates comparable with the typical rate in the ISM. However, in regions typical of the diffuse ISM, the radiation field is less important, $n_\mathrm{H}/G_0 \sim 1-10^{2}$, and the formation of H$_2$ on PAHs is less efficient \citep{2009ApJ...704..274L}. In these conditions, H$_2$ forms predominantly on dust grains \citep{boschman2015} through the Eley-Rideal mechanism involving chemisorbed H atoms. The possible mechanisms for the formation of H$_2$ in diffuse clouds with high $G_0$, shown in figure \ref{main_H2formation}, are either through abstraction of H$_2$ from PAHs, or through Eley Rideal mechanism on dust grains involving chemisorbed H atoms, or by photolysis of hydrogenated amorphous carbons \citep{2014A&A...569A.119A}.

\paragraph{Translucent gas}
In translucent clouds, the temperature of dust grains becomes lower than in diffuse clouds (15-18~K) and, above visual extinctions of around 3, ices start to cover the dust surfaces. The interactions of H atoms with ices are different than those with bare surfaces because (1) H atoms cannot chemisorb on ices and (2) H can only physisorb on icy surfaces with binding energies comparable, or slightly lower, than the binding energies on bare surfaces. These points imply that H$_2$ formation on icy dust can only involve physisorbed H atoms, and therefore that H$_2$ in translucent clouds is formed predominantly via the Langmuir-Hinshelwood mechanism. However, the efficiency of the H$_2$ formation on icy dust strongly depends on the porosity of the ices. If the ices are crystalline, the efficiency of forming H$_2$ can be 100$\%$ up to surface temperatures of $\sim$12 K. Above this surface temperature, the efficiency will drop exponentially, which implies that H$_2$ formation will be very inefficient at higher temperatures. If the ices are porous, the efficiency is 100$\%$ until the surface temperature reaches $\sim 19~\mathrm{K}$. In this case, the formation of H$_2$ is efficient for the range of temperatures encountered in translucent clouds. Figure \ref{main_H2formation} summarizes the  conditions met in translucent clouds as well as the main process for the formation of H$_2$, which is via the Langmuir-Hinshelwood mechanism involving two physisorbed H atoms.

\subsection{Dense PDRs}\label{dense_pdr} 

\begin{figure*}
   \centering
    \includegraphics[width=0.80\textwidth]{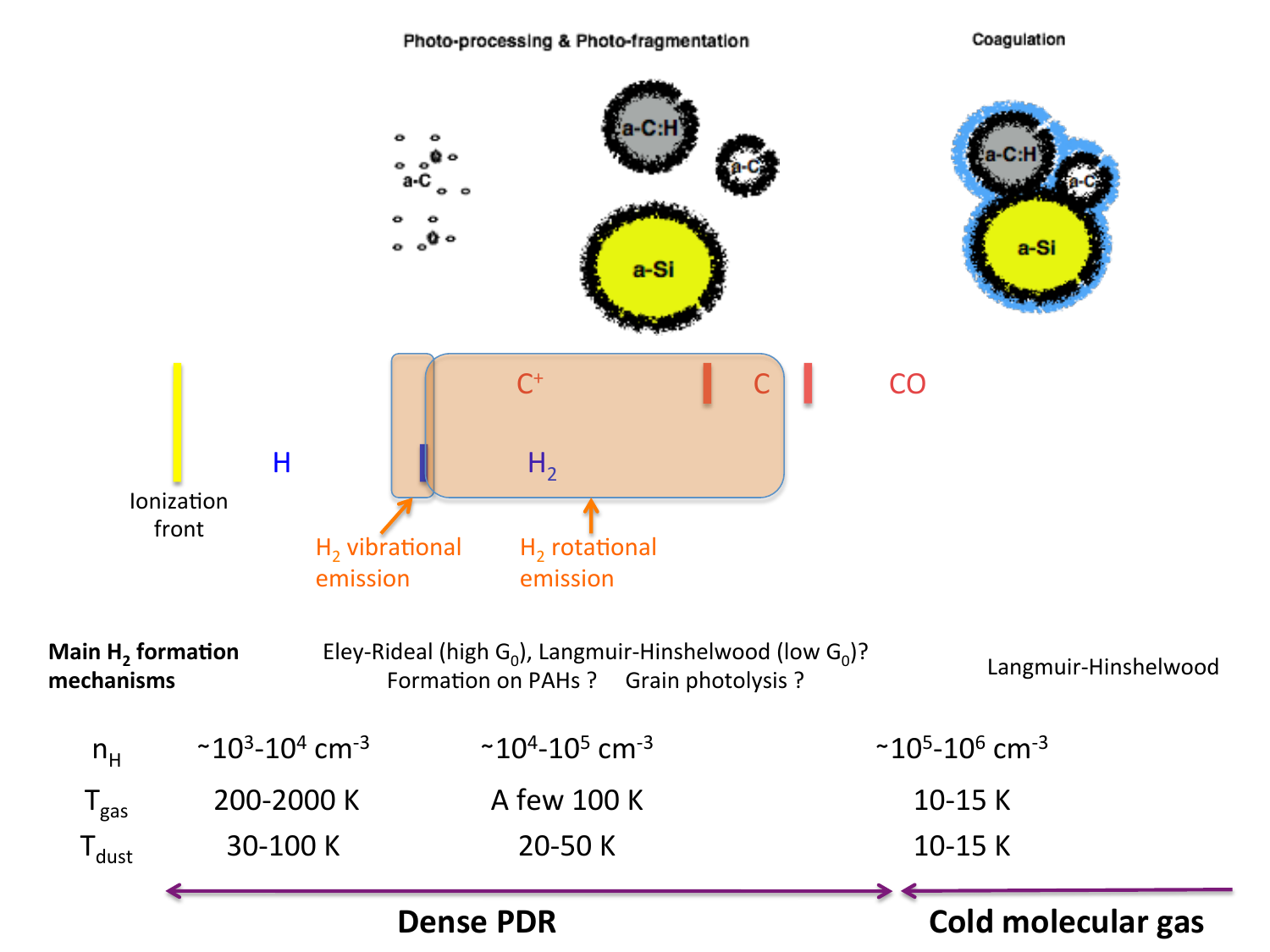}
      \caption{Schematic illustrating the main H$_2$ formation processes in PDRs.}
       \label{main_H2formation_PDR}
\end{figure*}

Photo-dissociation regions (PDRs) are regions of neutral interstellar clouds in which the heating and chemistry are dominated by the impact of stellar UV photons \citep{Hollenbach97,Hollenbach99}. This definition encompasses both diffuse neutral clouds (cf. previous subsection) in the ambient galactic UV field, and dense molecular clouds exposed to the radiation of nearby young massive stars. This second kind of environment, commonly called dense PDRs, is the focus of this section. 

\subsubsection{Physical conditions in PDRs}

The physics and chemistry of PDRs is controlled by the local gas density $n$ and the intensity $G_0$ of the UV radiation field. 
In dense PDRs, the gas densities can range from $10^3\,\mathrm{cm}^{-3}$ to $10^7\,\mathrm{cm}^{-3}$, with UV intensities $G_0$ from a few units to the order of $10^6$ \citep{Hollenbach99}. The UV field intensities impinging on dense PDRs seem to be correlated to the gas density of the dense structures \citep{YoungOwl02}, and directly proportional to their thermal pressure \citep{chevance2016,Joblininprep}, hinting at the role of radiative feedback from the stars in the formation of these dense structures in star forming regions.


The typical physical conditions in dense PDRs are given in Table~\ref{physical_parameters}. As the large ranges of UV fields and gas densities result in large variations of the gas conditions and dust properties (which can affect the possible H$_2$ formation mechanisms at work) we separate low-illumination PDRs ($G_0<1000$) from high illumination PDRs ($G_0>1000$). 
Typical examples of low illumination PDRs are the Horsehead \citep{habart2005b}, and the $\rho$ Oph.  \citep{Habart03} PDRs. Famous high illumination PDRs include NGC7023~NW \citep{Kohler14, Joblininprep}, and the Orion Bar \citep{Parmar91,Joblininprep}. 


The UV field is progressively extinguished by dust, and by the molecules it dissociates, as it penetrates into the cloud. As a result, the physical conditions vary with the optical depth $A_V$ into the cloud. As H$_2$ dissociation occurs through line absorption, H$_2$ self-shielding is important and results in a sharp transition from atomic to molecular hydrogen (see \citealt{Sternberg14} and \citealt{Bialy16} for an analytical theory of the H/H$_2$ transition), the depth of this transition being highly dependent on the ratio of the gas density to UV intensity (the optical depth of the transition goes from a few $10^{-4}$ to $\sim 1$). The resulting PDR structure is layered, with a hot atomic layer before the H/H$_2$ transition, a warm molecular layer between the H/H$_2$ transition and the C$^+$/C/CO transition, and colder molecular gas deeper inside the cloud. This structure is represented in Fig.~\ref{main_H2formation_PDR}, with a more precise description of the physical conditions in the different layers of the PDR. 


\subsubsection{Dust evolutionary processes in PDRs}

Due to the strong UV fields and gas density variations associated with PDRs, the nature of the dust in these regions evolves from the edge of the PDR to its center, as well as varying from one PDR to another \citep[e.g.,][]{2005A&A...429..193R,Rapacioli06,Berne07,Pilleri12,arab2012,Arabthesis,Kohler14,pilleri2015}.

 Studies involving the whole dust emission spectrum, from the infrared to the (sub-)mm, indicate significant variations of the dust properties compared with that in the diffuse ISM \citep[e.g.,][]{arab2012,Arabthesis}.
 In particular, a decrease in the abundance of small polyaromatic rich carbon grains (denoted as PAHs) by at least a factor of 2 is very likely. This reduction could result from photo-destruction due to the strong UV radiative energy input. 
Moreover, \cite{Pilleri12} found that the fraction of carbon locked in very small grains, relative to the total carbon in the IR band carriers (i.e., very small grains and PAHs), decreased with increasing UV radiation field, which was interpreted as evidence for photo-destruction of very small grains.  
In PDRs, such as NGC 7023, \cite{pilleri2015} also provide evidence for a change in the aliphatic to aromatic composition, most likely due to the strong UV radiative energy input. This deduction could suggest that photo-processing of very small grains produces PAHs.

Deeper inside the PDRs, with a decreased UV field and an increased gas density, small grains must coagulate onto the surface of big grains. Compared with the diffuse ISM, an increase in the emissivity of the big grains indicates coagulation \citep[e.g.,][]{kohler2011,Arabthesis}. Indirect determinations of the dust extinction properties in some PDRs have also been reported (e.g. determination of $R_{V}=5.62$ in the NGC7023~NW PDR by \citealt{Witt06}), hinting at increased grain sizes compared to the diffuse ISM.
However, we must emphasize that  dust properties pertinent to the diffuse ISM (such as the \citealt{Draine07} and \citealt{Compiegne11} models) remain used in most PDR models.
 
 Models indicate that the grains remain bare until a visual extinction of $\gtrsim 3$ \citep{Hollenbach09,Esplugues16}, so that bare surfaces are most relevant for understanding H$_2$ formation in PDRs. 
 
 \subsubsection{Dust temperatures in PDRs and size distribution}

Dust grains in the PDR are heated by the UV field and their temperature decreases from the edge to the inner part of the cloud. The grain temperature remains significantly lower than the gas temperature in most PDRs. 
For large grains, temperatures derived from modified blackbody fits and radiative transfer model are given in Table~\ref{physical_parameters}.

In high illumination PDRs \citep[such as the Orion Bar,][]{arab2012}, dust temperatures are found to decrease from $70\,\mathrm{K}$ to $35\,\mathrm{K}$, with values of $50-60\,\mathrm{K}$ in the H/H$_2$ transition region (where H$_2$ emission lines peak). Similar gradients but with overall lower temperatures (due to lower UV fields) are found in NGC7023~NW ($50-25\,\mathrm{K}$ with $30\,\mathrm{K}$ at the H/H$_2$ transition, \citealt{Kohler14}), and in the Horsehead PDR ($30-13\,\mathrm{K}$, $20-30\,\mathrm{K}$ at the H/H$_2$ transition). 
 
 The dust size distributions in the ISM (including diffuse and dense PDRs) result in most of the available dust surface being associated with small grains. The small grains are thus potentially the most important contributors to H$_2$ formation, in particular in warm regions. The temperature of small grains has a more complex behavior than that of larger grains. Due to the presence of UV photons, the temperature of small dust grains fluctuates constantly (spikes of a few 100~K for a grain of dimension a few nm absorbing a 912 {\AA} photon). This effect was first investigated in order to understand the IR emission of PAHs and very small grains \citep{Desert86, Draine01}, which is dominated by these high temperature spikes. These temperature fluctuations can also significantly affect the efficiency of the different H$_2$ formation mechanisms in PDRs, as small grains spend a large fraction of their time at low temperatures ($10-20\,\mathrm{K}$) between UV photon absorptions, even at the warm edge of PDRs. Here then, surface processes can proceed between temperature spikes, significantly increasing the efficiency of the Langmuir-Hinshelwood mechanism for H$_2$ formation \citep{cuppen2006,Bron14}, or of ortho-para conversion of H$_2$ on grain surfaces \citep{2016A&A...588A..27B}.

\subsubsection{Discussion of the relevant or possible formation mechanisms in PDRs}

In PDRs, the Langmuir-Hinshelwood formation mechanism for H$_2$, where two H atoms physisorbed on a grain surface meet and react, does not appear to be effective on big  grains at thermal equilibrium; here, the low physisorption energies lead to very short residence times at the dominant temperatures of the dust in these regions. 
In order to allow formation of H$_2$ at higher surface temperatures, mechanisms that involve chemisorbed  H atoms have been proposed by several authors \citep[e.g.,][]{Cazaux2004,lebourlot2012}. This alternative formation process is usually modeled using the Eley-Rideal (ER) mechanism. Here,  H atoms bonded chemically to the surface (chemisorbed), are stationary on the surface until an H atom from the gas-phase interacts with them to form an H$_2$ molecule.

The ER mechanism, which requires high gas temperatures to allow H atoms to enter the chemisorbed state, is predicted to be  efficient only in high illumination PDRs \citep[$G_0 > 10^3$, e.g.][]{lebourlot2012}. 
\cite{boschman2015} model the influence of this ER route to H$_2$ formation from PAHs, on total H$_2$ formation rates in PDR. They find that the photodesorption of H$_2$ from PAHs can reproduce the high H$_2$ formation rates derived in moderately/highly excited PDRs \citep[]{Habart04}. Nevertheless, the presence of the thermal barrier to chemisorption, observed experimentally \citep[when hydrogenating coronene cations,][]{ boschman2012}, limits this process to high gas temperatures ($>$ 200 K). 
For  PDRs with low/intermediate excitation ($G_0 < 200$ and $n_\mathrm{H} > 10^3$ cm$^{-3}$), \cite{Bron14} show that the LH mechanism on small grains with fluctuating temperatures could be an efficient route to H$_2$ formation. 

Other processes involving chemisorbed H atoms, such as the photon-processing of grains, are also of great interest in PDRs.
As shown experimentally by \cite{2014A&A...569A.119A}, UV photon-irradiation of a-C:H leads  to very efficient production of H$_2$ molecules with rates similar to the ones derived in moderately/highly excited PDRs. 
In line with the interstellar evolution of carbonaceous dust, H$_2$ formation may occur via such UV photon-driven C-H and C-C bond dissociations in a-C:H (nano-)grains and the associated decomposition of those grains observed in PDRs.
This process could also liberate (the precursors to) species such as C$_2$H, C$_3$H$_2$, C$_3$H$^+$, C$_4$H, which have been observed in PDRs \citep{Pety2005,Pety2012,Guzman2015,Cuadrado2015,Pilleriinprep}.
\cite{jones2015} theoretically investigate this UV-induced H$_2$ formation pathway, adopting the dust composition and size distribution from the \cite{jones2013} dust model, which is specifically tuned to the evolution of the optical and thermal properties of a-C:H grains in the ISM. 
They conclude that such a process would be sustainable as long as the radiation field is intense enough to photo-dissociate C-H bonds but not intense enough to break the C-C bonds in the aliphatic bridging structure; the latter process would photo-fragment the a-C(:H) grains. Thus, this H$_2$ formation mechanism will be inefficient deep inside the cloud, because there are too few extreme ultraviolet (EUV) photons left. Neither will this process be important in intense radiation fields, because of the rapid photo-fragmentation of the grains. These ideas appear to be in general agreement with the H$_2$ observations presented in \cite{Habart04}, who suggest an enhanced H$_2$ formation rate in moderately-excited PDRs. Because the small grains ($a\sim$0.5-5 nm) dominate the dust surface and the C-H bond photo-dissociation efficiency decreases with dust size (other channels such as thermal excitation or fluorescence begin to compete), these grains make the largest contribution to the total H$_2$ formation rate. However, either fast rehydrogenation of the small grains or advection of unprocessed grains from deeper inside the cloud are necessary to sustain a steady H$_2$ formation rate. This aspect of the UV promoted process still has to be investigated in more detail, by going beyond the steady-state approach of the current main PDR models.
These possible H$_2$ formation processes are summarized in Fig.\ref{main_H2formation_PDR}. 

\subsubsection{Impacts of H$_2$ formation in PDR studies}

The efficiency of H$_2$ formation controls the depth within the cloud at which the dominant form of hydrogen changes from atomic to molecular. As the gas temperature decreases, as we go from the PDR edge to the inner part of the cloud, the H$_2$ formation efficiency thus affects the gas temperature at which molecular gas begins to appear, and hence the amount of warm molecular material in PDRs. The warm molecular layer immediately following the H/H$_2$ transition is crucial to explain several PDR tracers, such as the rotational emission of H$_2$ \citep{Habart04}, high-$J$ CO lines \citep{Joblininprep}, and abundances of CH$^+$ (and other species resulting from warm molecular chemistry). These tracers are then used to understand energy transfer (e.g. radiative vs. mechanical) in extragalactic studies (e.g. \citealt{Kamenetzky14, Rosenberg15}).
A precise understanding of the H$_2$ formation mechanisms at play is thus crucial for the interpretation of these tracers.

\subsection{Cold shielded ISM} 

Here we define the cold shielded interstellar medium as any medium characterized by low temperatures (below 15~K), medium species densities (a few $10^4$), and high visual extinction (i.e., absence of directly incident UV photons). Such regions, like for example cold dense cores as defined by \citet{2007ARA&A..45..339B}, make up one stage of the star formation process. There is no observational evidence of small grains \citep{2005A&A...429..193R,2016MNRAS.456.2290T}, while grain growth is suspected to occur early in the process of cold core formation \citep{2010Sci...329.1622P,2015A&A...582A..70S,2016A&A...588A..44Y}. 

Observations indicate a log-normal distribution of dust mass, with a peak in the distribution around 0.5-1 $\mu$m \citep{2015A&A...582A..70S,2016A&A...588A..44Y}. These grains are covered by mantles of molecules, mostly composed of H$_2$O \citep{2015ARA&A..53..541B}. 
Therefore, in these regions, H$_2$ formation proceeds mostly via physisorption (LH) pathways. Due to low gas temperature, the sticking coefficient is close to unity. The low temperature of the grains prevents H atom desorption, and H atom diffusion allows  efficient H$_2$ recombination. H$_2$ formed will progressively desorb as its coverage increases and the average binding energy decreases \citep{Amiaud:2006}. Due to the interactions with other adsorbed H$_2$ molecules, the desorbing hydrogen molecules are expected not to be vibrationally excited  \citep{Congiu:2009}.

H$_2$ cannot be directly observed in the cold shielded ISM, but it is very likely that most of the hydrogen is in the molecular form.  This is because the destruction of H$_2$ by secondary UV radiation, induced by cosmic rays, is readily compensated for by a high formation efficiency. From the chemical point of view, H$_2$ is the most abundant molecule in cold cores; however, the influence of molecular hydrogen on the  surface chemistry of the cores is limited. Most hydrogenation reactions on the surfaces involving H$_2$ have significant barriers so that, even though the abundance of H$_2$ is larger than that of atomic hydrogen, the reactions with H are considered to be much faster. There are a few of these reactions that are, however,  exothermic and may proceed via tunneling; for example,  H$_2$ + OH \citep{Oba:2012,2017arXiv170805559M}.

Recently, astrophysicists have modified gas-grain codes to take into account the competition between reaction and diffusion on the surfaces \citep{2011ApJ...735...15G,2016MNRAS.459.3756R}. This diffusion increases significantly the efficiency of reactions with activation barriers. A simple way to understand this effect is to consider that a species (the most mobile one) reaching a site on the surface which is already occupied by another species, will have a certain probability to remain associated with that species for some time, rather than leaving it instantaneously. During this time, the probability for a reaction with barrier to occur increases. With this new generation of models, the abundance of H$_2$ on the surface is important. 

The crucial parameter to determine the surface abundance of H$_2$ is not its rate of formation (which is very fast in the cold shielded ISM), it is rather its binding energy. The binding energy of H$_2$ on water ices is such that at high densities (above $10^9$ cm$^{-3}$ at 10~K), the molecules begin to get depleted from the gas phase and start to dominate the coverage of the grains \citep{2015A&A...574A..24H,2016A&A...594A..35W}. The sticking of H$_2$ onto multi-layers, i.e. on to itself, is much less efficient \citep{2007ApJ...668..294C} slowing down the depletion of H$_2$ from the gas. The details of the sticking of H$_2$ on the surface has to be carefully considered in gas-grain models because it can affect the general grain chemistry as for instance shown by \citet{2016A&A...594A..35W} in protoplanetary disks.\\
It has been proposed in the literature that the formation of H$_2$ on dust surfaces would be so exothermic that each reaction forming H$_2$ would locally heat the surface of the grains, allowing for the evaporation of some light species such as CO \citep{1993MNRAS.260...37D}. Such non-thermal evaporation processes have been added to several gas-grain models, and desorption by such a route is claimed to be more efficient than desorption by cosmic-ray heating  \citep{1993MNRAS.260...37D,1994MNRAS.267..949W,2007MNRAS.382..733R}. However, in contrast, recent experiments have reported a negative effect on desorption rates due to H$_2$ formation \citep{Minissale:2016}. Here the energy release upon H$_2$ formation changes the morphology of the ice \citep{2011PCCP...13.8037A}. Calculations have shown that H$_2$ formation can heat nanograins up to 53 K \citep{Navarro-Ruiz:2014}. Taking into account that grains in dark clouds are larger and that water ice efficiently dissipate excess energy, it is unlikely that H$_2$ formation triggers desorption of other species.


\subsection{Hot shielded regions} 

In the process of forming a star, cold cores evolve towards warm/hot shielded regions (also known as hot cores or hot corinos for massive and low mass protostars respectively) where the dust and gas temperature is a few hundred Kelvin and the density of the gas is above $10^7$~cm$^{-3}$. Gas and dust are inherited from the previous growth phase, so that hydrogen is almost entirely molecular and grains are big. The ices that formed earlier have evaporated, so that the grains are bare. The H$_2$ molecules cannot be observed in these regions, and their abundance in itself does not have much impact on the chemistry. The H$_2$ formation rate however determines the abundance of atomic hydrogen, which is involved in many destruction reactions with activation barriers \citep{2010ApJ...721.1570H}. 

Rate equation simulations including both physisorption and chemisorption, (\citet{Cazaux:2005}) have shown that H$_2$ formation on carbonaceous surfaces could be efficient up to about 1000~K, although this efficiency would decrease with the temperature. Up to 300~K, both Langmuir-Hinshelwood and Eley-Rideal (with chemisorbed H) mechanisms would contribute. At higher temperatures, the chemisorbed H starts to move and the Langmuir-Hinshelwood mechanism dominates. 

\citet{Iqbal:2012} performed a detailed study of the temperature dependent H$_2$ formation efficiency, using continuous-time random-walk Monte Carlo simulations on both silicate and carbonaceous surfaces. Their results show that the H$_2$ formation rate estimated by the rate equation method was overestimated because the mean abundance of H atoms at high temperature can be smaller than 1 H per dust grain \citep[see][for more details on this well known problem associated with rate equation models]{1998ApJ...495..309C, Cazaux:2005, cuppen2013}. Assuming deep chemisorption sites, present together with shallower physisorption sites, the H$_2$ formation remains efficient up to 700 K on both surfaces without the Eley-Rideal mechanism. These models have however only used low H fluxes, simulating ISM regions with densities smaller than 100~cm$^{-3}$. At higher H fluxes, the efficiency of H$_2$ formation may remain significant at even higher temperatures.



\subsection{Shocked environments}\label{shocks}

\subsubsection{Shocks in the ISM and their role in H$_2$ formation}

In the multiphase environment of galaxies, different types of shocks are expected. Fast shocks, ranging over $100-1000$~km~s$^{-1}$, heat the gas to high temperatures ($T \approx 10^6 - 10^8$~K), and produce EUV and X-ray photons that photo-ionize the tenuous medium \citep{Allen2008}. These shocks provide one source of energy that produce the Hot Ionized Medium in galaxies, and in the circum-galactic medium when the accreted gas from cosmological filaments is shocked at the halo boundary \citep{2003MNRAS.345..349B, 2016arXiv160904405C}. On the other hand, low-velocity shocks (typically $\lesssim 50$~km~s$^{-1}$) are known to \textit{(i)} initiate the formation of molecules in the gas that cools behind the shock \citep[e.g][]{Hollenbach1979}, and \textit{(ii)} be a very efficient process for exciting molecules, especially H$_2$, via collisions in the dense gas compressed and heated by the shock. Depending on the physical conditions in the shocked and post-shock gas, the different H$_2$ formation mechanisms discussed  above may occur.  

\cite{jenkins.peimbert1997} describe a steady change in the line profile as $J$ increases from 0 to 5 in high-resolution IMAPS\footnote{IMAPS, Interstellar Medium Absorption Profile Spectrograph, on board the ORFEUS-SPAS I mission, is the only instrument which recorded the far-UV H$_2$ bands at high spectral resolution with R $>$ 100\,000} observations of what they interpret as H$_2$ forming in the post-shock zone of originally atomic gas.
They suggest that the H$_2$ seen in absorption in the higher $J$ levels (which are broader and slightly shifted toward negative velocities relative to the lowest $J$ levels) is initially produced via the formation of a negative hydrogen ion H$^-$ in the warm (2000 K $< T <$ 6500 K) and partially ionized medium just behind the shock, through the reactions: 
\begin{equation}
\rm H + e^-  \longrightarrow  \rm H^- + h\nu\\ and \quad 
\rm H^- + H \longrightarrow \rm H_2 + e^-
\end{equation}
Later, in zones where  the post-shock gas has cooled well below 2000 K and is almost fully recombined, the formation of H$_2$ on grains dominates and is more easily seen in the lowest $J$ levels. 

\subsubsection{Gas conditions in shocks}

The typical physical conditions in shocks are given in Table~\ref{physical_parameters}. Here we have separated observations of Galactic and extragalactic shocked regions. In Galactic regions, most of the constraints on the  physical conditions in shocked molecular regions come from CO, H$_2$, SiO and H$_2$O observations in supernovae remnants \citep{Hewitt2009, 2010ApJ...724...69N, Gusdorf2012} and outflows from young stellar objects \citep[e.g.][]{2015A&A...575A..98G, Podio2015}. In extragalactic shocked regions, like AGN- or starburst-driven outflows or galaxy interactions, observations of H$_2$ line emission show a distribution of temperatures much above the equilibrium temperature set by UV and cosmic ray heating. The same holds, on much smaller scales, in the case of the diffuse ISM in the Solar neighborhood \citep{Gry2002,2011ApJ...743..174I}. The range of gas temperatures can only be accounted for if supersonic turbulence dissipates and heats the gas through spatially localized events \citep[e.g.][]{Falgarone1990, Hily-Blant2009, Guillard2015}.

The formation of H$_2$ in shocked regions greatly affects the gas cooling efficiency \citep{Cuppen2010}. For $J$-shocks, the cooling is first dominated by H$_2$ for a very short period of time, at high temperatures where the molecule experiences dissociation. Then, at $\approx 8000$~K, the [O$\,${\sc i}]$\lambda =6300\,$\AA~line takes over. When the gas has cooled down sufficiently, H$_2$, H$_2$O, O and C$^+$ dominate the cooling. More precisely, at $T \approx 1000$~K, the fine-structure line [O$\,${\sc i}]$\lambda =63.2\,\mu$m dominates. Other important lines in this temperature range are [C$\,${\sc i}], [N$\,${\sc i}]$\lambda =5200\,$\AA, [C$\,${\sc ii}]$\lambda =158\,\mu$m, [N$\,${\sc ii}]$\lambda =121.8\,\mu$m, [S$\,${\sc ii}]$\lambda =34.82\,\mu$m and [Fe$\,${\sc ii}]$\lambda =25.99\,\mu$m. At lower temperatures, the oxygen is converted into CO, H$_2$O, and OH, which become the dominant coolants. 
For $C$-shocks, H$_2$ is the main coolant for $T > 300$~K, which provides  excitation of the mid-IR rotational lines. Remarkably, the cooling is dominated by H$_2$O emission at $T < 200$~K \citep{Flower2010}.

\subsubsection{Dust processing in shocks and impact on H$_2$ formation}
\label{subsubsec:shocks_dust}

Grain processing in shocks affects the formation of H$_2$ and arises from interactions between the grain and other ``particles'' (grain, atom, ion, atomic nucleus, electron, photon). These interactions can be \textit{destructive}, i.e. leading to  a transfer of grain atoms to the gas-phase, and thus to a net mass loss from the grain, or \textit{non-destructive} where the grain size distribution becomes modified but the total dust mass is conserved. These processes are counterbalanced by grain formation, specifically accretion and nucleation, but also coagulation between grains.

Destructive interactions can  directly eject atoms and ions from the grain surface (via \textit{sputtering}, ion field emission\footnote{direct expulsion of atoms and ions from the surface at extreme surface irregularities \citep{Draine1979}} or direct Coulomb explosion via extreme charging effects arising from electron-grain interactions or the photoelectric effect), or vaporize/sublime the grain, in case of $\gtrsim 20$~km~s$^{-1}$ grain-grain collisions.  Vaporisation/sublimation via absorption of energetic photons (UV, $\gamma$) or interaction with cosmic rays, may also be an important destruction mechanism for volatile material, such as ice mantles. 

Grain-grain collisions at velocities $\gtrsim 1-2$~km~s$^{-1}$ lead to non-destructive \textit{fragmentation} (or \textit{shattering}) of both grains \citep{Tielens1994}. The size distribution of the fragments follows a power-law, suggesting that dust fragmentation in shocks is responsible for the observed power-law size-distribution   in the diffuse ISM \citep{Jones1996, Weingartner2001, Clayton2003}. By definition, fragmentation is not a destructive process. It re-distributes the dust mass towards smaller grain sizes, increasing the total cross-section of the distribution of grains, and hence their optical and UV extinction. 

Dust grains, through gas-grain interactions, have an important impact on the physical and chemical properties of the shock. The grains participate in the formation of molecules and their removal, and the inertia of the grains can modify the propagation of the shock, especially in the case of C-shocks. A detailed review of the processes involving grains in shocks, as well as their impact on the structure of C-shocks, can be found in \citep{Flower2003a}. For detailed calculations of the evolution of the dust size distribution in MHD shocks, see \citet{2011A&A...527A.123G}.

\subsubsection{Impact of H$_2$ formation on shock dynamics}

In shocks, chemistry and hydrodynamics are coupled. Chemical processes, in particular H$_2$ formation, have a very important impact on the shock structure. H$_2$ formation plays a role in the thermal balance of the shock, which in turn impacts the number of reactions the shock can initiate and the consequent reaction rates. Without chemistry, the ionization fraction would vary only via the differential compression of the ionized and neutral fluids by the shock wave, and would have the same value in the pre-shock and post-shock regions. Allowing for chemistry, the ionization degree of the post-shock region is one order of magnitude lower. H$_2$ affects the ionization fraction because it essentially initiates the neutralization of C$^+$ via the reaction:
\begin{equation}
\rm C^+ + H_2  \longrightarrow  \rm CH^+ + H 
\end{equation}
followed by a chain of reactions, which balance overall to yield
 \begin{equation}
\rm CH^+ + e^-  \longrightarrow  \rm C + H
\end{equation}
Consequently, the ion-neutral coupling is weaker, hence broadening the size of the precursor (the width of the shock is $\approx 5$ times larger).
Since the energy of the shock is dissipated over a larger region, as compared to the hypothetical case without chemistry, the maximum temperature of the shock is lower \citep{PineaudesForets1997}.

\subsection{Mixed regions}

Pure shocked regions are difficult to isolate in Galactic and extragalactic sources \citep[e.g.][]{2011ApJ...743..174I, Guillard2009}. In addition to the shock excitation, a background UV field or cosmic rays are often present, which affect the  thermal and chemical characteristics of the gas, in particular H$_2$ formation. 
For instance, protostars generate a strong ultraviolet radiation field that ionizes their surroundings.  This field drives powerful shock waves in the neighboring medium in the form of jets and bipolar outflows, whose structure can be partially organized by a local, strong magnetic field. Such an ejection activity locally modifies the interstellar chemistry, contributing to the cycle of matter. Excitation of H$_2$ gas in galaxies is therefore a mix of processes, and the degeneracy of models makes the identification of the dominant processes a challenge for modern astrophysics \citep[e.g.][]{Appleton2013}.  

The treatment of irradiation, which includes UV heating but also its impact on the chemistry through photo-ionization and photo-dissociation, is crucial in shock models to interpret complex regions where a mix of excitation is observed \citep{Lesaffre2013}. 
Preliminary results comparing steady-state H$_2$ abundances with and without radiative transfer and shielding effects in high-resolution MHD simulations seem to show very important differences in H$_2$ abundances\citep{hartwig2015, richings2014}. When the protecting effect of the shielding is not taken into account, most of the molecular hydrogen is photo-dissociated and only about 3\% of the total mass of H$_2$ (compared to the case that includes shielding effects) survives in the densest regions where the formation is efficient enough to counterbalance the photo-destruction  \citep {valdiviaIAU2017inprep}.
%
More generally, (magneto)-hydrodynamical simulations of the ISM which includes coupling with radiation and radiative transfer effects will be the subject of intense developments in the next years.   

\section{Conclusions and perspectives}

In this paper, we present an overview of where we stand on the question: How does H$_2$ form in the interstellar medium? This report of our understanding of this fundamental scientific problem, of course, is moulded by  the opinions and experience of the authors of this paper.  These opinions were distilled by gathering together these scientists who are working in different fields of research which impinge on interstellar H$_2$ formation: experimental and theoretical physico-chemistry, observers and modelers of the interstellar medium.   \\

As far as chemical physics is concerned, the physisorption (LH) mechanism for H$_2$ formation has been extensively studied, and with some exceptions, conclusions have converged. Amorphous carbonaceous surfaces have been the least studied, and are probably the richest and most complex materials to investigate. When it comes to H$_2$ formation involving chemisorbed atoms (ER), molecular synthesis by this pathway has been investigated in much less detail ; specifically, no experiments have been performed on silicates although there are calculations that address this process and should be tested with experiments. Formation of molecular hydrogen on graphite surfaces (an idealized representation of an interstellar dust grain) has been well investigated both experimentally and theoretically.  This detailed investigation is probably a unique situation, and even then marginal discrepancies can still be found between different studies, especially for the case of computational work.  The characterization of the ro-vibrational excitation of nascent H$_2$ leaving from the surface of the dust grain has been performed for formation on graphite. These studies can guide searches for observational signatures of H$_2$ formation in space and, hence, should be extended over a wider range of environmental parameters and surface types. \\

In calculations of the dynamics of the formation process, the sticking coefficient of H atoms on a graphitic surface is difficult to determine. The current calculations have developed models to take into account energy exchange (via phonons) with the surface during the collision, either for the physisorption site or for the chemisorption site. These models are limited in not just the number of degrees of freedom they consider, but also the limited range of geometries investigated and the fact that the surface is modeled as ``perfect''. The challenge is to include, in the same model, physisorption sites and chemisorption sites when calculating the sticking coefficient of an H atom on a graphitic surface.\\

More generally, the coupling between physisorption and chemisorption is probably the most challenging to study, and the most pressing problem with which both experimentalists and quantum physicists are battling. It is noteworthy that most of the studies discussed above have focused on only one of these two formation mechanisms.  However,  the fact that H$_2$ is formed in a wide variety of astrophysical environments suggests that both these mechanisms  operate in parallel.
Furthermore, there is a need to transform the results of laboratory studies of surface processes (sticking, diffusion, desorption, energy partition, etc.) to the efficiency of formation of molecular hydrogen in actual ISM environments. This has been done to a certain extent for silicate surfaces, but more work is needed both on silicates and carbonaceous materials. 

New routes to H$_2$ formation, via UV irradiation of hydrogenated carbons, or via other energetic routes, have to be investigated in greater detail. It has been demonstrated that hydrogenated carbons can  efficiently produce H$_2$ following energetic events, but after this carbonaceous grains need to be resupplied with H atoms. This "refilling mechanism" is complex and probably couples diffusion via physisorbed states and storage via chemisorbed states. The existence of these potential H$_2$ reservoirs underlines the need to studying the links between physisorption and chemisorption on disordered materials.  In the specific case of astrophysical shocks,  the interaction of  high temperature gas with cold surfaces, and the impact of this interaction on H$_2$ formation, also still needs to be studied. 

Finally, on several relevant surfaces many aspects of the H$_2$ formation process, such as nuclear spin conversion and energy partitioning, remain open questions. The observed trends, revealed mostly by the study of H interaction with graphite, have to be confirmed. In the case of LH processes, the substrate retains a minor part of the energy released by H$_2$ formation.

For the chemical modeling of cold cores, the amount of H$_2$ on the surfaces can have a big impact on the  abundances of other species. Thus, the study of H$_2$ formation in the presence of co-adsorbed H$_2$ (with perhaps also H$_2$O, CO, CO$_2$) by experimentalists would be important.  Such experiments would perhaps reveal  the possibility of trapping nascent H$_2$ in ices.

In the near future, JWST (launch foreseen in late 2018) will allow medium resolution ($\lambda / \Delta \lambda \approx 3000$) spectroscopy of H$_2$ lines, with an increase of two orders of magnitude in spatial resolution and sensitivity compared to ISO or Spitzer. This mission will provide a fantastic set of new data which will present stringent tests of our current understanding of H$_2$ in space. The gain in sensitivity and spatial resolution (0.3'' at 10 $\mu$m) will generate high quality maps of the H$_2$ rotational and ro-vibrational transitions. A wealth of moderately/highly excited $J$ levels will be detectable for the first time. Combined with data probing the ro-vibrational levels, these observations will cover a huge range of excitation energies, making them excellent probes of the excitation of H$_2$ and the associated gas. 

In the nearby Galactic sources, the critical H/H$_2$ transition zones, that could be very sharp because of the H$_2$ self-shielding and strong gas density gradients, will be resolved via the H$_2$ ro-vibrational and rotational emission line profiles. This will constrain the possible H$_2$ formation mechanisms (ER, LH on small grains fluctuating in temperatures or UV processes) and ortho-to-para ratios associated with H$_2$ formation in a warm gas. The wavelength coverage of JWST will allow us to study in detail the links between H$_2$ and the properties of the small dust particles. How rapidly the dust properties change at the cloud edge, via photo-processing, and how these changes impact the H$_2$ formation process, are just two of the  key questions that JWST will address. 

Being the first instruments able to spatially and spectrally resolve H$_2$, and forbidden lines of the ionized gas, at rest-frame near-IR and mid-IR wavelengths out to $z = 1.5 - 3.5$, the JWST spectrometers will also allow the study of H$_2$ line emission in the distant Universe. These spectrometers will allow us to investigate the physical state, and the kinematics, of the warm ($> 50$~K) molecular gas.  these observations will constrain the impact of galaxy mergers and AGN feedback on the formation and evolution of massive galaxies. In $z \sim 2$ active star-forming galaxies, high gas velocity dispersions \citep[e.g.][]{Lehnert2009} suggest high turbulence, and we can expect bright H$_2$ emission \citep{Guillard2009,Guillard2015a}, as suggested by H$_2$ detections in $z \approx 2$ infrared-luminous galaxies \citep{Fiolet2010, Ogle2012}. Identifying the source of turbulence in high-redshift sources is a key step forward for studies of galaxy evolution.

The spectrometer EXES on SOFIA can probe several H$_2$ rotational lines with diffraction limited resolution of 1-5'' and a spectral resolution of $R\sim 10^4 - 10^5$ at 20 $\mu$m. This is the only instrument with sufficient spectral resolution and sensitivity to measure a pure rotational H$_2$ line profile from a relatively faint source. 
Studies, like that reported by \cite{jenkins.peimbert1997}, emphasize the interest of observing H$_2$ at high spectral resolution in order to study the velocity profiles of the different H$_2$ rotational excitation levels ($J$) to, for example:
\begin{enumerate}
\item Isolate components or cloud regions with different rotational excitation.
\item Determine the role of internal spatial structures in the self-shielding  of diffuse H$_2$ that strongly influences the HI $\rightarrow$ H$_2$ transition in galaxies.
\item Study the increase in velocity dispersion with $J$. Specifically, can we confirm the creation of H$_2$ in warm, partly-ionized post-shock zones proceeds via the formation of H$^-$?
\item	Measure the temperature and turbulence of H$_2$ in high-$J$ levels.  Is the excess  abundance in $J > 2$ in diffuse clouds due to shocks, or dissipation of turbulence in vortices?
\item	Find what fractions of the 4.5 eV of energy released upon H$_2$ formation are transferred to internal excitation of the molecule, kinetic motion of the molecule, and the thermal bath of the grain.
\end{enumerate}
The possibility of observing the H$_2$ electronic bands at high spectral resolution in absorption in the UV might be opened (in the distant future) by the LUVOIR project.  LUVOIR is one of four Decadal Survey Mission Concept Studies initiated by NASA in 2016, which should include a high-resolution UV spectrometer. This spectrometer would allow to observe the far-UV electronic bands of H$_2$ at high resolution (R $>$ 120\,000) toward many stars  in our Galaxy, as well as towards individual stars in the  Local Group galaxies. Such a capability would allow us to extend the study of H$_2$ formation to very different environments.
\paragraph{Acknowledgements}
All authors thank the EPOC lab, the LAB, and the Institute of Advanced Study of the University of Cergy Pontoise for their support during the meeting that catalyzed the writing of this manuscript.
VW's research is funded by an ERC Starting Grant (3DICE, grant agreement 336474). GV acknowledges financial support from the National Science Foundation's Astronomy \&
Astrophysics Division (Grants No. 1311958 and 1615897). LH acknowledges support from ERC Consolidator Grant GRANN (grant agreement no. 648551). GN acknowledges support from the Swedish Research Council. VW, FD and SM acknowledge the CNRS program "Physique et Chimie du Milieu Interstellaire" (PCMI) co-funded by the Centre National d'Etudes Spatiales (CNES).
SDP acknowledges funding from STFC, UK. FD thanks VW for accepting the difficult task of managing this project and SDP for his efforts in reading and re-writing the text. V.V acknowledges funding from the European Research Council (ERC) under the European UnionÕs Horizon 2020 research and innovation programme (MagneticYSOS project, grant agreement No 679937).

{\small
 \begin{landscape}
\begin{table*}
   \caption{Physical parameters of each type of astrophysical source.}
   \label{physical_parameters}
   \begin{center}
   \begin{tabular}{lccccccc}
    Source & $G_0$ (Habing unit) & n$_{\rm H}$ (cm$^{-3}$) & Gas T (K) & Dust T (K) & f$_{\rm H_2}$ & H$_2$ o/p & R$_{\rm H_2}$ ($\times 10^{-17}$) \\
    Dense PDR (HI) 	&  a few $10^3$ - $10^4$ $^{(a)}$ & $10^4$ - $10^6$ $^{(a,b,c)}$ & 10 - 3000 $^{(a)}$ & 25 - 70 $^{(g)}$ & 0 - 1 & 1 - 3 $^{(k,l)}$ & (3 - 15) $^{(r)}$ \\
    Dense PDR (LI) & a few 10 - 100 $^{(a)}$ & $10^3$ - $10^5$ $^{(d,e)}$ &  10 - 400 $^{(d)}$ &  13 - 30 $^{(g)}$ & 0 - 1 & 1 - 3 $^{(m,l,e)}$ & (10 - 30) $^{(r)}$ \\
    Diffuse clouds  & 1 & 1-100 & 50 - 130 & 20 $^{(h)}$ & $10^{-4} - 0.8$ $^{(j)}$ & 0.4 - 0.9 $^{(n,o)}$ & (3 - 5) $^{(s,n,o)}$ \\
    Cold Cores & 1 & $10^4$ - $10^5$ $^{(f)}$& 10 $^{(f)}$ & 6 - 15 $^{(i)}$ & 1 & 0 - 0.1 $^{(p,q)}$ & ? \\
    Hot Cores $^{(t)}$ &  1 & $> 10^6$ & 100 - 300 & 100 - 300 & ? & ? & ? \\
    Shocks (Galactic) & 1 - 10$^{(u)}$ & $10^3 - 10^4$ $^{(v)}$& 10 - $10^4$ & ? & 0.2 - 1 & 0.5 - 3 $^{(w)}$& ? \\
    Shocks (extra-Galactic) & 0 - 100$^{(x)}$& 10 - $10^5$ & 10 - $10^5$ & ? & $10^{-5}$ - 1 $^{(y)}$ & 1 - 3 $^{(z)}$ & ? \\
         \hline
   \end{tabular}
   \end{center}
   \medskip
   References: (a) \citet{Hollenbach99}, (b) \citet{Parmar91}, (c) \citet{Kohler14}, (d) \citet{habart2005b}, (e) \citet{habart2011}, (f) \citet{2007ARA&A..45..339B}, (g) \citet{arab2012}, (h) \citet{2014A&A...571A..11P}, (i) \citet{hocuk2016}, (j) \citet{2008ApJ...688.1124S},
   (k) \citet{fuente99}, (l) \citet{fleming2010}, (m) \citet{Habart03}, (n) \citet{Gry2002}, (o) \citet{2005ApJ...627..251L}, (p) \citet{2009A&A...506.1243T}, (q) \citet{2009A&A...494..623P}, 
   (r) \citet{Habart04}, (s) \citet{1974ApJ...191..375J}, (t) \citet{2009ARA&A..47..427H}, (u) \citet{Lesaffre2013}, (v), \citet{2015A&A...575A..98G}, (w) \citet{2010ApJ...724...69N}, (x) \citet{Guillard2009,Guillard2012} and \citet{2016ApJ...832..209U}, (y) \citet{2003MNRAS.341...70F}, (z) \citet{1998MNRAS.297..624H}.
\end{table*}
\end{landscape}
}




  \bibliographystyle{elsarticle-harv} 
\bibliography{complete_bib.bib}





\end{document}